\documentclass[11pt,a4paper]{article}

\usepackage[truedimen,margin=30mm]{geometry} 

\usepackage{mathrsfs}
\usepackage{amssymb}
\usepackage{amsmath}
\usepackage{ascmac}
\usepackage{amsthm}
\usepackage[pdftex]{graphicx}
\usepackage[pdftex]{color}
\usepackage{natbib}
\usepackage{setspace}
\usepackage{color}
\usepackage{url}
\usepackage{multirow}

\usepackage{titlesec}
\titleformat*{\section}{\large\bfseries}
\titleformat*{\subsection}{\it}

\newtheorem{theorem}{Theorem}
\newtheorem{lemma}{Lemma}

\newtheorem{algo}{Algorithm}
\newtheorem{assumption}{Assumption}
\newcommand{\argmin}{\mathop{\rm argmin}\limits}

\def\a{{\text{\boldmath $a$}}}
\def\A{{\text{\boldmath $A$}}}
\def\B{{\text{\boldmath $B$}}}
\def\D{{\text{\boldmath $D$}}}
\def\e{{\text{\boldmath $e$}}}
\def\G{{\text{\boldmath $G$}}}
\def\H{{\text{\boldmath $H$}}}
\def\I{{\text{\boldmath $I$}}}
\def\M{{\text{\boldmath $M$}}}
\def\R{{\text{\boldmath $R$}}}
\def\S{{\text{\boldmath $S$}}}
\def\U{{\text{\boldmath $U$}}}
\def\V{{\text{\boldmath $V$}}}
\def\x{{\text{\boldmath $x$}}}
\def\X{{\text{\boldmath $X$}}}
\def\y{{\text{\boldmath $y$}}}

\def\al{{\alpha}}
\def\be{{\beta}}
\def\ga{{\gamma}}
\def\de{{\delta}}
\def\ep{{\varepsilon}}
\def\la{{\lambda}}
\def\si{{\sigma}}

\def\th{{\theta}}

\def\bal{{\text{\boldmath $\alpha$}}}
\def\bbe{{\text{\boldmath $\beta$}}}
\def\bga{{\text{\boldmath $\gamma$}}}
\def\bep{{\text{\boldmath $\varepsilon$}}}
\def\bla{{\text{\boldmath $\lambda$}}}

\def\beh{{\widehat \be}}

\def\muh{{\widehat \mu}}

\def\balh{{\widehat \bal}}
\def\bbeh{{\widehat \bbe}}

\def\De{{\Delta}}
\def\Si{{\Sigma}}
\def\Ga{{\Gamma}}

\def\La{{\Lambda}}
\def\bDe{{\text{\boldmath $\De$}}}
\def\bde{{\text{\boldmath $\de$}}}
\def\bSi{{\text{\boldmath $\Si$}}}
\def\bLa{{\text{\boldmath $\La$}}}

\def\ph{{\widehat p}}

\def\Bc{{\cal B}}

\def\tr{{\rm tr}}
\def\Var{{\rm Var}}

\def\one{{\rm I}}

\def\gh{\widehat g}

\def\Zt{\widetilde Z}

\def\Ds{\mathscr D}

\def\Rr{\mathbb R}
\def\ept{\widetilde \varepsilon}

\def\bDs{{\text{\boldmath $\Ds$}}}

\def\zero{{\bf\text{\boldmath $0$}}}
\def\one{{\bf\text{\boldmath $1$}}}

\title{{\bf Grouped Generalized Estimating Equations for Longitudinal Data Analysis}}
\date{}
\author{}

\begin{document}

\maketitle
\doublespacing

\vspace{-1.5cm}
\begin{center}
{\large 
Tsubasa Ito$^1$ and Shonosuke Sugasawa$^2$
}
\end{center}

\noindent
$^1$Faculty of Economics and Business, Hokkaido University\\
$^2$Center for Spatial Information Science, The University of Tokyo

\medskip
\begin{center}
{\bf \large Abstract}
\end{center}

\vspace{-0cm}
Generalized estimating equation (GEE) is widely adopted for regression modeling for longitudinal data, taking account of potential correlations within the same subjects. Although the standard GEE assumes common regression coefficients among all the subjects, such an assumption may not be realistic when there is potential heterogeneity in regression coefficients among subjects. In this paper, we develop a flexible and interpretable approach, called grouped GEE analysis, to modeling longitudinal data with allowing heterogeneity in regression coefficients. The proposed method assumes that the subjects are divided into a finite number of groups and subjects within the same group share the same regression coefficient. We provide a simple algorithm for grouping subjects and estimating the regression coefficients simultaneously, and show the asymptotic properties of the proposed estimator. The number of groups can be determined by the cross-validation with averaging method. We demonstrate the proposed method through simulation studies and an application to a real dataset.


\newpage
\section{Introduction}
\label{int}

Longitudinal data where response variables (repeated measurements) within the same subject are correlated widely appears in biomedical studies. 
For analyzing longitudinal data, it is typically difficult to correctly specify the underlying correlation structures among response variables within the same subject, and one of the standard approaches is the generalized estimating equations (GEE) developed by \cite{LZ1986}, which uses ``working" correlation structures specified by users.
The advantage of the GEE approach is that the estimator is still consistent even when the working correlation is misspecified. 
However, the existing GEE methods assume homogeneous regression coefficients that are common to all the subjects, which could be restrictive in practical applications since there might be potential heterogeneity among subjects or clusters, as confirmed in several applications \citep{BB2012,Lin2012,Nagin2018}.
To address such heterogeneity, a crude approach is to apply a model separately to each subject, but the results are typically inaccurate and unstable due to small subject-wise sample sizes as often arise in real longitudinal data.  
Therefore, some compromised approach is required.

In this work, we extend the standard GEE analysis to take into account potential heterogeneity in longitudinal data.
Specifically, we develop grouped GEE analysis by adopting the grouping approach that is widely adopted in literature for panel data analysis \cite{BM2015,Liu2020,ZWZ2019}.
We assume that subjects in longitudinal data can be classified into a finite number of groups, and subjects within the same group share the same regression coefficients; that is, the regression coefficients are homogeneous over subjects in the same groups. 
Since the grouping assignment of subjects is unknown, we treat it as unknown parameters and estimate them and the group-wise regression coefficient simultaneously. 
Given the grouping parameters, the standard GEE can be performed to obtain group-wise estimators of regression coefficients. 
On the other hand, given the group-wise regression coefficients, we consider estimating the grouping parameters using a kind of Mahalanobis distance between response variables and predictors with taking account of potential correlations via a working correlation matrix.
In other words, we employ the working correlation not only in performing GEE analysis in each group but also in estimating the grouping assignment.
We will show that the grouped GEE method can be easily carried out by a simple iterative algorithm similar to the {\it $k$-means} algorithm that combines the existing algorithm for the standard GEE and simple optimization steps for grouping assignment. 
Moreover, we adopt the cross-validation
with the averaging method proposed in \cite{Wang2010} to carry out a data-dependent selection of the number of groups.

We derive the statistical properties of the grouped GEE estimator in an asymptotic framework where both $n$ (the number of subjects) and $T$ (the number of repeated measurements) tend to infinity, but we here allow $T$ to grow considerably slower than $n$, namely, $n/T^\nu\to0$ for some $\nu>0$.
Hence, our method can be applied when $T$ is much smaller than $n$ as observed in many applications using longitudinal data. 
As theoretical difficulties of the grouped estimation in longitudinal data analysis, the true correlations within the same subject can be considerably high, so the existing theoretical argument assuming negligibly small correlations imposed typically by mixing conditions \cite{BM2015,GV2019,ZWZ2019} for the true underlying correlations are no more applicable.
To overcome the limitation of the existing theoretical argument, we consider grouping assignment using a kind of Mahalanobis distance with working correlation. We will show that such a grouping strategy leads to the consistent estimation of the grouping parameters as long as the working correlation is reasonably close to the true one.
Therefore, even when the underlying correlations within the same subject are not weak, we can successfully estimate the grouping parameters using a reasonable working correlation matrix.  
Then, we will establish consistency and asymptotic normality of the grouped GEE estimator of the regression coefficients and provide a consistent estimator of asymptotic variances.

In the context of longitudinal data or clustered data analysis, several methods to take account of the potential heterogeneity among subjects have been proposed. 
\cite{Ng2014,Rubin1997,Sugasawa2019,Sun2007} proposed a mixture modeling based on random effects, but the estimation algorithms can be computationally very intensive since the algorithms include iteration steps that entail numerical integration.  
On the other hand, \cite{Rosen2000,Tang2016} proposed a mixture modeling based on the GEE, but the primary interest in these works is estimating the component distributions in the mixture rather than grouping subjects. 
\cite{Fokkema2018,Hajjem2011,Hajjem2017} employed regression tree techniques for grouping observations, but the tree-based methods can handle grouping based on covariate information rather than regression coefficients. 
Moreover, \cite{Coffey2014,Vogt2017,ZQ2018} proposed grouping methods for longitudinal curves, and \cite{Tang2020} developed covariate-specific grouping methods via regularization.
Lastly, \cite{Zhu2018} is similar to our work, which proposed the GEE-type loss functions penalizing pairwise distance of heterogeneous fixed effects, but computational cost rapidly becomes much larger as the sample size increases compared to the $k$-means method.
To the best of our knowledge, this paper is the first one to consider grouped estimation in the GEE analysis by the $k$-means algorithm with a quite small computational burden.

This paper is organized as follows.
In Section \ref{sec:gee}, we illustrate the proposed GEE analysis and provide an iterative estimation algorithm.
We also propose the averaging method for selecting the number of groups. 
In Section \ref{sec:ap}, we give the asymptotic properties of the grouped GEE estimator.
In Section \ref{sec:num}, we demonstrate the grouped GEE analysis through simulation studies and an application to a real longitudinal dataset. 
We give some discussions in Section \ref{sec:dis}.
All the technical details and the proofs of the theorems, additional numerical results, and data analyses are provided in the Supporting Information.
R code implementing the proposed method is available at Github repository (\url{https://github.com/sshonosuke/GGEE}).

\section{Grouped GEE Analysis}\label{sec:gee}

\subsection{Grouped models for longitudinal data}
For longitudinal data, let $y_{it}$ be the response of interest and $\x_{it}$ be a $p$-dimensional vector of covariate information of subject $i$ at time $t$, where $i=1,\ldots,n$ and $t=1,\ldots,T_i$.
For ease of notation, we set $T_i=T$ for all $i$, representing a balanced data case, but the extension to an unbalanced case is straightforward.
We consider a generalized linear model for $y_{it}$, given by 
\begin{equation}
f(y_{it}|\x_{it}; \bbe_i,\phi)=\exp\Big[\{y_{it}\th_{it}-a(\th_{it})+b(y_{it})\}/\phi\Big],
\label{eqn:md}
\end{equation}
where $a(\cdot)$ and $b(\cdot)$ are known functions, and $\th_{it}=u(\x_{it}^\top\bbe_i)$ for a known monotone function $u(\cdot)$.
A commonly used link function is the canonical link function, that is, $u(x)=x$.
Here $\bbe_i$ is the regression parameter of interest that can be heterogeneous among subjects, and $\phi$ is a known scale parameter common to all subjects.
Under the model (\ref{eqn:md}), the first two moments of $y_{it}$ are given by $m(\x_{it}^\top\bbe_i)=a'(\th_{it})$ and $\si^2(\x_{it}^\top\bbe_i)=a''(\th_{it})\phi$, respectively.
For example, under binary response, it follows that $a(x)=\log\{1+\exp(x)\}$, leading to the logistic model given by $m(\x_{it}^\top\bbe_i)=\{1+\exp(-\x_{it}^\top\bbe_i)\}^{-1}$.

In the standard GEE analysis, the regression parameters are homogeneous, that is, $\bbe_i=\bbe$, but we allow potential heterogeneity among the subjects. 
However, the number of $\bbe_i$ increases with the number of subjects, so $\bbe_i$ cannot be estimated with reasonable accuracy as long as $T$ is not large, which is the standard situation in longitudinal data analysis. 
Hence, we consider a grouped structure for the subjects, that is, the $n$ subjects are divided into $G$ groups, and subjects within the same group share the same regression coefficients. 
Specifically, we introduce an unknown grouping variable $g_i\in \{1,\ldots,G\}$ which determines the group that $i$th subject belongs to.
Then, we define $\bbe_i=\bbe_{g_i}$ under which the unknown regression parameters are $\bbe_1,\ldots,\bbe_G$.
Therefore, if $G$ is not large compared with $n$ and $T$, then $\bbe_1,\ldots,\bbe_G$ can be accurately estimated.
Moreover, due to the grouping nature, the estimation results of $g_i$ give grouping of subjects in terms of regression coefficients, so the estimation result is easily interpretable for users. 
We also treat $G$ as an unknown parameter, but we assume that $G$ is known for a while. 
The estimation will be discussed in Section \ref{sec:eng}.

\subsection{Estimation algorithm}\label{sec:alg}
Define $\y_i=(y_{i1},\ldots,y_{iT})^\top$ as a $T$-dimensional response vector, $\X_i=(\x_{i1},\ldots,\x_{iT})^\top$ as a $
T\times p$ covariate matrix.
We also define $m(\X_i\bbe_g)=(m(\x_{i1}^\top\bbe_g),\ldots,m(\x_{iT}^\top\bbe_g))^\top$, $\A_i(\bbe_g)={\rm diag}(\si^2(\x_{i1}^\top\bbe_g),\ldots,\si^2(\x_{iT}^\top\bbe_g))$, $\bDe_i(\bbe_g)={\rm diag}(u'(\x_{i1}^\top\bbe_g),\ldots,u'(\x_{iT}^\top\bbe_g))$, where ${\rm diag}(\a)$ is a diagonal matrix with a vector $\a$ as the diagonal elements, and $\D_i(\bbe_g)=\A_i(\bbe_g)\bDe_i(\bbe_g)\X_i$.
In what follows, we might abbreviate the explicit dependence on the parameters for notational simplicity when there seems to be no confusion.
We here introduce ``working" correlation matrix $\R(\bal)$ to approximate the true underlying correlation matrix of $\y_i$, which is 
assumed to be common across different subjects for simplicity.
This assumption can be easily extended to the heterogeneous correlation structures among different subjects.
The working correlation matrix can be chosen freely, where it might include the nuisance unknown parameter $\bal$.
Then, we define working covariance matrix $\V_i(\bbe)$ as $\V_i(\bbe)=\A_i^{1/2}(\bbe)\widehat\R \A_i^{1/2}(\bbe)$ with $\widehat\R=\R(\widehat\bal)$.
If $\widehat\R$ is consistent to the true correlation matrix $\R^0$, $\V_i(\bbe^0)$ with the true parameter $\bbe^0$ is also consistent to the true covariance matrix of $\y_i$.

Given the grouping parameter $\bga=(g_1,\ldots,g_n)$, we can estimate $\bbe_g$ by performing the standard GEE estimation \cite{LZ1986} for each group, namely, solving the following estimating equation:
\begin{align}\label{eqn:gee}
\begin{split}
&\S_g(\bbe_g)\equiv\sum_{i=1}^n \one(g_i=g)\S_i(\bbe_g)=\zero, \\
&{\rm s.t} \quad \S_i(\be_g) \equiv \D_i^\top(\bbe_g)\V_i^{-1}(\bbe_g)\{\y_i-m(\X_i\bbe_g)\},
\end{split}
\end{align} 
which is the GEE based on the subjects classified to the $g$th group. 
We can employ an existing numerical algorithm for the standard GEE to obtain the solution of (\ref{eqn:gee}).
On the other hand, given $\bbe=(\bbe_1^\top,\ldots,\bbe_G^\top)^\top$, it is quite reasonable to classify the subjects into groups having the most suitable regression structures to explain the variation of $\y_i$.
Thus, we propose estimating the unknown $\bga$ based on the following minimization problem: 
\begin{align}\label{eqn:cc}
\gh_i(\bbe)=\argmin_{g=1,\ldots,G} \ \ \{\y_i-m(\X_i\bbe_g)\}^\top\widehat\R^{-1}\{\y_i-m(\X_i\bbe_g)\}.
\end{align}
The objective function in (\ref{eqn:cc}) can be seen as a kind of the Mahalanobis distance with taking the working correlation structure into account.
Such estimation strategy for the grouping variable has not been paid attention to very much, but the use of the working correlation in the grouping step is shown to be quite important to expand our theoretical argument given in Section \ref{sec:ap}.
Note that the above minimization problem can be carried out separately for each subject; thus (\ref{eqn:cc}) can be easily solved by simply evaluating all the values of the objective function over $g\in\{1,\ldots,G\}$.

Regarding the estimation of the nuisance parameter $\bal$ in the working correlation, we suggest using a moment-based method.
Given $\bbe$ and $\bga$, one can estimate $\bal$ by solving the following minimization problem:  
\begin{align}\label{eqn:alpha}
\widehat\bal(\bbe,\bga)= \argmin_\bal \Big\| \R(\bal)-\frac1n\sum_{i=1}^n\A_i^{-1/2}\{\y_i-m(\X_i\bbe_{g_i})\}\{\y_i-m(\X_i\bbe_{g_i})\}^\top \A_i^{-1/2}\Big\|_F,
\end{align}
where $\|\cdot\|_F$ is the Frobenius norm. 
This method can be easily extended to the heterogeneous correlation structures among different groups.
Let $\bal_1,\ldots,\bal_G$ be different correlation parameters.
Then, $\bal_g$ can be estimated by minimizing 
\begin{align}
\Big\| \R(\bal_g)-n_g^{-1}\sum_{i=1}^n \one(g_i=g)\A_i^{-1/2}\{\y_i-m(\X_i\bbe_{g_i})\}\{\y_i-m(\X_i\bbe_{g_i})\}^\top \A_i^{-1/2}\Big\|_F,\nonumber
\end{align}
where $n_g$ is the number of subjects classified to the $g$th group.

The estimating equation (\ref{eqn:gee}) and two optimization problems (\ref{eqn:cc}) and (\ref{eqn:alpha}) define the grouped GEE estimator of $\bbe$ and $\bga$, and the estimator can be easily computed by the following iterative algorithm:

\begin{algo}[grouped GEE estimation] \ \ \\
Starting from some initial values $\bbe^{(0)}$, $\bga^{(0)}$ and $\bal^{(0)}$, we repeat the following procedure until algorithm converges: 
\begin{itemize}
\item[-]
Update $\bga^{(r)}$ to get $\bga^{(r+1)}$ by solving (\ref{eqn:cc}) with $\bbe=\bbe^{(r)}$ and $\bal=\bal^{(r)}$.

\item[-]
Update $\bbe^{(r)}$ to get $\bbe^{(r+1)}$ by solving (\ref{eqn:gee}) with $\bga=\bga^{(r+1)}$ and $\bal=\bal^{(r)}$.

\item[-]
Update $\bal^{(r)}$ to get $\bal^{(r+1)}$ by solving (\ref{eqn:alpha}) with $\bbe=\bbe^{(r+1)}$ and $\bga=\bga^{(r+1)}$.
\end{itemize}
\end{algo}

Since there might be multiple solutions for the grouped GEE estimator, the above algorithm might be sensitive to the setting of initial values.
A reasonable starting value for $\bal$ would induce an independent correlation matrix of $\R$, for example, $\bal=\zero$ in the exchangeable working correlation.
Regarding $\bbe$ and $\bga$, we suggest two simple methods to determine their initial values. 
First method is to apply the finite mixture models with $G$ components of the form: $y_{it}|(z_{it}=k)\sim h_k(y_{it}; \x_{it}^\top\bbe_k)$ and $P(z_{it}=k)=\pi_k$, for $k=1,\ldots,G$, where $h_k$ is the distribution having mean $m(\x_{it}^\top\bbe)$.
Then, we set the initial values of $\bbe_k$ and $g_i$ to the estimates of $\bbe_k$ and the maximizer of $\sum_{t=1}^{T}P_{\ast}(z_{it}=k)$ over $k\in \{1,\ldots,G\}$, respectively, where $P_{\ast}(z_{it}=k)$ is the conditional probability that $y_{it}$ belongs to the $k$th group. 
The second approach is separately fitting the regression model with mean structure $m(\x_{it}^\top\bbe_i)$ for each subject.
Based on the estimates $\bbeh_i$ of $\bbe_i$, we apply the $k$-means clustering algorithm with $G$ clusters to the $n$-points $\{\bbeh_1,\ldots,\bbeh_n\}$, and set the initial values of $\bbe_k$ and $g_i$ to the center of the resulting clusters and clustering assignment, respectively. 
Note that the second method is only applicable when $T$ is sufficiently larger than $p$ to get stable estimates of $\bbe_i$.

\subsection{Selecting the number of groups}\label{sec:eng}
Since the number of groups is typically unknown in practice, we need to estimate it based on appropriate criteria.
One possible strategy is to adopt a criterion using quasi-likelihood \cite{Wedd1974} and to use a penalty term in view of Bayesian-type information criterion in GEE analysis \cite{WQ2009}.
However, the theoretical asymptotic properties of such approaches are not necessarily clear even under the standard GEE settings so that the theoretical investigation would be more complicated under the grouping structure.  
Instead, we here adopt the cross-validation with averaging method (CVA) proposed in \cite{Wang2010}, which is shown to have the selection consistency when the clusters are properly separated into subgroups.
The same strategy is adopted in \cite{ZWZ2019} in the context of quantile regression for panel data.

The CVA criterion is concerned with clustering instability under given $G$.
For $c=1,\ldots,C$, we randomly divide $n$ subjects into three subsets: two training datasets with sizes $M$ and one testing set with size $n-2M$, where the subject indices included in the three subsets are denoted by $Z_1^c, Z_2^c$ and $Z_3^c$, respectively, that is, $|Z_1^c|=|Z_2^c|=M$, $|Z_3^c|=n-2M$, $Z_h^c \cap Z_{h'}^c=\emptyset$ for $h\neq h'$ and $\cup_{h=1,2,3} Z_h^c=\{1,\ldots,n\}$.
We first apply the proposed grouped GEE method to the two training datasets, which gives us the estimates of regression coefficients and working correlation matrices. 
Then, we can compute the optimal grouping assignment in the test data as 
$$
\gh_i^{(h)}=\argmin_{g=1,\ldots,G} \{\y_i-m(\X_i\bbeh_g^{(h)})\}^\top \{\widehat\R^{(h)}\}^{-1}\{\y_i-m(\X_i\bbeh_g^{(h)})\}, \ \ \ \ i\in Z_3^c,
$$
where $\bbeh_g^{(h)}$ and $\widehat\R^{(h)}$ are estimates of regression coefficients and working correlation based on the $h$th training data for $h=1,2$.  
Based on the grouping assignment, grouping instability can be quantified as 
$$
{\widehat s}^c(G)=\sum_{i,j\in Z_3^c} \one\Big\{\one(\gh_i^{(1)}=\gh_j^{(1)})+\one(\gh_i^{(2)}=\gh_j^{(2)})=1\Big\},
$$
since the summand of ${\widehat s}^c(G)$ takes the value $1$ when the $i$th and $j$th subjects in the testing set are classified into the same group if we use the estimators based on one training data, but they are classified into the different groups if we use the estimators based on the other training data, which implies that the grouping results are more unstable as ${\widehat s}^c(G)$ is large.
By averaging the above values over $c=1,\ldots,C$, we have ${\widehat s}(G)=C^{-1}\sum_{c=1}^C{\widehat s}^c(G)$, and we select $G$ as the minimizer of the criterion among some candidates of $G$. 
Finally, regarding the choice of $M$, we set $M=\lfloor n/3 \rfloor$, so the three subsets have almost the same numbers of subjects.


\section{Asymptotic Properties}\label{sec:ap}

We here provide the asymptotic properties of the grouped GEE estimators, that is, the grouping parameter $\bga$ can be consistently estimated, and $\bbeh_g$ admits both consistency and asymptotic normality. 
Our asymptotic framework is that both $n$ and $T$ tend to infinity, but we allow $T$ to grow considerably slower than $n$, as discussed later.

We first prepare some notations before assumptions.
Let $\M_g(\bbe_g)={\rm Cov}(\S_g(\bbe_g))=\sum_{i=1}^n\one\{g_i=g\}\D_i^\top \V_i^{-1}\bSi_i(\bbe_g)\V_i^{-1}\D_i$ and $\H_g(\bbe_g)=-{\rm E}[\partial \S_g(\bbe_g)/\partial\bbe_g]=\sum_{i=1}^n\one\{g_i=g\}\D_i^\top \V_i^{-1}\D_i$.
We here denote the working correlation matrix as $\R(\bal,\bbe,\bga)$ to emphasize its dependence on $\bal$, $\bbe$ and $\bga$, and let $\widehat\R(\bbe,\bga)=\R(\balh,\bbe,\bga)$.
We also let $\overline\R(\bbe,\bga)=\R(\overline{\bal},\bbe,\bga)$ be a constant positive definite matrix, where $\overline{\al}$ is a nonrandom constant to which $\balh$ converges.
We do not require $\overline\R(\bbe,\bga)$ to be the true correlation matrix $\R^0$.
Next, we denote $\overline\V_i(\bbe_g)$ by replacing $\widehat\R(\bbe,\bga)$ with $\overline\R(\bbe,\bga)$ in $\V_i(\bbe_g)$.
$\overline\S_i(\bbe_g)$, $\overline\S_g(\bbe_g)$, $\overline\M_g(\bbe_g)$ and $\overline\H_g(\bbe_g)$ are defined similarly.
To facilitate the Taylor expansion of the estimating function of GEE, we denote the negative gradient function of $\S_i(\bbe_g)$ as $\bDs_i(\bbe_g)=-{\partial \S_i(\bbe_g)}/{\partial\bbe_g^\top}$.
$\overline\bDs_i(\bbe_g)$ is defined as $\overline\V_i(\bbe_g)$.
For $g=1,\ldots,G$, let $\bbe_g^0$ be a true value of $\bbe_g$ and $g_i^0$ be a group variable which $i$th cluster actually belongs to.
Then, we also define the oracle score function for $\bbe_g$ under the true grouping assignment as $\S_g^{*}(\bbe_g)=\sum_{i=1}^n \one\{g_i^0=g\}\S_i(\bbe_g)$.
$\overline\S_g^{*}(\bbe_g)$, $\overline\M_g^{*}(\bbe_g)$ and $\overline\H_g^{*}(\bbe_g)$ are similarly defined.
As discussed in \cite{XY2003}, to prove the existence and weak consistency of the clustered GEE estimators, we need assumptions given later in Assumption (A3), that is, for all $g=1,\ldots,G$, $\overline\H_g^{*}(\bbe_g^0)$'s or $\la_{\min}(\overline\H^{*})\equiv\min_{1\leq g\leq G}\la_{\min}(\overline\H_g^{*}(\bbe_g^0))$ are divergent at a rate faster than $\tau\equiv\sup_{\bbe,\bga}\la_{\max}(\{\overline\R(\bbe,\bga)\}^{-1}\R^0)$.
To make further assumptions, we need to introduce some notations similar to those in \cite{Wang2011,XY2003}. 
We denote a local neighborhood of $\bbe^0=(\bbe_1^{0\top},\ldots,\bbe_G^{0\top})^\top$ as $\Bc_{nT}=\{\bbe=(\bbe_1^\top,\ldots,\bbe_G^\top)^\top : \max_{g=1,\ldots,G}||\{\overline\H_g^{*}(\bbe_g^0)\}^{1/2}(\bbe_g-\bbe_g^0)||\leq C\tau^{1/2}\}$.
Lastly, we denote $\ep_{it}=A_{it}^{-1/2}(\bbe_{g_i^0}^0)\{y_{it}-m(\x_{it}^\top\bbe_{g_i^0}^0)\}$ and $\bep_i=(\ep_{i1},\ldots,\ep_{iT})^\top$ for all $i=1,\ldots,n$ and $t=1,\ldots,T$.

We here give some regularity assumptions, and the other technical assumptions are given in Supporting Information Section S.1.

\begin{assumption}
\ \par
\noindent
(A1)
(i) For all $g=1,\ldots,G$, the unknown parameter $\bbe_g$ belongs to a compact subset $\Bc\in\Rr^p$, the true parameter value $\bbe_g^0$ lies in the interior of $\Bc$, 
(ii) the covariates $\{\x_{it}, i=1,\ldots,n, t=1,\ldots,T\}$ are in a compact set $\mathcal X$.

\noindent
(A2)
(i) For all $g=1,\ldots,G$, $\lim_{n\to\infty}(1/n)\sum_{i=1}^n \one\{g_i^0=g\}=\pi_g>0$ and (ii) for all $g,g'=1,\ldots,G$ such that $g\neq g'$ and $c>0$, $\min_{1\leq g,g' \leq G}||\bbe_g^0-\bbe_{g'}^0||>c$.

\noindent
(A3)
$\tau\la_{\min}^{-1}(\overline\H^{*})\to0$.

\noindent
(A4)
For all $i=1,\ldots,n$ and $t=1,\ldots,T$, $E[\ep_{it}^{2+2/\zeta}]\leq M$ for some $0<\zeta\leq1$.

\noindent
(A5)
The eigenvalues of the true correlation matrix $\R^0$ are bounded away from $0$, and the eigenvalues of $\overline\R(\bbe,\bga)$ are bounded away from 0 uniformly for any $\bbe$ and $\bga$.
All off-diagonal elements of $\R^0$ are uniformly bounded away from $1$.
\end{assumption}

Assumption (A1) seems to be slightly strict.
However, the compactness of the parameter space and the set of all possible covariates is required because in the proof of the consistency of our grouped GEE estimators, we need to bound $a''(\th_{it})$ and $u_{it}'(\eta_{it})$ uniformly on the whole parameter space for all $i=1,\ldots,n$ and $t=1,\ldots,T$.
Assumption (A2) is typically imposed in the literature on the grouping approach in panel data models (\cite{BM2015}), which ensures that the $G$ subgroups are well separated so that the parameters $\bbe_g$'s and $\bga$ can be identifiable.
Assumption (A3) is the same as the condition (L*) in \cite{XY2003}.
Assumption (A4) is slightly stronger than the condition in Lemma 2 of \cite{XY2003} since we require the fourth moment of error terms to be finite.
Assumption (A5) is the same assumption imposed well in the literature on GEE with large cluster sizes.
Assumption (A5) is a much weaker assumption than the one typically adopted in the existing literature on the grouped estimation \cite{BM2015,GV2019,ZWZ2019} in which $\{\ep_{it}\}_{t=1,\ldots,T}$ is assumed to satisfy some strong mixing conditions with a faster-than-polynomial decay rate.
Such assumptions are quite unrealistic in longitudinal data analysis, so we do not impose any restriction on the correlation strength of $\{\ep_{it}\}_{t=1,\ldots,T}$, which is essentially related to the use of a kind of Mahalanobis distance for grouping assignment given in (\ref{eqn:cc}).
Moreover, since we assume that true correlations are uniformly bounded away from $1$, we can estimate each $\bbe_i$ consistently by solving $\S_i(\bbe_i)=\zero$ from Assumptions in Supporting Information Section S.1, as argued in \cite{XY2003}.

We now give our main theorems.
We first establish the existence and weak consistency of the grouped GEE estimators and the classification consistency of the grouping variables.

\begin{theorem}\label{thm:cons}
Suppose the Assumptions (A1)-(A5) and the Assumptions in Supporting Information Section S.1 hold.
For all $g=1,\ldots,G$, $\S_g(\bbe_g)=\zero$ has a root $\bbeh_g$ such that $\bbeh_g\to\bbe_g^0$ in probability.
Moreover, as $n$ and $T$ tend to infinity such that $n/T^\nu\to0$ for some $\nu>0$, it holds that $P(\max_{1\leq i\leq n}|\gh_i(\bbeh)-g_i^0|>0)=o(1)+O(nT^{-\de})$ for all $\de>0$ for $\gh_i(\bbeh)$'s are obtained by (\ref{eqn:cc}).
\end{theorem}
Since the second part of Theorem \ref{thm:cons} holds for all $\de>0$, the probability of miss-clustering vanishes if we take $\de$ larger than $\nu$ in Assumption (A9) (iv) in Supporting Information Section S.1.

We next establish the asymptotic normality of $\bbeh_g$ for $g=1,\ldots,G$.
The following notations are similar to \cite{XY2003}: $c^*=\max_{1\leq g\leq G}\la_{\max}(\overline\M_g^{*}(\bbe_g^0)^{-1}\overline\H_g^{*}(\bbe_g^0))$ and
\begin{align*}
\ga^*=\max_{1\leq g\leq G}\max_{i:g_i^0=g}\la_{\max}(\overline\H_g^{*}(\bbe_g^0)^{-1/2}\D_i^\top(\bbe_g^0)\overline\V_i^{-1}(\bbe_g^0)\D_i(\bbe_g^0)\overline\H_g^{*}(\bbe_g^0)^{-1/2}).
\end{align*}
The following result is a direct consequence of Theorem 4 in \cite{XY2003} combined with Lemma S.9 in Supporting Information Section S.1.

\begin{theorem}\label{thm:an}
Suppose the Assumptions (A1)-(A5) and the Assumptions in Supporting Information Section S.1 hold.
Moreover, suppose that, for all $g=1,\ldots,G$, there exists a constant $\zeta$ such that $(c^*T)^{1+\zeta}\ga^*\to0$ as $n\to\infty$.
Moreover, suppose the marginal distribution of each observation has a density of the form from (\ref{eqn:md}).
Then, as $n$ and $T$ tend to infinity such that $n/T^\nu\to0$ for some $\nu>0$, we have $\overline\M_g^*(\bbe_g^0)^{-1/2}\overline\H_g^*(\bbe_g^0)(\bbeh_g-\bbe_g^0)\to N(0,\I_p)$ in distribution.
\end{theorem}

From Theorem \ref{thm:an}, it can be easily shown that for all $n$, $\{\overline\H_g^{*}(\bbe_g^0)\}^{-1}\overline\M_g^{*}(\bbe_g^0)\{\overline\H_g^{*}(\bbe_g^0)\}^{-1}$ is minimized in the matrix sense when $\overline\V_i=\bSi_i$ for all $i$.
This implies that the group GEE estimator becomes most efficient when we can specify the working correlation matrix correctly, and the corresponding asymptotic variance of $\bbeh_g$ is given by 
$\lim_{n\to\infty}[\sum_{i=1}^n\one\{g_i^0=g\}\D_i^\top \bSi_i^{-1}\D_i]^{-1}$.

Moreover, $\{\H_g(\bbeh_g)\}^{-1}\M_g(\bbeh_g)\{\H_g(\bbeh_g)\}^{-1}$ can be used as the estimator of the asymptotic variance of $\bbeh_g$.
Since this estimator of the asymptotic variance of $\bbeh_g$ involves $\M_g(\bbeh_g)$ depending on the unknown covariance matrix $\widehat\bSi_i=\bSi_i(\bbeh_{\gh_i})$ for $\gh_i=g$, following \cite{LZ1986}, we suggest obtaining $\M_g(\bbeh_g)$ by
\begin{align*}
\sum_{i=1}^n \one(\gh_i=g) \D_i^\top(\bbe_g)\V_i^{-1}(\bbe_g)\{\y_i-m(\X_i\bbe_g)\}\{\y_i-m(\X_i\bbe_g)\}^\top \V_i^{-1}(\bbe_g)\D_i(\bbe_g)\Big|_{\bbe_g=\bbeh_g},
\end{align*}
which is consistent to $\overline\M_g^*(\bbe_g^0)$ as $n\to\infty$ from Lemma 1 in Supporting Information Section S.2.
Similarly, we can show that $\H_g(\bbeh_g)$ is consistent to $\overline\H_g^*(\bbe_g^0)$, which implies that $\{\H_g(\bbeh_g)\}^{-1}\M_g(\bbeh_g)\{\H_g(\bbeh_g)\}^{-1}$ converges to the asymptotic variance of $\bbeh_g$.
Although the variability in the estimation of grouping parameters can be ignored according to Theorem \ref{thm:cons} and \ref{thm:an}, it can be considerable under finite sample sizes. 
As an alternative method, we also suggest using clustered bootstrap \citep[e.g.][]{field2007}. 
This approach generates the bootstrap sample $y_1^{\ast},\ldots,y_n^{\ast}$ from the distribution placing probability $1/n$ on each of $y_i=(y_{i1},\ldots,y_{iT})$.
Letting $\bbeh_g^{\ast}$ be the estimator obtained from the bootstrap sample $y_1^{\ast},\ldots,y_n^{\ast}$, the asymptotic variance of $\bbeh_g$ can be approximated by the sample variance of replications of $\bbeh_g^{\ast}$.

\section{Simulation studies}\label{sec:num}

We investigate the finite sample performance of the proposed grouped GEE method through simulation studies. 
First, we consider the estimation and classification accuracy of the grouped GEE estimator.
To this end, we generated two dimensional covariate vector $(x_{1it}, x_{2it})$ form a two-dimensional normal distribution with mean $\zero$, marginal variance $1$ and correlation $0.4$, for $i=1,\ldots,n$ and $t=1,\ldots,T$.  
We considered the logistic model for the marginal expectation of $Y_{it}$, namely, $Y_{it}\sim {\rm Ber}(\pi_{it})$ and ${\rm logit}(\pi_{it})=\X_{it}^\top\bbe_{g_i}$, where $\X_{it}=(1, x_{1it}, x_{2it})^\top$, $g_i\in \{1,\ldots,G\}$ and $\bbe_g=(\beta_{g0}, \beta_{g1}, \beta_{g2})^\top$ is a vector of unknown regression coefficients.
Here we set $G=3$ and $\bbe_1=(0, -2, 0)^\top$, $\bbe_2=(1, 1, 2)^\top$ and $\bbe_3=(-1, 1, -2)^\top$.
For the true grouping assignment, we set $g_i=1$ for $i=1,\ldots,n/3$, $g_2=2$ for $i=n/3+1,\ldots,2n/3$ and $g_i=3$ for $i=2n/3+1,\ldots,n$.
Based on the probability $\pi_{it}$, we generated $(Y_{i1},\ldots,Y_{iT})$ from a correlated binary vector using R package ``bindata" with two scenarios of correlation matrix, exchangeable correlation matrix with $0.5$ correlation parameter, and AR(1) correlation matrix with $0.7$ correlation parameter. 
We then applied the proposed grouped GEE method with $G=3$ and four options of correlation matrices, independent (ID), exchangeable correlation (EX), AR(1) correlation (AR), and unstructured correlation (US) matrices, and unknown parameters in these correlation matrices were also estimated. 
For comparison, we also applied the naive grouping (NG) method that first separately fits the logistic regression to each subject to estimate subject-specific regression coefficients, then group them via $k$-means clustering and re-estimate group-wise regression coefficients.

We evaluated the performance of the estimation of $\bbe_g$ by using the squared error loss defined as ${\rm SEL}_g=\sum_{k=0}^2(\beh_{gk}-\be_{gk})^2$, and assessed the classification accuracy via the classification error given by ${\rm CE}=n^{-1}\sum_{i=1}^n\one(\gh_i\neq g_i)$.
In Tables \ref{tab:beta} and \ref{tab:g}, we reported the average values of SEL and CE using 5000 Monte Carlo replications, respectively, under four combinations of $(n, T)$.
\begin{table}[htbp]
\caption{Average values of squared error loss of the regression coefficients in three groups based on the proposed grouped GEE method with independent (ID), exchangeable correlation (EX), first-order autoregressive (AR), and unstructured (US) working correlation matrices. 
The results of the naive grouping (NG) method using the subject-wise estimation of regression coefficients are also given for comparison. 
The reported values are averaged over 5000 Monte Carlo replications and are multiplied by 100.}
\label{tab:beta}
\begin{center}
\begin{tabular}{ccccccccccccccccc}
\hline
&&& \multicolumn{5}{c}{true correlation: EX} & & \multicolumn{5}{c}{true correlation: AR}\\ 
$(n, T)$ & Group & & ID & EX & AR & US & NG & & ID & EX & AR & US & NG \\
\hline
 & 1 & & 8.8 & 7.8 & 9.0 & 8.7 & 12.9 & & 7.9 & 7.4 & 7.2 & 7.3 & 10.2 \\
$(180,10)$ & 2 & & 9.3 & 8.3 & 9.1 & 8.6 & 12.7 & & 8.2 & 7.6 & 7.5 & 7.4 & 10.1 \\
 & 3 & & 9.3 & 7.8 & 9.3 & 8.7 & 12.7 & & 8.0 & 7.4 & 7.6 & 7.7 & 10.3 \\
 \hline
 & 1 & & 4.4 & 3.7 & 4.4 & 5.0 & 5.0 & & 3.1 & 2.9 & 2.8 & 3.2 & 3.1 \\
$(180,20)$ & 2 & & 4.3 & 3.8 & 4.3 & 5.1 & 4.9 & & 3.1 & 2.9 & 2.8 & 3.2 & 3.2 \\
 & 3 & & 4.3 & 3.8 & 4.4 & 5.3 & 5.0 & & 3.1 & 3.0 & 2.8 & 3.3 & 3.0 \\
 \hline
 & 1 & & 6.5 & 5.2 & 6.2 & 5.6 & 10.3 & & 5.8 & 5.0 & 5.0 & 4.7 & 7.4 \\
$(270,10)$ & 2 & & 6.4 & 5.4 & 6.1 & 5.6 & 10.1 & & 5.7 & 4.9 & 4.8 & 5.0 & 7.5 \\
 & 3 & & 6.8 & 5.2 & 6.3 & 5.5 & 10.4 & & 5.7 & 4.8 & 4.8 & 4.8 & 7.5 \\
 \hline
 & 1 & & 2.9 & 2.5 & 2.8 & 3.0 & 3.4 & & 2.1 & 1.9 & 1.8 & 2.0 & 2.1 \\
$(270,20)$ & 2 & & 2.9 & 2.5 & 2.9 & 2.9 & 3.4 & & 2.0 & 2.0 & 1.9 & 1.9 & 2.0 \\
 & 3 & & 2.8 & 2.5 & 2.9 & 3.0 & 3.4 & & 2.1 & 2.0 & 1.9 & 2.0 & 2.1 \\
\hline
\end{tabular}
\end{center}
\end{table}

\begin{table}[htbp]
\caption{Average values of classification error ($\%$) of the grouping parameters in the grouped GEE analysis with independent (ID), exchangeable correlation (EX), first-order autoregressive (AR) and unstructured (US) working correlation matrices, averaged over 5000 Monte Carlo replications. }
\label{tab:g}
\begin{center}
\begin{tabular}{ccccccccccccc}
\hline
&& \multicolumn{4}{c}{true correlation: EX} & & \multicolumn{4}{c}{true correlation: AR}\\
$(n, T)$ & & ID & EX & AR & US & & ID & EX & AR & US \\
 \hline
$(180, 10)$ & & 9.6 & 4.4 & 6.6 & 5.3 & & 6.5 & 4.8 & 4.0 & 4.8 \\
$(180, 20)$ & & 4.3 & 1.5 & 2.3 & 1.8 & & 1.9 & 1.6 & 1.2 & 1.5 \\
$(270, 10)$ & & 8.5 & 4.3 & 6.0 & 4.9 & & 6.1 & 4.6 & 4.0 & 4.4 \\
$(270, 20)$ & & 3.7 & 1.5 & 2.1 & 1.4 & & 1.8 & 1.4 & 1.3 & 1.4 \\
\hline
\end{tabular}
\end{center}
\end{table}

From Table \ref{tab:beta}, we can see that the correct specification of working correlation matrices induces the most efficient estimation of the regression coefficient. In contrast, using the other working correlations that are not necessarily equal to the true correlation structures can still provide a more efficient estimation than the independent working structure.
We also note that the US working correlation includes both EX and AR, although the number of unknown parameters is much larger than these structures. 
Hence, the estimation performance under the moderate sample size such as $(n,T)=(180, 10)$ is not very satisfactory, but the performance improves as the sample size increases.  
Regarding NG, the performance is comparable when $T$ is not small (e.g., $T=20$), while the performance gets worse as $T$ decreases.
This would be because the subject-wise fitting does not perform well when $T$ is not large, leading to poor grouping results.   
From Table \ref{tab:g}, it is observed that introducing working correlation structures in the classification step (\ref{eqn:cc}) achieves a more accurate classification than the common classification strategy using the standard sum of squared residuals as adopted in existing literature when observations within the same subject are correlated. 
Moreover, the results reveal that the correct specification of the working correlation leads to the most accurate classification. 
In Supplementary Material, we provide simulation results for $95\%$ confidence intervals of $\beta_1, \beta_2$ and $\beta_3$.

We next investigate the performance of the CVA selection strategy given in Section \ref{sec:eng} by adopting the same data generating process with an exchangeable correlation structure.
For the simulated dataset, we selected the number of components $G$ using the CVA criteria from the candidate $G\in \{2, 3, \ldots, 7\}$, noting that the true number of components is $3$.
We employed four working correlations, ID, EX, AR and US, to carry out the grouped GEE analysis for each $G$.
Based on Monte Carlo replications, we obtained selection probabilities of each $G$, which are reported in Table \ref{tab:sel}.
\begin{table}[htbp]
\caption{Selection probabilities ($\%$) of the number of groups ($G$) obtained from the CVA criteria in Section \ref{sec:eng} with independent (ID), exchangeable (EX), first-order autoregressive (AR) and unstructured (US) working correlation matrices, based on 200 Monte Carlo replications. }
\label{tab:sel}
\begin{center}
\begin{tabular}{ccccccccccccc}
\hline
 & working & & \multicolumn{6}{c}{$G$}\\
$(n, T)$ & correlation & & 2 & 3 & 4 & 5 & 6 & 7 \\
 \hline
\multirow{4}{*}{$(180,10)$} & ID & & 0.5 & 61.0 & 8.0 & 10.0 & 3.5 & 17.0 \\
 & EX & & 3.0 & 95.0 & 0.5 & 1.0 & 0.5 & 0.0 \\
 & AR & & 3.0 & 78.0 & 5.0 & 7.5 & 2.5 & 4.0 \\
 & US & & 0.0 & 89.5 & 2.0 & 2.5 & 1.0 & 5.0 \\
 \hline
\multirow{4}{*}{$(180,20)$} & ID & & 0.0 & 94.0 & 3.0 & 1.0 & 0.5 & 1.5 \\
 & EX & & 0.0 & 100.0 & 0.0 & 0.0 & 0.0 & 0.0 \\
 & AR & & 0.0 & 100.0 & 0.0 & 0.0 & 0.0 & 0.0 \\
 & US & & 0.0 & 93.5 & 5.5 & 0.0 & 0.0 & 1.0 \\
 \hline
\multirow{4}{*}{$(270,10)$} & ID & & 3.0 & 77.5 & 3.5 & 9.5 & 0.5 & 6.0 \\
 & EX & & 2.0 & 98.0 & 0.0 & 0.0 & 0.0 & 0.0 \\
 & AR & & 2.0 & 97.0 & 0.0 & 0.5 & 0.0 & 0.5 \\
 & US & & 0.5 & 98.5 & 0.5 & 0.0 & 0.5 & 0.0 \\
 \hline
\multirow{4}{*}{$(270,20)$} & ID & & 0.0 & 96.5 & 1.0 & 1.5 & 0.0 & 1.0 \\
 & EX & & 0.5 & 99.5 & 0.0 & 0.0 & 0.0 & 0.0 \\
 & AR & & 0.0 & 100.0 & 0.0 & 0.0 & 0.0 & 0.0 \\
 & US & & 0.0 & 100.0 & 0.0 & 0.0 & 0.0 & 0.0 \\
\hline
\end{tabular}
\end{center}
\end{table}  
It is observed that the use of independent working correlations under significant correlations within the same individual does not necessarily provide satisfactory selection performance when the number of samples is limited. 
We can also see that the selection probabilities of the true number of components based on EX and US working correlations tend to be larger than those of using the AR working correlation structure since the true correlation is EX.
Moreover, when the sample sizes are large, such as $(n, T)=(270,20)$, the adopted CVA strategy can select the true number of components with a probability of almost 1, which would be compatible with the selection consistency of the strategy.

Finally, we compare the proposed grouped GEE method with some existing methods under situations where the subjects do not necessarily admit perfect grouping. 
To this end, we considered the following underlying scenarios for the subject-specific regression coefficients: 
\begin{align*}
&{\rm (S1)} \ \ \bbe_i\sim \pi_1\bde(0, -2, 0)+\pi_2\bde(1,1,2) + \pi_3\bde(-1,1,-2), \ \ \ \ \pi_1=\pi_2=\pi_3=\frac13\\
&{\rm (S2)} \ \ \bbe_i=(0, -2, 0)\one(g_i=1)+(1,1,2) \one(g_i=2) + (-1,1,-2)\one(g_i=3) + U([-0.5, 0.5]^3)\\
&{\rm (S3)} \ \ \bbe_{i0}\sim U([-0.2, 0.2]), \ \ \beta_{i1}\sim U([-2,2]) , \ \ \ \beta_{i2}\sim U([0,2]),
\end{align*}
where $\bde(a_1,a_2,a_3)$ denotes a Dirac distribution on $(a_1,a_2,a_3)$, $U(A)$ denotes the uniform distribution on the region $A$, and $g_i$ is the grouping variable defined as $g_i=1$ for $i=1,\ldots,n/3$, $g_2=2$ for $i=n/3+1,\ldots,2n/3$ and $g_i=3$ for $i=2n/3+1,\ldots,n$.
Note that scenario (S1) is quite similar to the one used in the previous simulation study.
On the other hand, in scenarios (S2) and (S3), the subjects do not admit complete classification since the regression coefficients are different among subjects. 
We also note that in scenario (S2), the subjects may admit approximate classification based on $g_i$, but there seems to be no trivial classification in scenario (S3) as the regression coefficients are completely random. 
The binary response variable $Y_{it}$ in the same way as the previous study with the exchangeable correlation structure with $0.5$ correlation parameter.  
We generated a new vector of covariates $\X_{i,T+1}$ from the same data generating process, and the target to be estimated is the success probability of future observations, $\mu_i\equiv {\rm logit}^{-1}(\X_{i,T+1}^\top \bbe_i)$. 
For the simulated dataset, we applied the proposed grouped GEE (GGEE) method with the estimated number of groups to estimate $\bbe_i$ by $\bbeh_{\gh_i}$.
For comparison, we applied random coefficient models (RC), growth mixture models \citep[e.g.][]{Ram2009}, denoted by GMM, and pairwise penalization approaches \citep{Zhu2018}, denoted by PWL, to estimate the subject-specific coefficient $\bbe_i$, where the details of each method are provided in the Supplementary Material. 
Then, $\mu_i$ is estimated by $\X_{i,T+1}^\top \bbeh_i$. 
Furthermore, we also applied the generalized linear mixed model tree \citep{Fokkema2018,Hajjem2017}, denoted by GLMMT, to directly estimate $\mu_i$, for which we used the R package ``glmertree" \citep{Fokkema2018}.

The performance of estimating $\mu_i$ is measured by the square root of mean squared errors (RMSE), defined as $\{n^{-1}\sum_{i=1}^n(\muh_i-\mu_i)^2\}^{1/2}$.
The averaged values of RMSE based on 1000 Monte Carlo replications are presented in Table \ref{tab:comp}.

\begin{table}[htbp]
\caption{Squared root of mean squared error (RMSE) of estimators of the success probability of future observations, averaged over 1000 Monte Carlo replications, for the proposed method with two working correlation matrices (GGEE-EX and GGEE-US), and four competing methods. }
\label{tab:comp}
\begin{center}
\begin{tabular}{ccccccccccccc}
\hline
 && \multicolumn{2}{c}{(S1)} & \multicolumn{2}{c}{(S2)} & \multicolumn{2}{c}{(S3)}\\
Method & & $T=10$ & $T=20$ & $T=10$ & $T=20$ &$T=10$ & $T=20$ \\
\hline
CGEE-EX &  & 12.6 & 6.3 & 15.0 & 9.5 & 29.0 & 23.8 \\
CGEE-US &  & 13.8 & 8.9 & 16.1 & 11.4 & 29.1 & 24.2 \\
RC &  & 22.3 & 21.3 & 22.5 & 21.6 & 24.2 & 24.3 \\
LCM &  & 13.7 & 10.2 & 15.8 & 12.7 & 24.5 & 24.1 \\
MT &  & 32.7 & 33.3 & 34.0 & 35.4 & 20.5 & 21.3 \\
PWL &  & 19.9 & 15.9 & 20.8 & 16.8 & 21.3 & 21.5 \\
\hline
\end{tabular}
\end{center}
\end{table}

In scenario (S1), since the subject-specific regression coefficients can be perfectly grouped, the proposed methods provide better estimation accuracy than the other methods except for LCM. 
Moreover, in scenario (S2), the subjects do not hold exact grouping structures but can be approximately grouped, and the proposed method still works better than the other methods except for LCM. 
On the other hand, the regression coefficients are completely random in scenario (S3), and the results show that MT and PWL are appealing. 
It should be noted that the difference between the grouped GEE and RC methods are relatively comparable, which would indicate that the proposed grouped GEE method can reasonably approximate the subject-specific random coefficients by grouping subjects having similar regression coefficients. 
Finally, comparing the two working correlations, the EX correlation provides better performance than the US correlation since the EX is the true underlying correlation structure within the same subject. In contrast, the US correlation is quite comparable with EX.

\section{Application to the health and retirement study}\label{sec:app}

We apply the proposed method to the Health and Retirement Study (HRS) data, which come from the study conducted by the University of Michigan. 
This longitudinal panel study surveys adults over the age of 50 in the United States through detailed interviews once every two years for each participant and provides information on their health and economic circumstances.
For more details, see \cite{JS1995}.
The main goal of the study is to investigate the change in participants' health conditions in the HRS study over time and the relevant factors associated with their condition.
We used the data set from the HRS study, which can be obtained from an R Package ``LMest''.
The sample includes $n = 7074$ individuals followed at $T = 8$ approximately equally spaced occasions without missing responses or dropouts.
The response variable is the self-reported health status (named SHLT), in which five categories of statuses: `poor', `fair', `good', `very good, 'excellent', are recorded as an ordinal response variable from $1$ to $5$, noting that a smaller value corresponds to a high level of health condition.
We then dichotomized the response by setting values of $1$ or $2$ to ``healthy" (1) and the other values to ``unhealthy" (0).
As auxiliary information, we adopted indicator variables of gender (1:male, 0:female), indicators of black and others, respectively, indicators of two education levels, ``some college" (SC) and "college and above" (CAA), and age which is measured in years for each time occasion.
We also included a quadratic term age and seven time effects for $t=2,\ldots,8$. 
Among the individuals, it would be reasonable to assume that different types of individuals exist, that is, some individuals are always healthy, whereas some individuals are not, or their health condition changes during the term.
Therefore, instead of focusing on population-averaged regression coefficients, we here focus on such potential heterogeneity in the population to apply the proposed grouped GEE approach.

Let $y_{it}$ be the binary response variable, and $x_{it}$ be the vector of five covariates and an intercept, for $i=1,\ldots,n(=7074)$ and $t=1,\ldots,T(=8)$.
We consider the mean structure $E[y_{it}|\x_{it}]=m(\x_{it}^\top\bbe_{g_i})$ with $m(x)=\exp(x)/\{1+\exp(x)\}$ and $g_i\in\{1,\ldots,G\}$.
In this analysis, we use unstructured working correlation.
We first selected the number of groups, $G$, from candidates $\{2,3,\ldots, 10\}$, using the CVA value.
The CVA value for each $G$ is shown in Figure 1 in Supporting Information Section S.5.3, and the CVA value is minimized at $G=8$. 
Thus, we carried out the grouped GEE analysis with $G=8$ in what follows. 
The estimated regression coefficients and their standard errors in 8 groups are shown in Table \ref{tab:app}.

\begin{figure}
\centering
\includegraphics[width=16cm,clip]{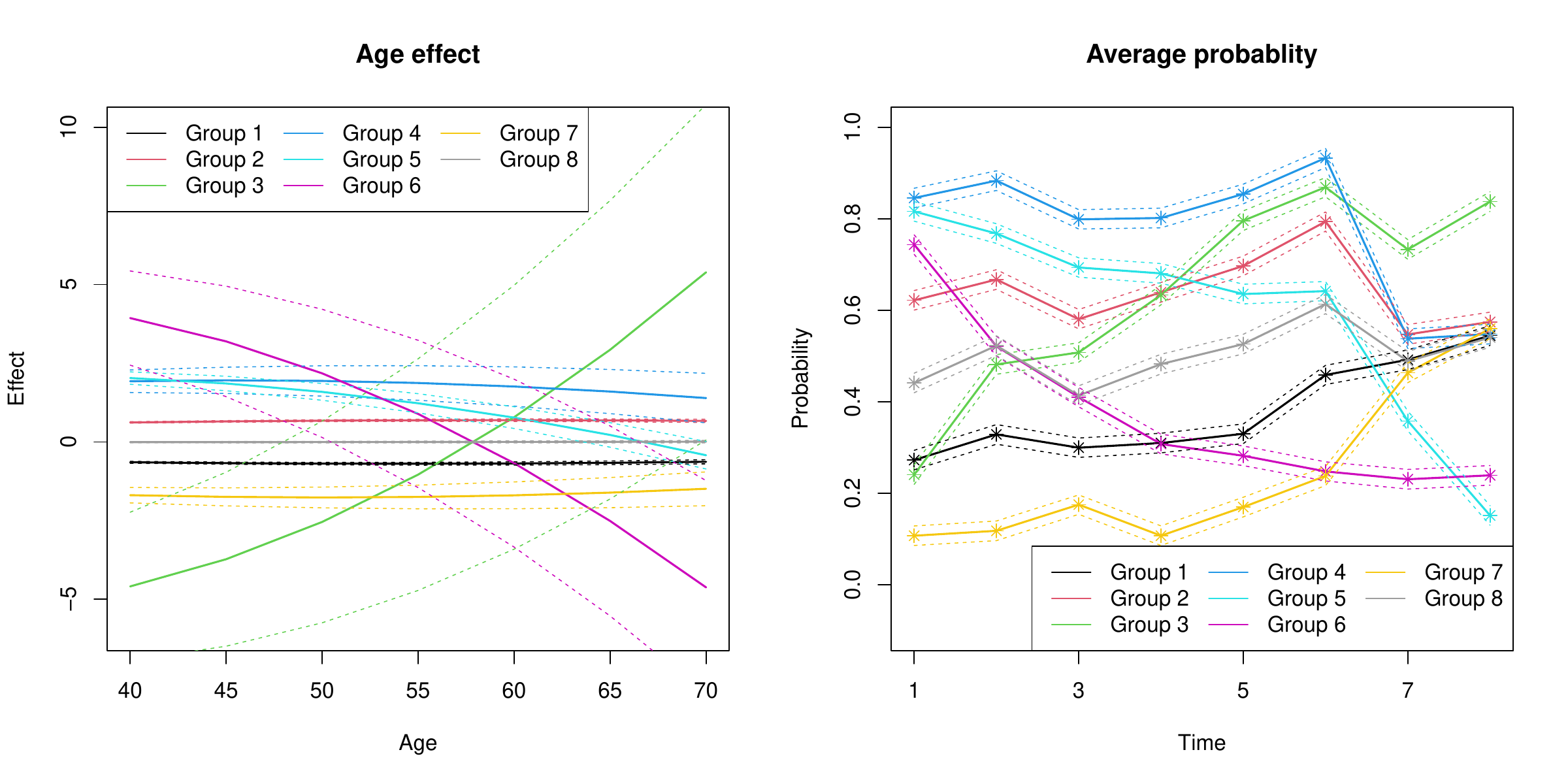}
\caption{Estimates (solid line) and $95\%$ point-wise confidence intervals (dotted line) of group-wise quadratic effects of age (left) and group-wise probability being ``unhealthy" (right) in detected eight heterogeneous groups.
This figure appears in color in the electronic version of this article, and any mention of color refers to that version.}
\label{fig:app}
\end{figure}

\begin{table}[htbp]
\caption{Point estimates (PE) and standard errors (SE) of group-specific regression coefficients, where the values of PE and SE are multiplied by 1000. }
\label{tab:app}
\begin{center}
\begin{tabular}{cccccccccccccccc}
\hline
&&& \multicolumn{8}{c}{Group}\\
 & & & 1 & 2 & 3 & 4 & 5 & 6 & 7 & 8 \\
\hline
group size & & & 1478 & 1650 & 191 & 686 & 310 & 117 & 559 & 2083 \\
\hline
Intercept & PE & & -0.87 & 0.79 & -7.64 & 2.87 & 4.33 & 7.02 & -2.29 & -0.02 \\
 & SE & & 0.01 & 0.01 & 0.83 & 0.13 & 0.07 & 0.54 & 0.09 & 0.01 \\
Gender & PE & & 4.25 & -1.50 & 1138.89 & -11.86 & 195.25 & -1455.80 & 34.66 & 2.33 \\
 & SE & & 2.07 & 1.66 & 75.91 & 27.53 & 10.40 & 75.20 & 29.57 & 1.54 \\
Black & PE & & 0.67 & -0.28 & -200.84 & -1.34 & 37.33 & -481.21 & 2.17 & 0.23 \\
 & SE & & 0.27 & 0.26 & 17.67 & 3.08 & 1.93 & 30.07 & 1.87 & 0.15 \\
Other & PE & & -0.16 & 0.06 & 42.70 & -0.10 & 0.54 & -41.30 & -1.64 & -0.03 \\
 & SE & & 0.04 & 0.01 & 4.10 & 0.30 & 0.02 & 2.25 & 1.10 & 0.02 \\
SC & PE & & 0.63 & -0.81 & -31.44 & 0.36 & 95.86 & 339.82 & 5.53 & 0.32 \\
 & SE & & 0.36 & 0.69 & 50.64 & 0.82 & 3.76 & 45.74 & 5.07 & 0.22 \\
CAA & PE & & -1.78 & 1.40 & 84.76 & 6.20 & -145.79 & 406.11 & -11.38 & -0.70 \\
 & SE & & 0.65 & 0.83 & 25.22 & 9.68 & 7.17 & 47.82 & 7.99 & 0.45 \\
Age & PE & & -25.82 & 23.28 & -370.78 & 85.88 & 126.41 & 317.68 & -70.48 & -0.75 \\
 & SE & & 0.31 & 0.24 & 24.73 & 3.89 & 2.12 & 15.64 & 2.68 & 0.19 \\
Age$^2$ & PE & & 0.24 & -0.20 & 6.40 & -0.94 & -1.89 & -5.48 & 0.70 & 0.01 \\
 & SE & & 0.01 & 0.00 & 0.43 & 0.06 & 0.03 & 0.27 & 0.04 & 0.00 \\
\hline
\end{tabular}
\end{center}
\end{table}
It is observed that estimated regression coefficients in the eight groups are very different from each other.
To visualize the difference, we computed the estimated quadratic function of the age effect in Figure \ref{fig:app}, which indicates that some groups have representative shapes of the age effect.

For example, the probability of ``health" of individuals classified in group 3 increases according to their age, while the opposite tendency is confirmed in group 6.
Although clear differences among four groups (groups 1,2,7, and 8) are not observed from Figure \ref{fig:app}, the regression coefficients of other covariates reported in Table \ref{tab:app} are quite different. 
Moreover, in each group, we computed average values of $y_{it}$ for $t=1,\ldots,T$, where the results are presented in the right panel in Figure \ref{fig:app}.
From the result, we can more directly understand the characteristics of the eight groups.
For example, individuals in groups 3 and 7 have a low probability of being ``healthy" at the earlier period, and the probability increases with the period. 
On the other hand, the probability in groups 5 and 6 decreases according to the period, but there is a difference in the shape of the decrease. 
Therefore, we can conclude that the classical GEE analysis assuming homogeneity in the regression coefficients is not an appropriate strategy for the dataset. In contrast, the proposed grouped GEE analysis can successfully capture the potential heterogeneity among individuals.

\section{Concluding Remarks}\label{sec:dis}
This paper developed a new statistical approach to analyzing longitudinal data. 
The proposed method called grouped GEE analysis carries out grouping subjects and estimating the regression coefficients simultaneously to take account of potential heterogeneity.
We employed working correlations in estimation and grouping steps and provided a simple iterative algorithm to obtain grouped GEE estimator. 
We also developed asymptotic properties of the proposed method.
The simulation studies and an application to the health and retirement study suggest the usefulness of the proposed approach.

The proposed method has some useful extensions.
First, we can introduce a penalty term in the grouping step as considered in \cite{Sugasawa2020}, which can make subjects have similar characteristics or covariates tend to be classified to the same group. 
This might make the estimation results more interpretable. 
Secondly, it would be possible to extend the proposed grouped GEE method for incomplete longitudinal data.
Since the grouped GEE separately applies the standard GEE to each group, we can employ existing methodology to handle missing data in the standard GEE method.  
Moreover, when the dimension of the regression coefficients is large, it would be better to conduct variable selection, which can be done by introducing a penalty function in the estimating equation as considered in \cite{Wang2012}.
Finally, instead of using working correlation matrices, it would be beneficial to consider quadratic inference functions \cite{QLL2000}, and develop the grouped GEE method with theoretical justifications. 
We leave the detailed investigation of these issues for interesting future works.

\section*{Acknowledgement}
This work is partially supported by the Japan Society for the Promotion of Science (JSPS KAKENHI) grant numbers: 18K12757 and 19K23242.

\vspace{0.5cm}
\bibliographystyle{chicago}
\bibliography{ref}

\begin{thebibliography}{}

\bibitem[\protect\citeauthoryear{Barban and Billari}{Barban and
  Billari}{2012}]{BB2012}
Barban, N. and F.~C. Billari (2012).
\newblock Classifying life course trajectories: a comparison of latent class
  and sequence analysis.
\newblock {\em Journal of the Royal Statistical Society, Series~C\/}~{\em 61},
  765--784.

\bibitem[\protect\citeauthoryear{Bates, Machler, Bolker, and Walker}{Bates
  et~al.}{2016}]{Bates2016}
Bates, D., M.~Machler, B.~Bolker, and S.~Walker (2016).
\newblock Fitting linear mixe-effects models using lme4.
\newblock {\em Journal of Statistical Software\/}~{\em 67}, 1--48.

\bibitem[\protect\citeauthoryear{Bonhomme and Manresa}{Bonhomme and
  Manresa}{2015}]{BM2015}
Bonhomme, S. and E.~Manresa (2015).
\newblock Grouped pattern of heterogeneity in panel data.
\newblock {\em Econometrica\/}~{\em 83}, 1147--1184.

\bibitem[\protect\citeauthoryear{Coffey, Hinde, and Holian}{Coffey
  et~al.}{2014}]{Coffey2014}
Coffey, N., J.~Hinde, and E.~Holian (2014).
\newblock Clustering longitudinal profiles using p-splines and mixed effects
  models applied to time-course gene expression data.
\newblock {\em Computational Statistics \& Data Analysis\/}~{\em 71}, 14--29.

\bibitem[\protect\citeauthoryear{Field and Welsh}{Field and
  Welsh}{2007}]{field2007}
Field, C.~A. and A.~H. Welsh (2007).
\newblock Bootstrapping clustered data.
\newblock {\em Journal of the Royal Statistical Society: Series B (Statistical
  Methodology)\/}~{\em 69\/}(3), 369--390.

\bibitem[\protect\citeauthoryear{Fokkema, Smits, Zeileis, Hothorn, and
  Kelderman}{Fokkema et~al.}{2018}]{Fokkema2018}
Fokkema, M., N.~Smits, A.~Zeileis, T.~Hothorn, and H.~Kelderman (2018).
\newblock Detecting treatment-subgroup interactions in clustered data with
  generalized linear mixed-effects model trees.
\newblock {\em Behavior research methods\/}~{\em 50\/}(5), 2016--2034.

\bibitem[\protect\citeauthoryear{Gu and Volgushev}{Gu and
  Volgushev}{2019}]{GV2019}
Gu, J. and S.~Volgushev (2019).
\newblock Panel data quantile regression with grouped fixed effects.
\newblock {\em Journal of Econometrics\/}~{\em 213}, 68--91.

\bibitem[\protect\citeauthoryear{Hajjem, Bellavance, and Larocque}{Hajjem
  et~al.}{2011}]{Hajjem2011}
Hajjem, A., F.~Bellavance, and D.~Larocque (2011).
\newblock Mixed effects regression trees for clustered data.
\newblock {\em Statistics \& Probability Letters\/}~{\em 81}, 451--459.

\bibitem[\protect\citeauthoryear{Hajjem, Larocque, and Bellavance}{Hajjem
  et~al.}{2017}]{Hajjem2017}
Hajjem, A., D.~Larocque, and F.~Bellavance (2017).
\newblock Generalized mixed effects regression trees.
\newblock {\em Statistics \& Probability Letters\/}~{\em 126}, 114--118.

\bibitem[\protect\citeauthoryear{Juster and Suzman}{Juster and
  Suzman}{1995}]{JS1995}
Juster, F.~T. and R.~Suzman (1995).
\newblock An overview of the health and retirement study.
\newblock {\em Journal of Human Resources\/}~{\em 30}, S7--S56.

\bibitem[\protect\citeauthoryear{Liang and Zeger}{Liang and
  Zeger}{1986}]{LZ1986}
Liang, L. and S.~L. Zeger (1986).
\newblock Longitudinal data analysis using generalized linear models.
\newblock {\em Biometrika\/}~{\em 73}, 13--22.

\bibitem[\protect\citeauthoryear{Lin and Ng}{Lin and Ng}{2012}]{Lin2012}
Lin, C.~C. and S.~Ng (2012).
\newblock Estimation of panel data models with parameter heterogeneity when
  group membership is unknown.
\newblock {\em Journal of Econometric Methods\/}~{\em 1}, 42--55.

\bibitem[\protect\citeauthoryear{Liu, Shang, Zhang, and Zhou}{Liu
  et~al.}{2020}]{Liu2020}
Liu, R., Z.~Shang, Y.~Zhang, and Q.~Zhou (2020).
\newblock Identification and estimation in panel models with overspecified
  number of groups.
\newblock {\em Journal of Econometrics\/}, to appear.

\bibitem[\protect\citeauthoryear{Nagin, Jones, Passos, and Tremblay}{Nagin
  et~al.}{2018}]{Nagin2018}
Nagin, D.~S., B.~L. Jones, V.~L. Passos, and R.~E. Tremblay (2018).
\newblock Group-based multi-trajectory modeling.
\newblock {\em Statistical Methods in Medical Research\/}~{\em 27}, 2015--2023.

\bibitem[\protect\citeauthoryear{Ng and McLachlan}{Ng and
  McLachlan}{2014}]{Ng2014}
Ng, S.~K. and G.~J. McLachlan (2014).
\newblock Mixture models for clustering multilevel growth trajectories.
\newblock {\em Computational Statistics \& Data Analysis\/}~{\em 71}, 43--51.

\bibitem[\protect\citeauthoryear{Qu, Lindsay, and Li}{Qu
  et~al.}{2000}]{QLL2000}
Qu, A., B.~G. Lindsay, and B.~Li (2000).
\newblock Improving generalised estimating equations using quadratic inference
  functions.
\newblock {\em Biometrika\/}~{\em 87\/}(4), 823--836.

\bibitem[\protect\citeauthoryear{Ram and Grimm}{Ram and Grimm}{2009}]{Ram2009}
Ram, N. and K.~J. Grimm (2009).
\newblock Methods and measures: Growth mixture modeling: A method for
  identifying differences in longitudinal change among unobserved groups.
\newblock {\em International journal of behavioral development\/}~{\em
  33\/}(6), 565--576.

\bibitem[\protect\citeauthoryear{Rio}{Rio}{2000}]{Rio2000}
Rio, E. (2000).
\newblock {\em {\it Th\'eorie asymptotique des processus al\'eatoires
  faiblement d\'ependants}}, pp.\  1158--1176.
\newblock Berlin: Springer.

\bibitem[\protect\citeauthoryear{Rosen, Jiang, and Tanner}{Rosen
  et~al.}{2000}]{Rosen2000}
Rosen, O., W.~Jiang, and M.~A. Tanner (2000).
\newblock Mixtures of marginal models.
\newblock {\em Biometrika\/}~{\em 87}, 391--404.

\bibitem[\protect\citeauthoryear{Rubin and Wu}{Rubin and Wu}{1997}]{Rubin1997}
Rubin, D.~B. and Y.~Wu (1997).
\newblock Modeling schizophrenic behavior using general mixture components.
\newblock {\em Biometrics\/}~{\em 53}, 243--261.

\bibitem[\protect\citeauthoryear{Sugasawa}{Sugasawa}{2021}]{Sugasawa2020}
Sugasawa, S. (2021).
\newblock Grouped heterogeneous mixture modeling for clustered data.
\newblock {\em Journal of the American Statistical Association\/}~{\em
  116\/}(534), 999--1010.

\bibitem[\protect\citeauthoryear{Sugasawa, Kobayashi, and Kawakubo}{Sugasawa
  et~al.}{2019}]{Sugasawa2019}
Sugasawa, S., G.~Kobayashi, and Y.~Kawakubo (2019).
\newblock Latent mixture modeling for clustered data.
\newblock {\em Statistics and Computing\/}~{\em 29}, 537--548.

\bibitem[\protect\citeauthoryear{Sun, Rosen, and Sampson}{Sun
  et~al.}{2007}]{Sun2007}
Sun, Z., O.~Rosen, and A.~R. Sampson (2007).
\newblock Multivariate bernoulli mixture models with application to postmortem
  tissue studies in schizophrenia.
\newblock {\em Biometrics\/}~{\em 63}, 901--909.

\bibitem[\protect\citeauthoryear{Tang and Qu}{Tang and Qu}{2016}]{Tang2016}
Tang, X. and A.~Qu (2016).
\newblock Mixture modeling for longitudinal data.
\newblock {\em Journal of Computational and Graphical Statistics\/}~{\em 25},
  1117--1137.

\bibitem[\protect\citeauthoryear{Tang, Xue, and Qu}{Tang
  et~al.}{2020}]{Tang2020}
Tang, X., F.~Xue, and A.~Qu (2020).
\newblock Individualized multidirectional variable selection.
\newblock {\em Journal of the American Statistical Association\/}, to appear.

\bibitem[\protect\citeauthoryear{Vogt and Linton}{Vogt and
  Linton}{2017}]{Vogt2017}
Vogt, M. and O.~Linton (2017).
\newblock Classification of non-parametric regression functions in longitudinal
  data models.
\newblock {\em Journal of the Royal Statistical Society: Series~B\/}~{\em 79},
  5--27.

\bibitem[\protect\citeauthoryear{Wang}{Wang}{2010}]{Wang2010}
Wang, J. (2010).
\newblock Consistent selection of the number of clusters via crossvalidation.
\newblock {\em Biometrika\/}~{\em 97}, 893--904.

\bibitem[\protect\citeauthoryear{Wang}{Wang}{2011}]{Wang2011}
Wang, L. (2011).
\newblock Gee analysis of clustered binary data with diverging number of
  covariates.
\newblock {\em The Annals of Statistics\/}~{\em 39}, 389--417.

\bibitem[\protect\citeauthoryear{Wang and Qu}{Wang and Qu}{2009}]{WQ2009}
Wang, L. and A.~Qu (2009).
\newblock Consistent model selection and data-driven smooth tests for
  longitudinal data in the estimating equations approach.
\newblock {\em Journal of the Royal Statistical Society: Series~B\/}~{\em
  71\/}(1), 177--190.

\bibitem[\protect\citeauthoryear{Wang, Zhou, and Qu}{Wang
  et~al.}{2012}]{Wang2012}
Wang, L., J.~Zhou, and A.~Qu (2012).
\newblock Penalized generalized estimating equations for high‐dimensional
  longitudinal data analysis.
\newblock {\em Biometreics\/}~{\em 68}, 353--360.

\bibitem[\protect\citeauthoryear{Wedderburn}{Wedderburn}{1974}]{Wedd1974}
Wedderburn, R.~W. (1974).
\newblock Quasi-likelihood functions, generalized linear models, and the
  {G}auss-newton method.
\newblock {\em Biometrika\/}~{\em 61}, 439--447.

\bibitem[\protect\citeauthoryear{Xie and Yang}{Xie and Yang}{2003}]{XY2003}
Xie, M. and Y.~Yang (2003).
\newblock Asymptotics for generalized estimating equations with large cluster
  sizes.
\newblock {\em The Annals of Statistics\/}~{\em 31\/}(1), 310--347.

\bibitem[\protect\citeauthoryear{Zhang, Wang, and Zhu}{Zhang
  et~al.}{2019}]{ZWZ2019}
Zhang, Y., J.~Wang, and Z.~Zhu (2019).
\newblock Quantile-regression-based clustering for panel data.
\newblock {\em Journal of Econometrics\/}~{\em 213}, 54--67.

\bibitem[\protect\citeauthoryear{Zhu and Qu}{Zhu and Qu}{2018}]{ZQ2018}
Zhu, X. and A.~Qu (2018).
\newblock Cluster analysis of longitudinal profiles with subgroups.
\newblock {\em Electronic Journal of Statistics\/}~{\em 12}, 171--193.

\bibitem[\protect\citeauthoryear{Zhu, Tang, and Qu}{Zhu et~al.}{2021}]{Zhu2018}
Zhu, X., X.~Tang, and A.~Qu (2021).
\newblock Longitudinal clustering for heterogeneous binary data.
\newblock {\em Statistica Sinica\/}, to appear.

\end{thebibliography}


\begin{thebibliography}{99}

\item
Bates, D., Machler, M., Bolker, B., and Walker, S. (2016). 
Fitting linear mixe-effects models using lme4. 
{\it Journal of Statistical Software} 67, 1--48.


\item
Bonhomme, S. and Manresa, E. (2015).
Grouped pattern of heterogeneity in panel data.
{\it Econometrica} 83,
1147--1184.


\item
Friedman, J., Hastie, T., and Tibshirani, R. (2010). 
Regularization paths for generalized linear models via coordinate descent. 
{\it Journal of statistical software} 33, 1.


\item
Rio, E. (2000).
{\it Th\'eorie asymptotique des processus al\'eatoires faiblement d\'ependants}.
Berlin: Springer.
1158--1176.

\item
Wang, L. (2011). 
GEE analysis of clustered binary data with diverging number of covariates.
{\it The Annals of Statistics} 39,
389--417.

\item
Xie, M. and Yang, Y. (2003). 
Asymptotics for generalized estimating equations with large cluster sizes. {\it The Annals of Statistics} 31, 310--347.


\item
Zhu, X., Tang, X., and Qu, A. (2018). 
Longitudinal clustering for heterogeneous binary data. 
{\it Statistica Sinica}, to appear.

\end{thebibliography}

\newpage

\vspace{0.5cm}
\begin{center}
{\LARGE {\bf Supplementary Materials for
``Grouped Generalized Estimating Equations for Longitudinal Data Analysis"}}

\bigskip

{\large {\bf Tsubasa Ito and Shonosuke Sugasawa}}
\end{center}

\setcounter{equation}{0}
\renewcommand{\theequation}{S\arabic{equation}}
\setcounter{section}{0}
\renewcommand{\thesection}{S\arabic{section}}
\setcounter{lemma}{0}
\renewcommand{\thelemma}{S\arabic{lemma}}
\setcounter{table}{0}
\renewcommand{\thetable}{S\arabic{table}}

\vspace{0.5cm}
\section{Additional assumptions}\label{sec:as}

We give the following notations similar to those in \cite{XY2003}, which are needed to provide assumptions assuring a sufficient conditions for the conditions (I*), (L*) and (CC) in \cite{XY2003}, under which the existence, weak consistency and asymptotic normality of the GEE estimator hold:
\begin{align*}
\pi=\sup_{\bbe,\bga}\frac{\la_{\max}(\overline\R^{-1}(\bbe,\bga))}{\la_{\min}(\overline\R^{-1}(\bbe,\bga))}, \quad
\xi=\tau \max_{1\leq i\leq n,1\leq t\leq T}\max_{1\leq g\leq G} \x_{it}^\top\{\overline\H_g^{*}(\bbe_g^0)\}^{-1}\x_{it}.
\end{align*}
In addition to the Assumption (A1)-(A5), we assume the following regularity assumptions for the grouped GEE:

\begin{assumption}
\ \par
\noindent
(A6)
For all $i=1,\ldots,n$ and $t=1,\ldots,T$, $a'(\th_{it})$ is uniformly three times continuously differentiable, $a''(\th_{it})$ is uniformly bounded away from $0$, and $u(\eta_{it})$ is uniformly four times continuously differentiable and $u'(\eta_{it})$ is uniformly bounded away from $0$.

\noindent
(A7)
For all $i=1,\ldots,n$, there exist positive constants, $b_1$, $b_2$ and $b_3$, such that $b_1\leq\la_{\min}((nT)^{-1}\sum_{i=1}^n\X_i^\top \X_i)\leq\la_{\max}((nT)^{-1}\sum_{i=1}^n\X_i^\top \X_i)\leq b_2$
and
$\la_{\max}(T^{-1}\X_i^\top \X_i)\leq b_3$.
For all $i$, there is $q$ such that $x_{itq}\neq x_{it'q}$ for some $t\neq t'$.

\noindent
(A8)
(i) $\pi^2\xi\to0$ and (ii) $v\pi\xi\to0$ for $v=(\sqrt{nT}\wedge T\pi/\min_{1\leq i\leq n,1\leq t\leq T}\{\si^2(\x_{it}^\top\bbe_{g_i^0}^0)\})$.

\noindent
(A9)
(i) $\sup_{\bbe\in\Bc_{nT}} \max_{1\leq k,l\leq T}|\{\R(\bal,\bbe,\bga)-\R(\bal,\bbe^0,\bga)\}_{k.l}|=O_p(\la_{\min}^{-1/2}(\overline\H^{*})\tau^{1/2})$ for any $\bal$ and $\bga$, (ii) for any $\bga$, $\sup_{\bbe\in\Bc_{nT}} \max_{1\leq k,l\leq T}|\{\widehat\R(\bbe,\bga)-\overline\R(\bbe,\bga)\}_{k.l}|=O_p(n^{-1/2}\vee\la_{\min}^{-1/2}(\overline\H^{*})\tau^{1/2})$ and $\max_{1\leq k,l\leq T}|\{\widehat\R(\bbe^0,\bga)-\overline\R(\bbe^0,\bga)\}_{k.l}|=O_p(n^{-1/2})$, and (iii) for any $\bal$, $\bbe$ and $\bga_{i*}$ whose only $i$th component differs from that of $\bga$,
$\max_{1\leq k,l\leq T}|\{R(\bal,\bbe,\bga_{i*})-R(\bal,\bbe,\bga)\}_{kl}|=O_p(1/n)$.
(iv) for any $\bbe\in\Bc$ and all $\de>0$, $\max_{1\leq k,l\leq T}|\{\widehat\R(\bbe,\bga)-\widehat\R(\bbe,\bga^0)\}_{kl}|=o_p(T^{-\de})$ for $\bga\in\Ga$, where $\Ga=\{\bga=(g_1,\ldots,g_n) :  n^{-1}\sum_{i=1}^n \one\{g_i\neq g_i^0\}=o_p(T^{-\de}) \ \  {\rm for \ all} \ \ \de>0\}$.

\end{assumption}
\vspace{0.5cm}

Assumption (A6) requires that the marginal variance of $y_{it}$ is uniformly larger than $0$ for any $\bbe\in\Bc$ and $\x_{it}\in{\mathcal X}$ for all $i=1,\ldots,n$ and $t=1,\ldots,T$.
The boundedness of $a^{(k)}(\th_{it})$ and $u^{(k)}(\eta_{it})$ for $\bbe_g$'s in a local neighborhood around $\bbe_g^0$ is also required to ensure the asymptotic properties of GEE estimators, which is satisfied from Assumptions (A1).
Assumption (A7) is also imposed well and ensures combined with Assumptions (A2) (i) that $\overline\H_g(\bbe_g)$, $\overline\M_g(\bbe_g)$ and so on are invertible when $n$ or $T$ is sufficiently large.
Assumption (A8) is the technical assumption similar to the assumptions in Lemma A.2 (ii), and A.3 (ii) of \cite{XY2003}, which ensure the sufficient conditions for the conditions (I*) and (CC) in \cite{XY2003}.
The idea behind Assumption (A9) is similar to that of the condition (A4) in \cite{Wang2011}, that is, it is essential to approximate $\S_g(\bbe_g)$ by $\overline\S_g^{*}(\bbe_g)$ whose moments are easier to evaluate.
For this, Assumption (A9) (i) and (ii) say that the estimated working correlation matrix can be approximated by $\overline\R(\bbe^0,\bga)$ in a local neighborhood of $\bbe_g^0$'s and $\overline\bal$.
Assumption (A9) (iii) says that each cluster is linearly additive for estimating the working correlation matrix.
Then, this is an intuitively reasonable assumption that most of the working correlation matrix estimators satisfy.
Assumption (A9) (iv) says that the estimated working correlation matrix can be approximated by $\overline\R(\bbe,\bga^0)$ if groups are consistently classified to their true groups on average.
In Section \ref{sec:pr}, we provide the accuracy of these approximations under the unstructured working correlation matrix.

We use the following notations.
The notation $a_{nT}\lesssim b_{nT}$ means that $a_{nT}\leq Cb_{nT}$ for all $n$ and $T$, for some constant $C$ that does not depends on $n$ and $T$.
For a column vector $\a$, we use $\a^\top$ to denote the transpose of $\a$ and $||\a||$ to denote the Euclidean norm of $\a$.
For a matrix $\A$, $\{\A\}_{kl}$ denotes the $(k,l)$-element of $\A$, $\la_{\min}(\A)$ $(\la_{\max}(\A))$ denotes the smallest (largest) eigenvalue of $\A$, $\A^\top$ denotes the transpose of $\A$ and $||\A||_F=\{\tr(\A^\top \A)\}^{1/2}$ is the Frobenius norm of $\A$.
We use the notation $a\vee b=\max(a,b)$ and $a\wedge b=\min(a,b)$.

\section{Proof of Theorem 1}\label{sec:pcons}

First of all, we need to show the next lemma.

\begin{lemma}\label{lem:gc}
Suppose the Assumptions (A1)-(A9).
If $n/T^\nu\to0$ for some $\nu>0$, it holds that for all $\de>0$,
\begin{align*}
\sup_{\bbe\in\Bc_{nT}} \frac1n\sum_{i=1}^n \one\{\gh_i(\bbe)\neq g_i^0\}=o_p(T^{-\de}),
\end{align*}
where $\gh_i(\bbe)$ is obtained by (2.3) in the main text.
\end{lemma}
\begin{proof}
For any $\bga$, $\bga_{i0}$ is obtained by replacing only its $i$th element with $g_i^0$, that is $\bga_{i0}=(g_1,\ldots,g_{i-1},g_i^0,g_{i+1},\ldots,g_n)$.
Note that, from the definition of $\gh_i(\bbe)$, we have, for all $g=1,\ldots,G$,
\begin{align*}
\one\{\gh_i(\bbe)= g\}
\leq&\one\Big\{ \{\y_i-m(\X_i\bbe_g)\}^\top\widehat\R^{-1}(\bbe,\bga)\{\y_i-m(\X_i\bbe_g)\}\\
&\ \ \ \ \ \ \ \ \ \leq \{\y_i-m(\X_i\bbe_{g_i^0})\}^\top\widehat\R^{-1}(\bbe,\bga_{i0})\{\y_i-m(\X_i\bbe_{g_i^0})\}\Big\}.
\end{align*}
Then, we can write
\begin{align*}
\frac1n\sum_{i=1}^n \one\{\gh_i(\bbe)\neq g_i^0\}=\sum_{g=1}^G\frac1n\sum_{i=1}^n \one\{g_i^0\neq g\} \one\{\gh_i(\bbe)= g\}
\leq \sum_{g=1}^G\frac1n\sum_{i=1}^nZ_{ig}(\bbe_g),
\end{align*}
where
\begin{align*}
Z_{ig}(\bbe_g)
=&\one\{g_i^0\neq g\} \one\Big\{\{\y_i-m(\X_i\bbe_g)\}^\top\widehat\R^{-1}(\bbe,\bga)\{\y_i-m(\X_i\bbe_g)\} \\
&\ \ \ \ \ \ \ \ \ \ \ \ \ \ \ \ \ \ \ \ \ \ \ \ \ \ \leq\{\y_i-m(\X_i\bbe_{g_i^0})\}^\top\widehat\R^{-1}(\bbe,\bga_{i0})\{\y_i-m(\X_i\bbe_{g_i^0})\}\Big\}.
\end{align*}
Similar to the proof of Lemma B.4 in \cite{BM2015}, we start by bounding $Z_{ig}(\bbe_g)$ on $\bbe\in\Bc_{nT}$ by a quantity that does not depend on $\bbe$.
Denote
\begin{align*}
W_{ig}(\bbe)
=&\{\y_i-m(\X_i\bbe_g)\}^\top\widehat\R^{-1}(\bbe,\bga)\{\y_i-m(\X_i\bbe_g)\}\\
&\ \ \ \ \ \ \ \ \ \ -\{\y_i-m(\X_i\bbe_{g_i^0})\}^\top\widehat\R^{-1}(\bbe,\bga_{i0})\{\y_i-m(\X_i\bbe_{g_i^0})\},
\end{align*}
then we have
\begin{align*}
Z_{ig}(\bbe_g)=\one\{g_i^0\neq g\}\one\{W_{ig}(\bbe)\leq0\}
\leq \one\{g_i^0\neq g\}\one\{W_{ig}(\bbe^0)\leq|W_{ig}(\bbe^0)-W_{ig}(\bbe)|\}.
\end{align*}
We have
\begin{align*}
|W_{ig}(\bbe^0)-W_{ig}(\bbe)|
\leq&\Big|\{\y_i-m(\X_i\bbe_{g_i^0}^0)\}^\top\widehat\R^{-1}(\bbe^0,\bga_{i0})\{\y_i-m(\X_i\bbe_{g_i^0}^0)\}\\
&\ \ \ \ -\{\y_i-m(\X_i\bbe_{g_i^0})\}^\top\widehat\R^{-1}(\bbe,\bga_{i0})\{\y_i-m(\X_i\bbe_{g_i^0})\}\Big|\\
&+\Big|\{\y_i-m(\X_i\bbe_g^0)\}^\top\widehat\R^{-1}(\bbe^0,\bga)\{\y_i-m(\X_i\bbe_g^0)\}\\
&\ \ \ \ -\{\y_i-m(\X_i\bbe_g)\}^\top\widehat\R^{-1}(\bbe,\bga)\{\y_i-m(\X_i\bbe_g)\}\Big|\\
\equiv&K_{ig}^{(1)}(\bbe)+K_{ig}^{(2)}(\bbe).
\end{align*}
We can write
\begin{align*}
K_{ig}^{(1)}(\bbe)
\leq&
|\{\y_i-m(\X_i\bbe_{g_i^0}^0)\}^\top\{\widehat\R^{-1}(\bbe^0,\bga_{i0})-\widehat\R^{-1}(\bbe,\bga_{i0})\}\{\y_i-m(\X_i\bbe_{g_i^0}^0)\}|\\
&+2|\{m(\X_i\bbe_{g_i^0}^0)-m(\X_i\bbe_{g_i^0})\}^\top\widehat\R^{-1}(\bbe,\bga_{i0})\{\y_i-m(\X_i\bbe_{g_i^0}^0)\}|\\
&+\{m(\X_i\bbe_{g_i^0}^0)-m(\X_i\bbe_{g_i^0})\}^\top\widehat\R^{-1}(\bbe,\bga_{i0})\{m(\X_i\bbe_{g_i^0}^0)-m(\X_i\bbe_{g_i^0})\}\\
\equiv&
\sum_{j=1}^3I_j.
\end{align*}
Since $A_{it}(\bbe_g)<\infty$ for all $i=1,\ldots,n$ and $t=1,\ldots,T$, for $I_1$, we can write
From Assumption (A1) , (A5) and (A9) (i), there is a constant $C_1$, independent of $n$ and $T$ such that
\begin{align*}
\sup_{\be\in\Bc_{nT}}I_1=C_1CT\la_{\min}^{-1/2}(\overline\H^{*})\tau^{1/2} \Big(\frac1T\sum_{j=1}^T \ep_{it}^2\Big).
\end{align*}
For $I_2$, from Taylor expansion, for $\bbe_{g_i^0}^*$ between $\bbe_{g_i^0}^0$ and $\bbe_{g_i^0}$, we have
\begin{align}\label{eqn:tem}
m(\X_i\bbe_{g_i^0}^0)-m(\X_i\bbe_{g_i^0})=
\phi \A_i(\bbe_{g_i^0}^*)\bDe_i(\bbe_{g_i^0}^*)\X_i(\bbe_{g_i^0}^0-\bbe_{g_i^0}).
\end{align}
Since $\max_{1\leq i\leq n}\max_{1\leq t\leq T}u'(\x_{it}^\top\bbe_g)<\infty$ from Assumptions (A1) and (A6), we have
\begin{align*}
I_2
\lesssim&
||\widehat\R^{-1}(\bbe^0,\bga_{i0})\{m(\X_i\bbe_{g_i^0}^0)-m(\X_i\bbe_{g_i^0})\}||\cdot||\bep_i||\\
\lesssim&
\la_{\max}(\widehat\R^{-1}(\bbe^0,\bga_{i0})) \{(\bbe_{g_i^0}^0-\bbe_{g_i^0})\X_i^\top\bDe_i(\bbe_{g_i^0}^*)\A_i^2(\bbe_{g_i^0}^*)\bDe_i(\bbe_{g_i^0}^*)\X_i(\bbe_{g_i^0}^0-\bbe_{g_i^0})\}^{1/2}||\bep_i||\\
\lesssim&
\la_{\max}(\widehat\R^{-1}(\bbe^0,\bga_{i0})) \la_{\max}^{1/2}(\X_i^\top \X_i)||\bbe_{g_i^0}-\bbe_{g_i^0}^0||(\bep_i^\top\bep_i)^{1/2}.
\end{align*}
Then, from Assumptions (A5), (A7) there is a constant $C_2$, independent of $n$ and $T$ such that
\begin{align*}
\sup_{\bbe\in\Bc_{nT}}I_2\leq C_2CT\la_{\min}^{-1/2}(\overline\H^{*})\tau^{1/2} \Big(\frac1T\sum_{t=1}^T \ep_{it}^2\Big)^{1/2}.
\end{align*}
As is the case with $I_2$, there is a constant $C_3$, independent of $n$ and $T$ such that
$\sup_{\bbe\in\Bc_{nT}}I_3\leq C_3C^2T\la_{\min}^{-1}(\overline\H^{*})\tau$.
For $K_{ig}^{(2)}(\bbe)$, we can write
\begin{align*}
K_{ig}^{(2)}(\bbe)
\leq&
|\{\y_i-m(\X_i\bbe_g^0)\}^\top\{\widehat\R^{-1}(\bbe^0,\bga)-\widehat\R^{-1}(\bbe,\bga)\}\{\y_i-m(\X_i\bbe_g^0)\}|\\
&+2|\{m(\X_i\bbe_g^0)-m(\X_i\bbe_g)\}^\top\widehat\R^{-1}(\bbe,\bga)\{\y_i-m(\X_i\bbe_g^0)\}|\\
&+\{m(\X_i\bbe_g^0)-m(\X_i\bbe_g)\}^\top\widehat\R^{-1}(\bbe,\bga)\{m(\X_i\bbe_g^0)-m(\X_i\bbe_g)\}.
\end{align*}
From the similar argument for $K_{ig}^{(1)}(\bbe)$, we can bound $K_{ig}^{(2)}(\bbe)$ by $C_4(CT\la_{\min}^{-1/2}(\overline\H^{*})\tau^{1/2}+C^2T\la_{\min}^{-1}(\overline\H^{*})\tau)$ for some $C_4>0$.
Next, we will bound $W_{ig}(\bbe^0,\bga)$ from below.
It can be written as
\begin{align*}
W_{ig}(\bbe^0,\bga)
=&
\{\y_i-m(\X_i\bbe_{g_i^0}^0)\}^\top\{\widehat\R^{-1}(\bbe^0,\bga)-\widehat\R^{-1}(\bbe^0,\bga_{i0})\}\{\y_i-m(\X_i\bbe_{g_i^0}^0)\}\\
&+\{m(\X_i\bbe_{g_i^0}^0)-m(\X_i\bbe_g^0)\}^\top\widehat\R^{-1}(\bbe^0,\bga)\{m(\X_i\bbe_{g_i^0}^0)-m(\X_i\bbe_g^0)\}\\
&+2\{m(\X_i\bbe_{g_i^0}^0)-m(\X_i\bbe_g^0)\}^\top\widehat\R^{-1}(\bbe^0,\bga)\{\y_i-m(\X_i\bbe_{g_i^0}^0)\}\\
\equiv&
\sum_{j=1}^3 J_j.
\end{align*}
From Assumption (A1) , (A5) and (A9) (iii),, there is a constant $C_5$, independent of $C$ and $T$, such that
$J_1\geq -C_5(T/n) (\sum_{t=1}^T\ep_{it}^2/T)$.
For $J_2$, we have
\begin{align*}
J_2
=&
\{m(\X_i\bbe_{g_i^0}^0)-m(\X_i\bbe_g^0)\}^\top\overline\R^{-1}(\bbe^0,\bga)\{m(\X_i\bbe_{g_i^0}^0)-m(\X_i\bbe_g^0)\}\\
&+\{m(\X_i\bbe_{g_i^0}^0)-m(\X_i\bbe_g^0)\}^\top\{\widehat\R^{-1}(\bbe^0,\bga)-\overline\R^{-1}(\bbe^0,\bga)\}\{m(\X_i\bbe_{g_i^0}^0)-m(\X_i\bbe_g^0)\}\\
\equiv&
J_{21}+J_{22}.
\end{align*}
For $J_{21}$, by using (\ref{eqn:tem}), we have for $\bbe_{g_i}^*$ between $\bbe_{g_i^0}^0$ and $\bbe_g^0$,
\begin{align*}
J_{21}
=
(\bbe_{g_i^0}^0-\bbe_g^0)^\top\X_i^\top\bDe_i(\bbe_{g_i}^*)\A_i(\bbe_{g_i}^*)\overline\R^{-1}(\bbe^0,\bga)\A_i(\bbe_{g_i}^*)\bDe_i(\bbe_{g_i}^*)\X_i(\bbe_{g_i^0}^0-\bbe_g^0).
\end{align*}
From Assumption (A7), $J_{21}$ is at least of order $O_p(T)$.
Then, from Assumption (A2) (ii) there is a constant $C_6^*$, independent of $C$ and $T$, such that $J_{21}\geq C_6^*T$.
From Assumptions (A5) and (A9) (ii), it can be shown that $J_{22}$ is dominated by $J_{21}$, then there is a constant $C_6$, independent of $C$ and $T$, such that 
$J_2\geq C_6T$.
Denote $\widetilde\bep_i=(\R^0)^{-1/2}\bep_i$.
For $J_3$, we have
\begin{align*}
J_3=&
2\{m(\X_i\bbe_{g_i^0}^0)-m(\X_i\bbe_g^0)\}^\top\overline\R^{-1}(\bbe^0,\bga)\A_i^{1/2}(\bbe_{g_i^0}^0)(\R^0)^{1/2}\widetilde\bep_i\\
&+2\{m(\X_i\bbe_{g_i^0}^0)-m(\X_i\bbe_g^0)\}^\top\{\widehat\R^{-1}(\bbe^0,\bga)-\overline\R^{-1}(\bbe^0,\bga)\}\{\y_i-m(\X_i\bbe_{g_i^0}^0)\}\\
\equiv&J_{31}+J_{32}.
\end{align*}
From Assumption (A9) (ii), $J_{32}$ is dominated by $J_{31}$.
Let $\U\bLa \U^\top$ be the eigendecomposition of $\overline\R^{-1/2}(\bbe^0,\bga)\A_i^{1/2}(\bbe_{g_i^0}^0)(\R^0)^{1/2}$, where $\bLa={\rm diag}(\la_1,\ldots,\la_T)$ for $\la_1\geq,\ldots,\la_T$ is a diagonal matrix formed from the eigenvalues and $\U$ is the corresponding eigenvectors of $\overline\R^{-1/2}(\bbe^0,\bga)\A_i^{1/2}(\bbe_{g_i^0}^0)(\R^0)^{1/2}$.
Then we can write
\begin{align*}
J_3=&
\{m^*(\X_i\bbe_{g_i^0}^0)-m^*(\X_i\bbe_g^0)\}^\top\bLa\widetilde\bep_i^*(1+o_p(1))\\
=&
C_7\sum_{t=1}^T \la_t\{m^*(\x_{it}^\top\bbe_{g_i^0}^0)-m^*(\x_{it}^\top\bbe_g^0)\}\ept_{it}^*(1+o_p(1)),
\end{align*}
for $m^*(\X_i\bbe_g)=\U\overline\R^{-1/2}(\bbe^0,\bga)m(\X_i\bbe_g)$ and $\widetilde\bep_i^*=\U\widetilde\bep_i$.
Combined with the above results, we thus obtain
\begin{align*}
\sup_{\bbe\in\Bc_{nT}}&Z_{ig}(\bbe_g)\\
\leq&
\one\{g_i^0\neq g\}\\
&\times \one\Big\{-C_5\frac{T}n\Big(\frac1T\sum_{t=1}^T \ep_{it}^2\Big)+C_6T+C_7\sum_{t=1}^T \la_t\{m^*(\x_{it}^\top\bbe_{g_i^0}^0)-m^*(\x_{it}^\top\bbe_g^0)\}\ept_{it}^*(1+o_p(1))\\
&\ \ \ \ \ \ \ \leq  C_1CT\la_{\min}^{-1/2}(\overline\H^{*})\tau^{1/2} \Big(\frac1T\sum_{t=1}^T \ep_{it}^2\Big)
+C_2CT\la_{\min}^{-1/2}(\overline\H^{*})\tau^{1/2}\Big(\frac1T\sum_{t=1}^T  \ep_{it}^2\Big)^{1/2}\\
&\ \ \ \ \ \ \ \ \ \ \ +C_3C^2T\la_{\min}^{-1}(\overline\H^{*})\tau+C_4(CT\la_{\min}^{-1/2}(\overline\H^{*})\tau^{1/2}+C^2T\la_{\min}^{-1}(\overline\H^{*})\tau)\Big\}.
\end{align*}
Since the right-hand side of the above inequality does not depend on $\bbe_g$ for $g=1,\ldots,G$, we can denote it as $\Zt_{ig}$.
As a result, we have
\begin{align*}
\sup_{\bbe\in\Bc_{nT}} \frac1n\sum_{i=1}^n \one\{\gh_i(\bbe)\neq g_i^0\} \leq \frac1n\sum_{i=1}^n\sum_{g=1}^G\Zt_{ig}.
\end{align*}
Using standard probability algebra, we have for all $g$ and $M$ in Assumption (A4) and for any $0<c<1$,
\begin{align*}
P&(\Zt_{ig}=1)\nonumber\\
\leq&
P\Big(-C_5\frac1n\Big(\frac1T\sum_{t=1}^T\ep_{it}^2\Big)+C_6+\frac1TC_7\sum_{t=1}^T \la_t\{m^*(\x_{it}^\top\bbe_{g_i^0}^0)-m^*(\x_{it}^\top\bbe_g^0)\}\ept_{it}^*(1+o_p(1))\\
&\ \ \ \ \ \ \ \leq  C_1C\la_{\min}^{-1/2}(\overline\H^{*})\tau^{1/2} \Big(\frac1T\sum_{t=1}^T\ep_{it}^2\Big)
+C_2C\la_{\min}^{-1/2}(\overline\H^{*})\tau^{1/2}\Big(\frac1T\sum_{t=1}^T\ep_{it}^2\Big)^{1/2}\\
&\ \ \ \ \ \ \ \ \ \ \ +C_3C^2\la_{\min}^{-1}(\overline\H^{*})\tau+C_4(C\la_{\min}^{-1/2}(\overline\H^{*})\tau^{1/2}+C^2\la_{\min}^{-1}(\overline\H^{*})\tau)\Big)\\
\leq&
P\Big(\frac1T\sum_{t=1}^T\ep_{it}^2\geq n^{1-c}M\Big)+P\Big(\frac1T\sum_{t=1}^T\ep_{it}^2\geq \la_{\min}^{1/2}(\overline\H^{*})\tau^{-1/2}M\Big)\\
&+P\Big(\frac1T\sum_{t=1}^T\ep_{it}^2\geq \la_{\min}(\overline\H^{*})\tau^{-1}M\Big)\\
&\ \ \ \ \ \ +P\Big(\frac1TC_7\sum_{t=1}^T \la_t\{m^*(\x_{it}^\top\bbe_{g_i^0}^0)-m^*(\x_{it}^\top\bbe_g^0)\}\ept_{it}^*(1+o_p(1))\\
&\ \ \ \ \ \ \ \ \ \ \ \ \ \ \ \leq C_5n^{-c}M-C_6+C_1CM+C_2C\sqrt{M}\\
&\ \ \ \ \ \ \ \ \ \ \ \ \ \ \ \ \ \ \  +C_3C^2\la_{\min}^{-1}(\overline\H^{*})\tau+C_4(C\la_{\min}^{-1/2}(\overline\H^{*})\tau^{1/2}+C^2\la_{\min}^{-1}(\overline\H^{*})\tau)\Big).
\end{align*}
From Markov's inequality, we have for any $\de>0$,
\begin{align*}
P\Big(\frac1T\sum_{t=1}^T\ep_{it}^2\geq n^{1-c}M\Big)
\leq \exp\Big(-n^{1-c}M\Big)E\Big[\exp\Big(\frac1T\sum_{t=1}^T\ep_{it}^2\Big)\Big].
\end{align*}
Since $E[T^{-1}\sum_{t=1}^T\ep_{it}^2]=1$ and $\Var(T^{-1}\sum_{t=1}^T\ep_{it}^2)<\infty$ from Assumption (A4), we have $T^{-1}\sum_{t=1}^T\ep_{it}^2=O_p(1)$.
Then, we have $P(T^{-1}\sum_{t=1}^T\ep_{it}^2\geq n^{1-c}M)=o_p(T^{-\de})$ for any $\de>0$.
Similarly, we have
\begin{align*}
P\Big(&\frac1T\sum_{t=1}^T\ep_{it}^2\geq \la_{\min}^{1/2}(\overline\H^{*})\tau^{-1/2}M\Big)\\
\leq& \exp\Big(-\la_{\min}^{1/2}(\overline\H^{*})\tau^{-1/2}M\Big)E\Big[\exp\Big(\frac1T\sum_{t=1}^T\ep_{it}^2\Big)\Big]=o_p(T^{-\de}),
\end{align*}
where the second inequality follows from Assumption (A3).
Similarly, we have
\begin{align*}
P\Big(\frac1T\sum_{t=1}^T\ep_{it}^2\geq \la_{\min}(\overline\H^{*})\tau^{-1}M\Big)=o_p(T^{-\de}).
\end{align*}
For the last probability,
\begin{align*}
P\Big(&\frac1TC_7\sum_{t=1}^T \la_t\{m^*(\x_{it}^\top\bbe_{g_i^0}^0)-m^*(\x_{it}^\top\bbe_g^0)\}\ept_{it}^*(1+o_p(1))\\
&\ \ \ \ \leq C_5n^{-c}M-C_6+C_1CM+C_2C\sqrt{M}\\
&\ \ \ \ \ \ +C_3C^2\la_{\min}^{-1}(\overline\H^{*})\tau+C_4(C\la_{\min}^{-1/2}(\overline\H^{*})\tau^{1/2}+C^2\la_{\min}^{-1}(\overline\H^{*})\tau)\Big),
\end{align*}
the right-hand side of the inequality in the probability, the first and the last two terms are dominated by other terms as $n,T\to\infty$.
Then, by taking a sufficiently small $C$, for $\eta>0$, the probability can be bounded above by
\begin{align*}
P\Big(&\Big|C_7\sum_{t=1}^T \la_t\{m^*(\x_{it}^\top\bbe_{g_i^0}^0)-m^*(\x_{it}^\top\bbe_g^0)\}\ept_{it}^*\Big|\geq T\eta\Big).
\end{align*}
Moreover, it is noted that $m^*(\x_{it}^\top\bbe_{g_i^0}^0)-m^*(\x_{it}^\top\bbe_g^0)=O_p(1)$ for all $i$ and $t$, and $\la_t$'s can be bounded by the eigenvalues of $\overline\R^{-1/2}(\bbe^0,\bga)(\R^0)^{1/2}$ multiplied by a constant.
Then, the left-hand side of the inequality is a linear combination of $\ept_{it}^*$, and its expectation is $0$, and the order of its variance is at most $O(T+\tau)$.
Since $\ept_{it}^*$ for $t=1,\ldots,T$ are uncorrelated, we can use Theorem 6.2 in \cite{Rio2000}, in which the second term of the right-hand side of the equation (6.5) vanishes in this case due to the uncorrelatedness of $\ept_{it}$'s.
Thus, by using the consequence of Theorem 6.2 in \cite{Rio2000} for $\la=T\eta/4$, $r=T^{1/2}$ and $s_n^2=T+\tau$, the probability above is bounded above by $4\{1+T^2\eta^2/(16T^{1/2}(T+\tau)))\}^{-T^{1/2}/2}=o(T^{-\de})$ for any $\de>0$.
This ends the proof.
\end{proof}

\bigskip

Similar to Wang (2011), in order to prove the consistency it is essential to approximate $\S_g(\bbe_g)$, $\H_g(\bbe_g)$ and so on by $\overline\S_g^{*}(\bbe_g)$ and $\overline\H_g^{*}(\bbe_g)$ whose moments are easier to evaluate.
The following lemmas \ref{lem:so} - \ref{lem:he} establish the accuracy of these approximations, which play important roles in deriving the asymptotic normality.

\begin{lemma}\label{lem:so}
Suppose the Assumptions (A1)-(A9).
If $n/T^\nu\to0$ for some $\nu>0$, it holds that, for all $g=1,\ldots,G$ and all $\de>0$, 
\begin{align*}
&\sup_{\bbe\in\Bc_{nT}, \bga\in\Ga}||\{\overline\H_g^{*}(\bbe_g^0)\}^{-1/2}\{\S_g(\bbe_g)-\S^*_g(\bbe_g)\}||=O_p(\la_{\min}^{-1/2}(\overline\H^{*})nT)o_p(T^{-\de}),\\
&\sup_{\bbe\in\Bc_{nT}, \ga\in\Ga}||\{\overline\H_g^{*}(\bbe_g^0)\}^{-1/2}\{\overline\S_g(\bbe_g)-\overline\S_g^{*}(\bbe_g)\}||=O_p(\la_{\min}^{-1/2}(\overline\H^{*})nT)o_p(T^{-\de}).
\end{align*}
\end{lemma}
\begin{proof}
We will show the second part of the lemma.
Form Assumption (A9) (ii), the first part of the lemma can be shown similarly by replacing $\overline\R(\bbe,\bga)$ and $\overline\R(\bbe,\bga^0)$ with $\widehat\R(\bbe,\bga)$ and $\widehat\R(\bbe,\bga^0)$ respectively.
It can be written as
\begin{align*}
&\overline\S_g(\bbe_g)-\overline\S_g^{*}(\bbe_g)\\
=&
\sum_{i=1}^n \one\{g_i=g\}\X_i^\top\bDe_i(\bbe_g)\A_i^{1/2}(\bbe_g)\overline\R^{-1}(\bbe,\bga)\A_i^{-1/2}(\bbe_g)\{\y_i-m(\X_i\bbe_g)\}\\
&-\sum_{i=1}^n \one\{g_i^0=g\}\X_i^\top\bDe_i(\bbe_g)\A_i^{1/2}(\bbe_g)\overline\R^{-1}(\bbe,\bga^0)\A_i^{-1/2}(\bbe_g)\{\y_i-m(\X_i\bbe_g)\}\\
=&
\sum_{i=1}^n \one\{g_i^0=g\}\X_i^\top\bDe_i(\bbe_g)\A_i^{1/2}(\bbe_g)\{\overline\R^{-1}(\bbe,\bga)-\overline\R^{-1}(\bbe,\bga^0)\}\A_i^{-1/2}(\bbe_g)\{\y_i-m(\X_i\bbe_g)\}\\
&+\sum_{i=1}^n (\one\{g_i=g\}-\one\{g_i^0=g\})\X_i^\top\bDe_i(\bbe_g)\A_i^{1/2}(\bbe_g)\overline\R^{-1}(\bbe,\bga)\A_i^{-1/2}(\bbe_g)\{\y_i-m(\X_i\bbe_g)\}\\
\equiv&I_1+I_2.
\end{align*}
For $I_1$, we have
\begin{align*}
I_1
&=\sum_{i:g_i^0=g}\sum_{t_1, t_2=1}^T \{\overline\R^{-1}(\bbe,\bga)-\overline\R^{-1}(\bbe,\bga^0)\}_{t_1,t_2}A_{it_1}^{1/2}(\bbe_g)A_{it_2}^{-1/2}(\bbe_g)\{y_{it_2}-m(\x_{it_2}^\top\bbe_g)\}\x_{it_1}\\
&=
\sum_{t_1=1}^T\sum_{t_2=1}^T \{\overline\R^{-1}(\bbe,\bga)-\overline\R^{-1}(\bbe,\bga^0)\}_{t_1,t_2}\\
&\ \ \ \ \ \ \ \ \ \ \times\Big[\sum_{i:g_i^0=g}A_{it_1}^{1/2}(\bbe_g)A_{it_2}^{-1/2}(\bbe_g)\{A_{it_2}^{1/2}(\bbe_g^0)\ep_{it_2}+m(\x_{it_2}^\top\bbe_g^0)-m(\x_{it_2}^\top\bbe_g)\}\x_{it_1}\Big].
\end{align*}
It is noted that we have
\begin{align*}
E\Big[\Big|\Big|\sum_{i:g_i^0=g}A_{it_1}^{1/2}(\bbe_g)A_{it_2}^{-1/2}(\bbe_g)A_{it_2}^{1/2}(\bbe_g^0)\ep_{it_2}\x_{it_1}\Big|\Big|^2\Big]
\lesssim
\sum_{i:g_i^0=g}\x_{it_1}^\top \x_{it_1}=O(n),
\end{align*}
and
\begin{align*}
\sup_{\bbe\in\Bc_{nT}}&\Big|\Big|\sum_{i:g_i^0=g}A_{it_1}^{1/2}(\bbe_g)A_{it_2}^{-1/2}(\bbe_g)\{m(\x_{it_2}^\top\bbe_g^0)-m(\x_{it_2}^\top\bbe_g)\}\x_{it_1}\Big|\Big|^2\\
\leq&
\sup_{\bbe\in\Bc_{nT}}\sum_{i:g_i^0=g}\Big|\Big|A_{it_1}^{1/2}(\bbe_g)A_{it_2}^{-1/2}(\bbe_g){\dot m}(\{\x_{it_2}^\top\bbe_g\}^*)\x_{it_2}^\top(\bbe_g^0-\bbe_g)\x_{it_1}\Big|\Big|^2\\
\lesssim&
\sup_{\bbe\in\Bc_{nT}}\sum_{i:g_i^0=g}(\bbe_g^0-\bbe_g)\x_{it_2}\x_{it_2}^\top(\bbe_g^0-\bbe_g)\x_{it_1}^\top \x_{it_1}\\
=&O_p(n\la_{\min}^{-1}(\overline\H^{*})\tau).
\end{align*}
It is noted that
$\max_{1\leq,k,l\leq T}|\{\widehat\R^{-1}(\bbe,\bga^0)-\widehat\R^{-1}(\bbe,\bga)\}_{kl}|=o_p(T^{-\de})$ for $\bga\in\Ga$ from Assumption (A9) (iv).
Then, we have $\sup_{\bbe\in\Bc_{nT}}||I_1||=O_p(n^{1/2}T^2)o_p(T^{-\de})$.
For $I_2$, we have from the triangle inequality
\begin{align*}
||I_2||^2\leq
\sum_{i=1}^n \one\{g_i\neq g_i^0\}\sum_{i=1}^n ||\X_i^\top\bDe_i(\bbe_g)\A_i^{1/2}(\bbe_g)\overline\R^{-1}(\bbe,\bga)\A_i^{-1/2}(\bbe_g)\{\y_i-m(\X_i\bbe_g)\}||^2
\end{align*}
Since we have
\begin{align*}
||\X_i^\top&\bDe_i(\bbe_g)\A_i^{1/2}(\bbe_g)\overline\R^{-1}(\bbe,\bga)\A_i^{-1/2}(\bbe_g)\{\y_i-m(\X_i\bbe_g)\}||^2\\
\lesssim&
\la_{\max}(\X_i^\top \X_i)||\y_i-m(\X_i\bbe_g)||^2=O_p(T^2),
\end{align*}
we have $\sup_{\bga\in\Ga}||I_2||=O_p(nT)o_p(T^{-\de})$, which ends the proof.
\end{proof}

\begin{lemma}\label{lem:s}
Suppose the Assumptions (A1)-(A9).
It holds that, for all $g=1,\ldots,G$,
\begin{align*}
\sup_{\bbe\in\Bc_{nT}}\||\{\overline\H_g^{*}(\bbe_g^0)\}^{-1/2}\{\S_g(\bbe_g)-\overline\S_g(\bbe_g)\}||
=
O_p(\la_{\min}^{-1/2}(\overline\H^{*}) T^2).
\end{align*}
\end{lemma}
\begin{proof}
From Lemma \ref{lem:so}, it is enough to show that 
\begin{align*}
||\{\overline\H_g^{*}(\bbe_g^0)\}^{-1/2}\{ \S^{*}_g(\bbe_g^0)-\overline\S_g^{*}(\bbe_g^0)\}||=O_p(\la_{\min}^{-1/2}(\overline\H^{*}) T^2).
\end{align*}
The proof is almost the same as that of Lemma 3.1 in \cite{Wang2011}.
Let $Q=\{q_{j_1,j_2}\}_{1\leq j_1,j_2\leq T}$ denote the matrix $\widehat\R^{-1}(\bbe,\bga^0)-\overline\R^{-1}(\bbe,\bga^0)$.
Then,
\begin{align*}
&\S^{*}_g(\bbe_g)-\overline\S_g^{*}(\bbe_g)\\
=&
\sum_{i=1}^n\sum_{t_1=1}^T\sum_{t_2=1}^T \{\widehat\R^{-1}(\bbe,\bga^0)-\overline\R^{-1}(\bbe,\bga^0)\}_{t_1,t_2}A_{it_1}^{1/2}(\bbe_g)A_{it_2}^{-1/2}(\bbe_g)\{y_{it_2}-m(\x_{it_2}^\top\bbe_g)\}\x_{it_1}\\
=&
\sum_{t_1=1}^T\sum_{t_2=1}^T\{\widehat\R^{-1}(\bbe,\bga^0)-\overline\R^{-1}(\bbe,\bga^0)\}_{t_1,t_2}\\
&\ \ \ \ \ \ \ \ \ \ \ \ \ \times\Big[\sum_{i=1}^nA_{it_1}^{1/2}(\bbe_g)A_{it_2}^{-1/2}(\bbe_g)\{A_{it_2}^{1/2}(\bbe_g^0)\ep_{it_2}+m(\x_{it_2}^\top\bbe_g^0)-m(\x_{it_2}^\top\bbe_g)\}\x_{it_1}\Big]
\end{align*}
Note that 
\begin{align*}
E\Big[\Big|\Big|\sum_{i=1}^nA_{it_1}^{1/2}(\bbe_g)A_{it_2}^{-1/2}(\bbe_g)A_{it_2}^{1/2}(\bbe_g^0)\ep_{it_2}\x_{it_1}\Big|\Big|^2\Big]
\lesssim
\sum_{i=1}^n\x_{it_1}^\top \x_{it_1}=O(n),
\end{align*}
and
\begin{align*}
&\sup_{\bbe\in\Bc_{nT}}\Big|\Big|\sum_{i=1}^nA_{it_1}^{1/2}(\bbe_g)A_{it_2}^{-1/2}(\bbe_g)\{m(\x_{it_2}^\top\bbe_g^0)-m(\x_{it_2}^\top\bbe_g)\}\x_{it_1}\Big|\Big|^2\\
=&
\sup_{\bbe\in\Bc_{nT}}\sum_{i=1}^n\Big|\Big|A_{it_1}^{1/2}(\bbe_g)A_{it_2}^{-1/2}(\bbe_g){\dot m}(\{\x_{it_2}^\top\bbe_g\}^*)\x_{it_2}^\top(\bbe_g^0-\bbe_g)\x_{it_1}\Big|\Big|^2\\
\lesssim&
\sup_{\bbe\in\Bc_{nT}}\sum_{i=1}^n(\bbe_g^0-\bbe_g)^\top \x_{it_2}\x_{it_2}^\top(\bbe_g^0-\bbe_g)\x_{it_1}^\top \x_{it_1}\\
=&
C^2\la_{\min}^{-1}(\overline\H^{*})\tau O_p(n).
\end{align*}
Similar to the proof of Lemma \ref{lem:so}, we have $\max_{1\leq k,l\leq T}\{\widehat\R^{-1}(\bbe,\bga^0)-\overline\R^{-1}(\bbe,\bga^0)\}_{kl}=O_p(n^{-1/2})$ from Assumption (A9) (ii).
Then, we have $\sup_{\bbe\in\Bc_{nT}}||\S^{*}_g(\bbe_g)-\overline\S_g^{*}(\bbe_g)||=O_p(T^2)$, which proves the lemma.
\end{proof}

\bigskip

The following Lemma is from Remark 1 in \cite{XY2003}.
\begin{lemma}\label{lem:dhbe}
It holds that, for all $i=1,\ldots,n$,
\begin{align*}
\overline\bDs_i(\bbe_g)=\overline\H_i(\bbe_g)+\overline\B_i(\bbe_g)+{\overline{{\text{\boldmath $\mathcal E$}}}}_i(\bbe_g),
\end{align*}
for $\overline\B_i(\bbe_g)=\overline\B_i^{[1]}(\bbe_g)+\overline\B_i^{[2]}(\bbe_g)$ and ${\overline{{\text{\boldmath $\mathcal E$}}}}_i(\bbe_g)={\overline{{\text{\boldmath $\mathcal E$}}}}_i^{[1]}(\bbe_g)+{\overline{{\text{\boldmath $\mathcal E$}}}}_i^{[2]}(\bbe_g)$, where 
\begin{align*}
&\overline\B_i^{[1]}(\bbe_g)=\X_i^\top{\rm diag}[\overline\R^{-1}\A_i^{-1/2}(\bbe_g)\{m(\X_i\bbe_{g_i^0}^0)-m(\X_i\bbe_g)\}]\G_i^{[1]}(\bbe_g)\X_i,\\
&\overline\B_i^{[2]}(\bbe_{g_i})=\X_i^\top\bDe_i(\bbe_g)\A_i^{1/2}(\bbe_g)\overline\R^{-1}{\rm diag}[m(\X_i\bbe_{g_i^0}^0)-m(\X_i\bbe_g)]\G_i^{[2]}(\bbe_g)\X_i,\\
&{\overline{{\text{\boldmath $\mathcal E$}}}}_i^{[1]}(\bbe_g)=\X_i^\top{\rm diag}[\overline\R^{-1}\A_i^{-1/2}(\bbe_g)\A_i^{1/2}(\bbe_{g_i^0}^0)\bep_i]\G_i^{[1]}(\bbe_g)\X_i,
\end{align*}
and
\begin{align*}
{\overline{{\text{\boldmath $\mathcal E$}}}}_i^{[2]}(\bbe_{g_i})=\X_i^\top\bDe_i(\bbe_g)\A_i^{1/2}(\bbe_g)\overline\R^{-1}{\rm diag}[\A_i^{1/2}(\bbe_{g_i^0}^0)\bep_i]\G_i^{[2]}(\bbe_g)\X_i.
\end{align*}
Here, $\G_i^{[\ell]}(\bbe_g)={\rm diag}({q'}_{it}^{[\ell]}(\bbe_g),\ldots,{q'}_{it}^{[\ell]}(\bbe_g)$, for $\ell=1,2$, where
\begin{align*}
q_{it}^{[1]}(\bbe_g)=[a''(\th_{it}]^{-1/2}m'(\eta_{it}), \quad q_{it}^{[2]}(\bbe_g)=[a''(\th_{it})]^{-1/2},
\end{align*}
and
\begin{align*}
{q'}_{it}^{[1]}(\bbe_g)=-\frac12 \frac{a^{(3)}(\th_{it})}{[a''(\th_{it})]^{5/2}}\{m'(\eta_{it})\}^2+\frac{m''(\eta_{it})}{[a''(\th_{it})]^{1/2}}, \quad
{q'}_{it}^{[2]}(\bbe_{g_i})=-\frac12 \frac{a^{(3)}(\th_{it})}{[a''(\th_{it})]^{5/2}}m'(\eta_{it}).
\end{align*}
\end{lemma}

\begin{lemma}\label{lem:dr}
Suppose the Assumptions (A1)-(A9).
It holds that, for any $\bla\in\Rr^p$ and $g=1,\ldots,G$,
\begin{align*}
\sup_{\bbe\in\Bc_{nT}}\sup_{||\bla||=1} |\bla^\top[\bDs_g^{*}(\bbe_g)-\overline\bDs_g^{*}(\bbe_g)]\bla|=O_p(\{\la_{\min}^{-1/2}(\overline\H^{*})\tau^{1/2}\vee n^{-1/2}\}T^2n).
\end{align*}
\end{lemma}
\begin{proof}
By Lemma \ref{lem:dhbe}, it is sufficient to prove the following three results:
\begin{align*}
&\sup_{\bbe\in\Bc_{nT}}\sup_{||\bla||=1} |\bla^\top[\H_g^{*}(\bbe_g)-\overline\H_g^{*}(\bbe_g)]\bla|=O_p(\{\la_{\min}^{-1/2}(\overline\H^{*})\tau^{1/2}\vee n^{-1/2}\}T^2n),\\
&\sup_{\bbe\in\Bc_{nT}}\sup_{||\bla||=1} |\bla^\top[\B_g^{*}(\bbe_g)-\overline\B_g^{*}(\bbe_g)]\bla|=O_p(\{\la_{\min}^{-1/2}(\overline\H^{*})\tau^{1/2}\vee n^{-1/2}\}T^2n),
\end{align*}
and
\begin{align*}
\sup_{\bbe\in\Bc_{nT}}\sup_{||\bla||=1} |\bla^\top[{\overline{{\text{\boldmath $\mathcal E$}}}}_g^{*}(\bbe_g)-{\overline{{\text{\boldmath $\mathcal E$}}}}_g^{*}(\bbe_g)]\bla|=O_p(\{\la_{\min}^{-1/2}(\overline\H^{*})\tau^{1/2}\vee n^{-1/2}\}T^2n).
\end{align*}
We have
\begin{align*}
|\bla^\top[\H_g^{*}(\bbe_g)-\overline\H_g^{*}(\bbe_g)]\bla|
=&\Big|\sum_{i:g_i^0=g}\bla^\top \X_i^\top\bDe_i(\bbe_g^0)\A_i^{1/2}(\bbe_g^0)\widehat\R^{-1}(\bbe,\bga)\{\widehat\R(\bbe,\bga)-\overline\R(\bbe,\bga)\}\\
&\ \ \ \ \ \ \ \ \times\overline\R^{-1}(\bbe,\bga)\A_i^{1/2}(\bbe_g^0)\bDe_i(\bbe_g^0)\X_i\bla\Big|\\
\lesssim& ||\widehat\R(\bbe,\bga)-\overline\R(\bbe,\bga)||_F\la_{\max}\Big(\sum_{i:g_i^0=g}\X_i^\top \X_i\Big),
\end{align*}
which implies that $\sup_{\bbe\in\Bc_{nT}}\sup_{||\bla||=1} |\bla^\top[\H_g^{*}(\bbe_g)-\overline\H_g^{*}(\bbe_g)]\bla|=O_p(\{\la_{\min}^{-1/2}(\overline\H^{*})\tau^{1/2}\vee n^{-1/2}\}T^2n)$ from Assumptions (A2) (i), (A7) and (A9) (ii).
Next, we will verify
\begin{align*}
\sup_{\bbe\in\Bc_{nT}}\sup_{||\bla||=1} |\bla^\top[\B_g^{[1]*}(\bbe_g)-\overline\B_g^{[1]*}(\bbe_g)]\bla|=O_p(\{\la_{\min}^{-1/2}(\overline\H^{*})\tau^{1/2}\vee n^{-1/2}\}T^2n),
\end{align*}
and
\begin{align*}
\sup_{\bbe\in\Bc_{nT}}\sup_{||\bla||=1} |\bla^\top[\B_g^{[2]*}(\bbe_g)-\overline\B_g^{[2]*}(\bbe_g)]\bla|=O_p(\{\la_{\min}^{-1/2}(\overline\H^{*})\tau^{1/2}\vee n^{-1/2}\}T^2n).
\end{align*}
We have from Cauchy-Schwarz inequality
\begin{align*}
|&\bla^\top[\B_g^{[1]*}(\bbe_g)-\overline\B_g^{[1]*}(\bbe_g)]\bla|\\
=&
\Big|\sum_{i:g_i^0=g}\bla^\top \X_i^\top{\rm diag}[\{\widehat\R^{-1}(\bbe,\bga)-\overline\R^{-1}(\bbe,\bga)\}\A_i^{-1/2}(\bbe_g)\\
&\ \ \ \ \ \ \ \times\{m(\X_i\bbe_g^0)-m(\X_i\bbe_g)\}]\G_i^{[1]}(\bbe_g)\X_i\bla\Big|\\
=&
\Big|\sum_{i:g_i^0=k}\bla^\top \X_i^\top \G_i^{[1]}(\bbe_g){\rm diag}[\X_i\bla]\{\widehat\R^{-1}(\bbe,\bga)-\overline\R^{-1}(\bbe,\bga)\}\A_i^{-1/2}(\bbe_g)\\
&\ \ \ \ \ \ \ \times\{m(\X_i\bbe_g^0)-m(\X_i\bbe_g)\}\Big|\\
\leq&
\sum_{i:g_i^0=g}||{\rm diag}[\X_i\bla]\G_i^{[1]}(\bbe_g)\X_i\bla||\\
&\ \ \ \ \ \ \ \times||\{\widehat\R^{-1}(\bbe,\bga)-\overline\R^{-1}(\bbe,\bga)\}\A_i^{-1/2}(\bbe_g)\{m(\X_i\bbe_g^0)-m(\X_i\bbe_g)\}||.
\end{align*}
We have
\begin{align*}
\bla^\top \X_i^\top \G_i^{[1]}(\bbe_g){\rm diag}^2[\X_i\bla]\G_i^{[1]}(\bbe_g)\X_i\bla
\leq \max_{1\leq t\leq T}|\x_{it}^\top\bla|^2 \max_{1\leq t\leq T}|{q'}_{it}^{[1]}(\bbe_g)|^2 \la_{\max}(\X_i^\top \X_i),
\end{align*}
and, by using (\ref{eqn:tem}), we have for $\bbe_g^*$ between $\bbe_g^0$ and $\bbe_g$,
\begin{align*}
\{&m(\X_i\bbe_g^0)-m(\X_i\bbe_g)\}^\top \A_i^{-1/2}(\bbe_g)\{\widehat\R^{-1}(\bbe,\bga)-\overline\R^{-1}(\bbe,\bga)\}^2\\
&\times \A_i^{-1/2}(\bbe_g)\{m(\X_i\bbe_g^0)-m(\X_i\bbe_g)\}\\
=&
(\bbe_g^0-\bbe_g)^\top \X_i^\top \bDe(\bbe_g^*)\A_i(\bbe_g^*)\A_i^{-1/2}(\bbe_g)[\widehat\R^{-1}(\bbe,\bga)\{\overline\R^{-1}(\bbe,\bga)-\widehat\R^{-1}(\bbe,\bga)\}\\
&\times\overline\R^{-1}(\bbe,\bga)]^2 \A_i^{-1/2}(\bbe_g)\A_i(\bbe_g^*)\bDe(\bbe_g^*)\X_i(\bbe_g^0-\bbe_g)\\
\lesssim&
||\widehat\R(\bbe,\bga)-\overline\R(\bbe,\bga)||_F^2 \la_{\max}(\X_i^\top \X_i) \la_{\min}^{-1}(\overline\H^{*})||\{\overline\H_g^{*}(\bbe_g^0)\}^{1/2}(\bbe_g-\bbe_g^0)||.
\end{align*}
Then, from Assumptions (A7) and (A9) (ii), we have 
\begin{align*}
\sup_{\bbe_k\in\Bc_{nT}}&\sup_{||\bla||=1} |\bla^\top[\B_{nk}^{[1]*}(\bbe_k)-\overline\B_{nk}^{[1]*}(\bbe_k)]\bla|\nonumber\\
=&
nO_p(T^{1/2})O_p(\{T\la_{\min}^{-1/2}(\overline\H^{*})\tau^{1/2}\vee Tn^{-1/2}\})O_p(T^{1/2})\la_{\min}^{-1/2}(\overline\H^{*})\tau^{1/2}\\
=&
O_p(\{\la_{\min}^{-1/2}(\overline\H^{*})\tau^{1/2}\vee n^{-1/2}\}T^2n)\la_{\min}^{-1/2}(\overline\H^{*})\tau^{1/2},
\end{align*}
which proves $\sup_{\bbe\in\Bc_{nT}}\sup_{||\bla||=1} |\bla^\top[\B_g^{[1]*}(\bbe_g)-\overline\B_g^{[1]*}(\bbe_g)]\bla|=O_p(\{\la_{\min}^{-1/2}(\overline\H^{*})\tau^{1/2}\vee n^{-1/2}\}T^2n)$ since $\la_{\min}^{-1}(\overline\H^{*})\tau \to0$.
Moreover, we have from Cauchy-Schwarz inequality
\begin{align*}
|\bla^\top&[\B_g^{[2]*}(\bbe_g)-\overline\B_g^{[2]*}(\bbe_g)]\bla|\\
=&
\Big|\sum_{i:g_i^0=g}\bla^\top \X_i^\top\bDe_i(\bbe_g)\A_i^{1/2}(\bbe_g)\{\widehat\R^{-1}(\bbe,\bga)-\overline\R^{-1}(\bbe,\bga)\}\\
&\ \ \ \ \ \times {\rm diag}[m(\X_i\bbe_g^0)-m(\X_i\bbe_g)]\G_i^{[2]}(\bbe_g)\X_i\bla\Big|\\
=&
\Big|\sum_{i:g_i^0=g}\bla^\top \X_i^\top\bDe_i(\bbe_g)\A_i^{1/2}(\bbe_g)\{\widehat\R^{-1}(\bbe,\bga)-\overline\R^{-1}(\bbe,\bga)\}\G_i^{[2]}(\bbe_g)\\
&\ \ \ \ \ \times {\rm diag}[\X_i\bla]\{m(\X_i\bbe_g^0)-m(\X_i\bbe_g)\}\Big|\\
\leq&
\sum_{i:g_i^0=g}||{\rm diag}[\X_i\bla]\G_i^{[2]}(\bbe_g)\{\widehat\R(\bbe,\bga)-\overline\R(\bbe,\bga)\}\A_i^{1/2}(\bbe_g)\bDe_i(\bbe_g)\X_i\bla||\\
&\ \ \ \ \ \times ||m(\X_i\bbe_g^0)-m(\X_i\bbe_g)||.
\end{align*}
We have
\begin{align*}
\bla^\top&\X_i^\top\bDe_i(\bbe_g)\A_i^{1/2}(\bbe_g)\{\widehat\R^{-1}(\bbe,\bga)-\overline\R^{-1}(\bbe,\bga)\}\G_i^{[2]}(\bbe_g){\rm diag}^2[\X_i\bla]\G_i^{[2]}(\bbe_g)\\
&\times \{\widehat\R^{-1}(\bbe,\bga)-\overline\R^{-1}(\bbe,\bga)\}\A_i^{1/2}(\bbe_g)\bDe_i(\bbe_g)\X_i\bla\\
\lesssim&
\max_{1\leq t\leq T}|\x_{it}^\top\bla|^2 \max_{1\leq t\leq T}|{q'}_{it}^{[2]}(\bbe_g)|^2 ||\widehat\R^{-1}(\bbe,\bga)-\overline\R^{-1}(\bbe,\bga)||_F^2 \la_{\max}(\X_i^\top \X_i),
\end{align*}
and for $\bbe_g^*$ between $\bbe_g$ and $\bbe_g^0$, we have
\begin{align*}
||m(\X_i\bbe_g^0)-m(\X_i\bbe_g)||^2
=&(\bbe_g^0-\bbe_g)^\top \X_i^\top \A_i(\bbe_g^*)\bDe_i^2(\bbe_g^*)\A_i(\bbe_g^*)\X_i(\bbe_g^0-\bbe_g)\\
\lesssim&
\la_{\max}(\X_i^\top \X_i) \la_{\min}^{-1}(\overline\H^{*}) ||\{\overline\H_g^{*}(\bbe_g^0)\}^{1/2}(\bbe_g-\bbe_g^0)||.
\end{align*}
Then, from Assumption (A7) and (A9) (ii) we have
\begin{align*}
\sup_{\bbe\in\Bc_{nT}}&\sup_{||\bla||=1} |\bla^\top[\B_g^{[2]*}(\bbe_g)-\overline\B_g^{[2]*}(\bbe_g)]\bla|\nonumber\\
=&
nO_p(\{T\la_{\min}^{-1/2}(\overline\H^{*})\tau^{1/2}\vee Tn^{-1/2}\}) O_p(T^{1/2})O_p(T^{1/2})\la_{\min}^{-1/2}(\overline\H^{*})\tau^{1/2}\\
=&
O_p(\{\la_{\min}^{-1/2}(\overline\H^{*})\tau^{1/2}\vee n^{-1/2}\}T^2n)\la_{\max}^{-1/2}(\overline\H^{*})\tau^{1/2},
\end{align*}
which proves $\sup_{\bbe\in\Bc_{nT}}\sup_{||\bla||=1} |\bla^\top[\B_g^{[2]*}(\bbe_g)-\overline\B_g^{[2]*}(\bbe_g)]\bla|=O_p(\{\la_{\min}^{-1/2}(\overline\H^{*})\tau^{1/2}\vee n^{-1/2}\}T^2n)$ since $\la_{\min}^{-1}(\overline\H^{*})\tau\to0$.
Lastly, we will verify
\begin{align*} 
\sup_{\bbe\in\Bc_{nT}}\sup_{||\bla||=1} |\bla^\top[{\text{\boldmath $\mathcal E$}}_g^{[1]*}(\bbe_g)-{\overline{{\text{\boldmath $\mathcal E$}}}}_g^{[1]*}(\bbe_g)]\bla|=O_p(\{\la_{\min}^{-1/2}(\overline\H^{*})\tau^{1/2}\vee n^{-1/2}\}T^2n),
\end{align*}
and
\begin{align*}
\sup_{\bbe\in\Bc_{nT}}\sup_{||\bla||=1} |\bla^\top[{{\text{\boldmath $\mathcal E$}}}_g^{[2]*}(\bbe_g)-{\overline{{\text{\boldmath $\mathcal E$}}}}_g^{[2]*}(\bbe_g)]\la|=O_p(\{\la_{\min}^{-1/2}(\overline\H^{*})\tau^{1/2}\vee n^{-1/2}\}T^2n).
\end{align*}
We have  from Cauchy-Schwarz inequality
\begin{align*}
|\bla^\top[&{\text{\boldmath $\mathcal E$}}_g^{[1]*}(\bbe_g)-{\overline{\text{\boldmath $\mathcal E$}}}_g^{[1]*}(\bbe_g)]\bla|\\
=&\Big|\sum_{i:g_i^0=g}\bla^\top \X_i^\top \G_i^{[1]}(\bbe_g){\rm diag}[\X_i\bla]\{\widehat\R^{-1}(\bbe,\bga)-\overline\R^{-1}(\bbe,\bga)\}\A_i^{-1/2}(\bbe_g)\A_i^{1/2}(\bbe_g^0)\bep_i\Big|\\
\leq&
\sum_{i:g_i^0=g}||\G_i^{[1]}(\bbe_g){\rm diag}[\X_i\bla]\X_i\bla||\cdot||\{\widehat\R^{-1}(\bbe,\bga)-\overline\R^{-1}(\bbe,\bga)\}\A_i^{-1/2}(\bbe_g)\A_i^{1/2}(\bbe_g^0)\bep_i||\\
\lesssim&
\sum_{i:g_i^0=g}\max_{1\leq j\leq T}\{||\x_{it}^\top\bla||\} 
\la_{\max}^{1/2}(\X_i^\top \X_i)||\widehat\R(\bbe,\bga)-\overline\R(\bbe,\bga)||_F ||\bep_i||.
\end{align*}
Then, from Assumption (A7) and (A9) (ii) we have we have
\begin{align*}
\sup_{\bbe\in\Bc_{nT}}&\sup_{||\bla||=1} |\bla^\top[{\text{\boldmath $\mathcal E$}}_{nk}^{[1]}(\bbe_k)-{\overline{\text{\boldmath $\mathcal E$}}}^{[1]*}(\bbe_k)]\bla|\\
=&
nO_p(T^{1/2})O_p(\{T\la_{\min}^{-1/2}(\overline\H^{*})\tau^{1/2}\vee Tn^{-1/2}\}) O_p(T^{1/2})\\
=&
O_p(\{\la_{\min}^{-1/2}(\overline\H^{*})\tau^{1/2}\vee n^{-1/2}\}T^2n),
\end{align*}
which proves $\sup_{\bbe\in\Bc_{nT}}\sup_{||\bla||=1} |\bla^\top[{\text{\boldmath $\mathcal E$}}_g^{[1]*}(\bbe_g)-{\overline{\text{\boldmath $\mathcal E$}}}_g^{[1]*}(\bbe_g)]\bla|=O_p(\{\la_{\min}^{-1/2}(\overline\H^{*})\tau^{1/2}\vee n^{-1/2}\}T^2n)$.
Moreover, we have from Cauchy-Schwarz inequality
\begin{align*}
|\bla^\top&[{\text{\boldmath $\mathcal E$}}_g^{[2]*}(\bbe_g)-{\overline{\text{\boldmath $\mathcal E$}}}_g^{[2]*}(\bbe_g)]\bla|\\
=&
\Big|\sum_{i:g_i^0=g}\bla^\top \X_i^\top\bDe_i(\bbe_g)\A_i^{1/2}(\bbe_g)\{\widehat\R^{-1}(\bbe,\bga)-\overline\R^{-1}(\bbe,\bga)\}{\rm diag}[\A_i^{1/2}(\bbe_{g_i^0}^0)\bep_i]\G_i^{[2]}(\bbe_g)\X_i\bla\Big|\\
\leq&
\Big(\sum_{i:g_i^0=g}||\{\widehat\R^{-1}(\bbe,\bga)-\overline\R^{-1}(\bbe,\bga)\}\A_i^{1/2}(\bbe_g)\bDe_i(\bbe_g)\X_i\bla||^2\Big)^{1/2}\\
&\ \ \ \times\Big(\sum_{i:g_i^0=g}||{\rm diag}[\A_i^{1/2}(\bbe_{g_i^0}^0)\bep_i]\G_i^{[2]}(\bbe_g)\X_i\bla||^2\Big)^{1/2}\\
\lesssim&
||\widehat\R(\bbe,\bga)-\overline\R(\bbe,\bga)||_F \max_{1\leq j\leq T}\{|A_{it}^{1/2}(\bbe_{g_i^0}^0)\ep_{it}|\}\la_{\max}\Big(\sum_{i:g_i^0=g}\X_i^\top \X_i\Big).
\end{align*}
Then, from Assumption (A7) and (A9) (ii) we have
\begin{align*}
\sup_{\bbe\in\Bc_{nT}}&\sup_{||\bla||=1}|\bla^\top[{\text{\boldmath $\mathcal E$}}_g^{[2]*}(\bbe_g)-{\overline{\text{\boldmath $\mathcal E$}}}_g^{[2]*}(\bbe_g)]\bla|\\
=&
O_p(\{T\la_{\min}^{-1/2}(\overline\H^{*})\tau^{1/2}\vee Tn^{-1/2}\})O_p(nT)\\
=&
O_p(\{\la_{\min}^{-1/2}(\overline\H^{*})\tau^{1/2}\vee n^{-1/2}\}T^2n),
\end{align*}
which proves $\sup_{\bbe\in\Bc_{nT}}\sup_{||\bla||=1} |\bla^\top[{\text{\boldmath $\mathcal E$}}_g^{[2]*}(\bbe_g)-{\overline{\text{\boldmath $\mathcal E$}}}_g^{[2]*}(\bbe_g)]\bla|=O_p(\{\la_{\min}^{-1/2}(\overline\H^{*})\tau^{1/2}\vee n^{-1/2}\}T^2n)$.
\end{proof}

\bigskip

The following three lemmas are from Lemma A.1. (ii), Lemma A.2. (ii), Lemma A.3. (ii) in \cite{XY2003}, respectively.
These three lemmas are hold under the assumption (AH) in \cite{XY2003}, which is satisfied in our problem from Assumptions (A1).

\begin{lemma}\label{lem:hh}
Suppose Assumption (A1) and (A8) (i) hold.
It holds that, for any $\bla\in\Rr^p$ and $g=1,\ldots,G$,
\begin{align*}
\sup_{\bbe\in\Bc_{nT}}\sup_{||\bla||=1} |\bla^\top\{\overline\H_g^{*}(\bbe_g^0)\}^{-1/2}\overline\H_g^{*}(\bbe_g)\{\overline\H_g^{*}(\bbe_g^0)\}^{-1/2}\bla-1|=o_p(1).
\end{align*}
\end{lemma}

\begin{lemma}\label{lem:hb}
Suppose Assumptions  (A1) and (A8) (i) hold.
It holds that, for any $\bla\in\Rr^p$ and $g=1,\ldots,G$,
\begin{align*}
\sup_{\bbe\in\Bc_{nT}}\sup_{||\bla||=1} |\bla^\top\{\overline\H_g^{*}(\bbe_g^0)\}^{-1/2}\overline\B_g^{*}(\bbe_g)\{\overline\H_g^{*}(\bbe_g^0)\}^{-1/2}\bla|=o_p(1).
\end{align*}
\end{lemma}

\begin{lemma}\label{lem:he}
Suppose Assumptions  (A1) and (A8) (ii) hold.
It holds that, for any $\bla\in\Rr^p$ and $g=1,\ldots,G$,
\begin{align*}
\sup_{\bbe_g\in\Bc_{nT}}\sup_{||\bla||=1} |\bla^\top\{\overline\H_g^{*}(\bbe_g^0)\}^{-1/2}{\overline{\text{\boldmath $\mathcal E$}}}_g^{*}(\bbe_g)\{\overline\H_g^{*}(\bbe_g^0)\}^{-1/2}\bla|=o_p(1).
\end{align*}
\end{lemma}

\bigskip

The proof is based on that of Theorem 3.6 in \cite{Wang2011}.
We will verify the following condition:
for any $\epsilon>0$, there exists a constant $C>0$ such that for all $n$ and $T$ sufficiently large,
\begin{align*}
P\Big(\sup_{\bbe\in\Bc_{nT},\bga\in\Ga}(\bbe_g-\bbe_g^0)^\top \S_g(\bbe_g)<0\Big)\geq 1-\epsilon,
\end{align*}
where $\Bc_{nT}=\{\bbe : \max_{g=1,\ldots,G}||\{\overline\H_g^{*}(\bbe_g^0)\}^{1/2}(\bbe_g-\bbe_g^0)||= C\tau^{1/2}\}$ and $\Ga=\{\bga=(g_1,\ldots,g_n) : n^{-1}\sum_{i=1}^n \one\{g_i\neq g_i^0\}=o_p(T^{-\de}) \ \  {\rm for \ all} \ \ \de>0\}$.
This is a sufficient condition to ensure the existence of a sequence of roots $\bbeh_g$ of the equation $\S_g(\bbe_g)=0$ for $g=1,\ldots,G$ such that $\bbeh\in\Bc_{nT}$ for $\bga\in\Ga$.
This is because from Assumption (A5) and (A7), we can estimate each $\bbe_i$ consistently by solving $\S_i(\bbe_i)=\zero$, and then, $P(\bga\notin\Ga)=o_p(1)$ from Lemma \ref{lem:gc}.

From Taylor expansion, we can write
\begin{align*}
(\bbe_g-\bbe_g^0)^\top \S_g(\bbe_g)
=&(\bbe_g-\bbe_g^0)^\top \S_g(\bbe_g^0)-(\bbe_g-\bbe_g^0)^\top\sum_{i=1}^n \one\{g_i=g\}\bDs_i(\bbe_{g_i}^*)(\bbe_{g_i}-\bbe_g^0)\\
\equiv& I_1+I_2,
\end{align*}
where $\bbe_{g_i}^*$ lies between $\bbe_{g_i}$ and $\bbe_g^0$ for $i=1,\ldots,n$.
Next, we write
\begin{align*}
I_1=(\bbe_g-\bbe_g^0)^\top\overline\S_g^{*}(\bbe_g^0)+(\bbe_g-\bbe_g^0)^\top\{\S_g(\bbe_g^0)-\overline\S_g^{*}(\bbe_g^0)\}
\equiv I_{11}+I_{12}.
\end{align*}
For $\ell=1,\ldots,p$, denote $\e_\ell\in{\mathbb R}^p$ with $\ell$th element equal to $1$ and the others equal to $0$.
Then, we have
\begin{align*}
E[&\{\e_\ell^\top\{\overline\H_g^{*}(\bbe_g^0)\}^{-1/2}\overline\S_g^{*}(\bbe_g^0)\}^2]\\
=&
\e_\ell^\top\{\overline\H_g^{*}(\bbe_g^0)\}^{-1/2}\sum_{i=1}^n\one\{g_i^0=g\}\X_i^\top\bDe_i(\bbe_g^0)\A_i^{1/2}(\bbe_g^0)\overline\R^{-1}(\bbe^0,\bga^0)\R^0\overline\R^{-1}(\bbe^0,\bga^0)\\
&\ \times \A_i^{1/2}(\bbe_g^0)\bDe_i(\bbe_g^0)\X_i\{\overline\H_g^{*}(\bbe_g^0)\}^{-1/2}\e_\ell\\
\leq&
\la_{\max}(\R^0\overline\R^{-1}(\bbe^0,\bga^0)).
\end{align*}
Thus, we can bound $|I_{11}|$ by
\begin{align*}
\sup_{\bbe\in\Bc_{nT}}|I_{11}|\leq||\{\overline\H_g^{*}(\bbe_g^0)\}^{1/2}(\bbe_g-\bbe_g^0)||\cdot||\{\overline\H_g^{*}(\bbe_g^0)\}^{-1/2}\overline\S_g^{*}(\bbe_g^0)||
\leq C\tau.
\end{align*}
From the Lemma \ref{lem:so} and \ref{lem:s}, we have
\begin{align*}
\sup_{\bbe\in\Bc_{nT}}|I_{12}|\leq&||\{\overline\H_g^{*}(\bbe_g^0)\}^{1/2}(\bbe_g-\bbe_g^0)||\cdot||\{\overline\H_g^{*}(\bbe_g^0)\}^{-1/2}\{\S_g(\bbe_g^0)-\overline\S_g^{*}(\bbe_g^0)\}||\\
\leq&
\tau^{1/2}O_p(\la_{\min}^{-1/2}(\overline\H^{*}) T^2).
\end{align*}
Since $\tau^{-1/2}\la_{\min}^{-1/2}(\overline\H^{*}) T^2\to0$ from Assumption (A3), $\sup_{\bbe\in\Bc_{nT}}|I_{12}|=o_p(\tau)$.
Hence, we have $\sup_{\bbe\in\Bc_{nT}}|I_1|\leq C\tau$.
In what follows, we will evaluate $I_2$.
It can be written as
\begin{align*}
I_2=&-(\bbe_g-\bbe_g^0)^\top\sum_{i=1}^n \one\{g_i=g\}\overline\bDs_i(\bbe_{g_i}^*)(\bbe_{g_i}-\bbe_g^0)\\
&-(\bbe_g-\bbe_g^0)^\top\sum_{i=1}^n \one\{g_i=g\}\{\bDs_i(\bbe_{g_i}^*)-\overline\Ds_i(\bbe_{g_i}^*)\}(\bbe_{g_i}-\bbe_g^0)\\
\equiv&I_{21}+I_{22}.
\end{align*}
For $g_i^0=g_i=g$, $\bbe_{g_i}^*$ lies between $\bbe_g$ and $\bbe_g^0$, and then we write $\bbe_{g_i}^*\equiv\bbe_g^*$ for such $i$.
Hence, we can write
\begin{align*}
I_{21}=&
-(\bbe_g-\bbe_g^0)^\top\overline\Ds_g^{*}(\bbe_g^*)(\bbe_g-\bbe_g^0)\\
&-(\bbe_g-\bbe_g^0)^\top\sum_{i=1}^n (\one\{g_i=g\}-\one\{g_i^0=g\})\overline\Ds_i(\bbe_{g_i}^*)(\bbe_{g_i}-\bbe_g^0)\\
\equiv&
I_{211}+I_{212}.
\end{align*}
For $I_{211}$, we write
\begin{align*}
I_{211}=&
-(\bbe_g-\bbe_g^0)^\top\overline\H_g^{*}(\bbe_g^*)(\bbe_g-\bbe_g^0)
-(\bbe_g-\bbe_g^0)^\top\{\overline\bDs_g^{*}(\bbe_g^*)-\overline\H_g^{*}(\bbe_g^*)\}(\bbe_g-\bbe_g^0)\\
\equiv&
I_{2111}+I_{2112}.
\end{align*}
For $I_{2111}$, we can write
\begin{align*}
I_{2111}=&
-(\bbe_g-\bbe_g^0)^\top\overline\H_g^{*}(\bbe_g^0)(\bbe_g-\bbe_g^0)\\
&-(\bbe_g-\bbe_g^0)^\top\{\overline\H_g^{*}(\bbe_g^0)\}^{1/2} \Big[\{\overline\H_g^{*}(\bbe_g^0)\}^{-1/2}\overline\H_g^{*}(\bbe_g^*)\{\overline\H_g^{*}(\bbe_g^0)\}^{-1/2}- \I_p\Big]\\
&\ \ \ \ \ \ \times \{\overline\H_g^{*}(\bbe_g^0)\}^{1/2}(\bbe_g-\bbe_g^0)\\
\equiv&
I_{21111}+I_{21112}.
\end{align*}
For $\bbe\in\Bc_{nT}$, we have $I_{21111}=-C^2\tau$.
Moreover, for $g_i^0=g_i=g$, $\bbe_{g_i}^*\equiv\bbe_g^*$ is contained in a local neighborhood of $\bbe_g^0$.
Then, for $I_{21112}$, we have from Lemma \ref{lem:hh},
\begin{align*}
|I_{21112}|
\leq&\sup_{\bbe\in\Bc_{nT}} \max\Big\{\Big|\la_{\min}\Big(\Big[\{\overline\H_g^{*}(\bbe_g^0)\}^{-1/2}\overline\H_g^{*}(\bbe_g^*)\{\overline\H_g^{*}(\bbe_g^0)\}^{-1/2}- \I_p\Big]\Big)\Big|,\\
&\ \ \ \ \ \ \ \ \ \ \ \ \ \ \ \ \ \ \ \ \ \ \ \ \Big|\la_{\max}\Big(\Big[\{\overline\H_g^{*}(\bbe_g^0)\}^{-1/2}\overline\H_g^{*}(\bbe_g^*)\{\overline\H_g^{*}(\bbe_g^0)\}^{-1/2}- \I_p\Big]\Big)\Big|\Big\}\\
&\ \ \ \ \ \ \ \times ||\{\overline\H_g^{*}(\bbe_g^0)\}^{1/2}(\bbe_g-\bbe_g^0)||^2\\
=&o(1) C^2\tau,
\end{align*}
which is dominated by $I_{21111}$.
Hence, for $\bbe\in\Bc_{nT}$ we have $I_{2111}=-C^2\tau$.
Next, we verify $I_{2112}$.
For $g_i^0=g_i=g$, we have from Lemma \ref{lem:dhbe}, \ref{lem:hb} and \ref{lem:he}
\begin{align*}
|I_{2112}|
=&
|(\bbe_g-\bbe_g^0)^\top\{\overline\B_g^{*}(\bbe_g^*)+{\overline{\text{\boldmath $\mathcal E$}}}_g^{*}(\bbe_g^*)\}(\bbe_g-\bbe_g^0)|\\
\leq&\sup_{\bbe\in\Bc_{nT}}\{\la_{\max}(\{\overline\H_g^{*}(\bbe_g^0)\}^{-1/2}\overline\B_g^{*}(\bbe_g^*)\{\overline\H_g^{*}(\bbe_g^0)\}^{-1/2})\\
&\ \ \ \ \ \ +\la_{\max}(\{\overline\H_g^{*}(\bbe_g^0)\}^{-1/2}{\overline{\text{\boldmath $\mathcal E$}}}_g^{*}(\bbe_g^*)\{\overline\H_g^{*}(\bbe_g^0)\}^{-1/2})\}||\{\overline\H_g^{*}(\bbe_g^0)\}^{1/2}(\bbe_g-\bbe_g^0)||^2\\
=&o(1) C^2\tau,
\end{align*}
which is dominated by $I_{2111}$.
Hence, for $\bbe\in\Bc_{nT}$ we have $I_{211}=-C^2\tau$.
Next, we verify $I_{212}$.
\begin{align*}
|I_{212}|\leq&
\Big|(\bbe_g-\bbe_g^0)^\top\sum_{i=1}^n (\one\{g_i=g\}-\one\{g_i^0=g\})\overline\H_i(\bbe_{g_i}^*)(\bbe_{g_i}-\bbe_{g_i^0}^0)\Big|\\
&+\Big|(\bbe_g-\bbe_g^0)^\top\sum_{i=1}^n (\one\{g_i=g\}-\one\{g_i^0=g\})\overline\B_i(\bbe_{g_i}^*)(\bbe_{g_i}-\bbe_{g_i^0}^0)\Big|\\
&+\Big|(\bbe_g-\bbe_g^0)^\top\sum_{i=1}^n (\one\{g_i=g\}-\one\{g_i^0=g\}){\overline{\text{\boldmath $\mathcal E$}}}_i(\bbe_{g_i}^*)(\bbe_{g_i}-\bbe_{g_i^0}^0)\Big|\\
\equiv&
I_{2121}+I_{2122}+I_{2123}.
\end{align*}
From Cauchy-Schwarz inequality, $\bbe\in\Bc_{nT}$ we have
\begin{align*}
|I_{2121}|
\lesssim&
\la_{\min}^{-1/2}(\overline\H^{*})||\{\overline\H_g^{*}(\bbe_g^0)\}^{1/2}(\bbe_g-\bbe_g^0)||\sum_{i=1}^n\one\{g_i\neq g_i^0\}n\{\max_{1\leq i\leq n}\sup_{\bbe\in\Bc}\la_{\max}(\overline\H_i(\bbe_{g_i}))\}\\
\lesssim&
C\la_{\min}^{-1/2}(\overline\H^{*})\tau^{1/2}n\Big(\frac1n\sum_{i=1}^n \one\{g_i\neq g_i^0\}\Big) n\{\max_{1\leq i\leq n}\sup_{\bbe\in\Bc}\la_{\max}(\overline\H_i(\bbe_{g_i}))\}.
\end{align*}
From Assupmtions (A1) and (A6), for $i=1,\ldots,n$ we have 
\begin{align*}
\max_{\bbe\in\Bc}\{\la_{\max}(\overline\H_i(\bbe_{g_i}))\}
\lesssim
\max_{\bbe\in\Bc}\max_{t=1,\ldots,T}[a''(\th_{it}(\bbe_{g_i}))\{u'(\x_{it}^\top\bbe_{g_i})\}^2]\la_{\max}(\X_i^\top \X_i)
=
O_p(T),
\end{align*}
for $\bbe_{g_i}^*$ between $\bbe_{g_i^0}^0$ and $\bbe_{g_i}$, which implies that
\begin{align*}
\sup_{\bbe\in\Bc_{nT}, \bga\in\Ga}|I_{2121}|=C\la_{\min}^{-1/2}(\overline\H^{*})\tau^{1/2}n^2To_p(T^{-\de})=o_p(\tau).
\end{align*}
Similarly, Cauchy-Schwarz inequality we have
\begin{align*}
\sup_{\bbe\in\Bc_{nT}, \bga\in\Ga}|I_{2122}|
\lesssim&
C\la_{\min}^{-1/2}(\overline\H^{*})\tau^{1/2}n\Big(\frac1n\sum_{i=1}^n \one\{g_i\neq g_i^0\}\Big)\\
&\times n[\{\max_{1\leq i\leq n}\sup_{\bbe\in\Bc}||\overline\B_i^{[1]}(\bbe_{g_i})||_F\}+\{\max_{1\leq i\leq n}\sup_{\bbe\in\Bc}||\overline\B_i^{[2]}(\bbe_{g_i})||_F\}].
\end{align*}
It is noted that we have from Cauchy-Schwarz inequality
\begin{align*}
\{&\overline\B_i^{[1]}(\bbe_{g_i})\}_{jk}\\
=&\e_j^\top \X_i^\top{\rm diag}[\overline\R^{-1}(\bbe,\bga)\A_i^{-1/2}(\bbe_{g_i})\{m(\X_i\bbe_{g_i^0}^0)-m(\X_i\bbe_{g_i})\}]\G_i^{[1]}(\bbe_{g_i})\X_i\e_k\\
\leq&
\la_{\max}(\X_i^\top \X_i)\la_{\max}({\rm diag}[\overline\R^{-1}(\bbe,\bga)\A_i^{-1/2}(\bbe_{g_i})\{m(\X_i\bbe_{g_i^0}^0)-m(\X_i\bbe_{g_i})\}])\la_{\max}(\G_i^{[1]}(\bbe_{g_i}))\\
=&
\la_{\max}(\X_i^\top \X_i)
\max_{1\leq k\leq T}\Big\{\sum_{t=1}^T\{\overline\R^{-1}(\bbe,\bga)\}_{kj}A_{it}^{-1/2}(\bbe_{g_i})\{m(\x_{it}^\top\bbe_{g_i^0}^0)-m(\x_{it}^\top\bbe_{g_i})\}\Big\}\\
&\times\la_{\max}(\G_i^{[1]}(\bbe_{g_i}))\\
=&O_p(T^2).
\end{align*}
Similarly $\{\overline\B_i^{[2]}(\bbe_{g_i})\}_{jk}=O_p(T^2)$, then we have
\begin{align*}
\sup_{\bbe\in\Bc_{nT}, \bga\in\Ga}|I_{2122}|=
\la_{\min}^{-1/2}(\overline\H^{*})\tau^{1/2}no_p(T^{-\de})nT^{5/2}=o_p(\tau).
\end{align*}
Similarly, we have
\begin{align*}
\sup_{\bbe\in\Bc_{nT}}|I_{2123}|
\lesssim&
C\la_{\min}^{-1/2}(\overline\H^{*})\tau^{1/2}n\Big(\frac1n\sum_{i=1}^n \one\{g_i\neq g_i^0\}\Big)\\
&\times n[\{\max_{1\leq i\leq n}\sup_{\bbe\in\Bc}||{\overline{\text{\boldmath $\mathcal E$}}}_i^{[1]}(\bbe_{g_i})||_F\}+\{\max_{1\leq i\leq n}\sup_{\bbe\in\Bc}||{\overline{\text{\boldmath $\mathcal E$}}}_i^{[2]}(\bbe_{g_i})||_F\}].
\end{align*}
It is noted that we have
\begin{align*}
E&[||{\overline{\text{\boldmath $\mathcal E$}}}_i^{[1]}(\bbe_{g_i})||_F^2]\\
=&\sum_{\ell=1}^T E[\e_\ell^\top{\overline{\text{\boldmath $\mathcal E$}}}_i^{[1]}(\bbe_{g_i})^\top{\overline{\text{\boldmath $\mathcal E$}}}_i^{[1]}(\bbe_{g_i})\e_\ell]\\
=&
\sum_{\ell=1}^T E\Big[\bep_i^\top \A_i^{1/2}(\bbe_{g_i^0}^0)\A_i^{-1/2}(\bbe_{g_i})\overline\R^{-1}(\bbe,\bga){\rm diag}[\X_i\e_\ell]\G_i^{[1]}(\bbe_{g_i})\X_i\\
&\ \ \ \ \ \ \ \ \ \ \times \X_i^\top \G_i^{[1]}(\bbe){\rm diag}[\X_i\e_\ell]\overline\R^{-1}(\bbe,\bga)\A_i^{-1/2}(\bbe_{g_i})\A_i^{1/2}(\bbe_{g_i^0}^0)\bep_i\Big]\\
\leq&
\sum_{\ell=1}^T \la_{\max}(\X_i^\top \X_i) \max_{1\leq i\leq n,1\leq t\leq T} \max_{\be\in\Bc}|{q'}_{it}^{[1]}(\bbe_{g_i})| \max_{1\leq i\leq n,1\leq t\leq T}|\x_{it}^\top \e_\ell|^2\\
&\ \ \ \ \times\max_{\bbe\in\Bc}\{ \max_{1\leq i\leq n,1\leq t\leq T}A_{it}^{-1}(\bbe_{g_i})A_{it}(\bbe_{g_i^0}^0)\} E[\bep_i^\top\bep_i]\\
=&
O(T^3),
\end{align*}
which implies that $||{\overline{\text{\boldmath $\mathcal E$}}}_i^{[1]}(\bbe_{g_i})||_F=O_p(T^{3/2})$.
Similarly $||{\overline{\text{\boldmath $\mathcal E$}}}_i^{[2]}(\bbe_{g_i})||_F=O_p(T^{3/2})$, then we have
\begin{align*}
\sup_{\bbe\in\Bc_{nT}, \bga\in\Ga}|I_{2123}|=
\la_{\min}^{-1/2}(\overline\H^{*})\tau^{1/2}no_p(T^{-\de})nT^{3/2}=o_p(\tau).
\end{align*}
Thus, $I_{2121}$, $I_{2122}$ and $I_{2123}$ are dominated by $I_{211}$ for $\bbe\in\Bc_{nT}$ and $\bga\in\Ga$.
Hence $I_{21}=-C^2 \tau$ for $\bbe\in\Bc_{nT}$ and $\bga\in\Ga$.
Lastly, we verify $I_{22}$.
We can write
\begin{align*}
I_{22}=&
-(\bbe_g-\bbe_g^0)^\top\sum_{i=1}^n \one\{g_i=g\}\one\{g_i=g_i^0\}\{\bDs_i(\bbe_{g_i}^*)-\overline\bDs_i(\bbe_{g_i}^*)\}(\bbe_{g_i}-\bbe_g^0)\\
&-(\bbe_g-\bbe_g^0)^\top\sum_{i=1}^n \one\{g_i=g\}\one\{g_i\neq g_i^0\}\{\bDs_i(\bbe_{g_i}^*)-\overline\bDs_i(\bbe_{g_i}^*)\}(\bbe_{g_i}-\bbe_g^0)\\
\equiv&I_{221}+I_{222}.
\end{align*}
For $I_{221}$, we can write, from Lemma \ref{lem:dr},
\begin{align*}
|I_{221}|
\leq&
\sup_{\bbe\in\Bc_{nT}}\max\{|\la_{\max}(\bDs_i(\bbe_{g_i}^*)-\overline\bDs_g^{*}(\bbe_g))|,|\la_{\min}(\bDs_i(\bbe_{g_i}^*)-\overline\bDs_g^{*}(\bbe_g))|\}\\
&\times \la_{\min}^{-1}(\overline\H^{*}) ||\{\overline\H_g^{*}(\bbe_g^0)\}^{1/2}(\bbe_g-\bbe_g^0)||^2\\
=&
O_p(\{\la_{\min}^{-1/2}(\overline\H^{*})\tau^{1/2}\vee n^{-1/2}\}T^2n) C^2\la_{\min}^{-1}(\overline\H_g^{*})\tau.
\end{align*}
Since $\la_{\min}(\overline\H^{*})$ is at least of order larger than $O_p(nT)$, and from definition, we have $\tau=\sup_{\bbe\in\Bc,\bga}\la_{\max}(\{\overline\R(\bbe,\bga)\}^{-1}\R^0)\leq\sup_{\bbe\in\Bc,\bga}\la_{\max}(\{\overline\R(\bbe,\bga)\}^{-1})\la_{\max}(\R^0)\leq O_p(T)$ form Assumption (A5), the order of $\tau\la_{\min}^{-2}(\overline\H^*)n^2$ is at most $O_p(T^{-1})$.
Then, from Assumption (A3) we have $\sup_{\bbe\in\Bc_{nT}}|I_{221}|=\tau o_p(1)$.
As for $I_{222}$, we have
\begin{align*}
|I_{222}|
\leq&
\la_{\min}^{-1/2}(\overline\H^{*}) ||\{\overline\H_g^{*}(\bbe_g^0)\}^{1/2}(\bbe_g-\bbe_g^0)||\\
&\times \sum_{i=1}^n \one\{g_i=g\}\one\{g_i\neq g_i^0\}\cdot||\bDs_i(\bbe_{g_i}^*)-\overline\bDs_i(\bbe_{g_i}^*)||_F\cdot||\bbe_{g_i}-\bbe_g^0||\\
\leq&
\la_{\min}^{-1/2}(\overline\H^{*})\tau n \Big(\sum_{i=1}^n \one\{g_i\neq g_i^0\}\Big) \sum_{i=1}^n ||\bDs_i(\bbe_{g_i}^*)-\overline\bDs_i(\bbe_{g_i}^*)||_F\cdot||\bbe_{g_i}-\bbe_g^0||.
\end{align*}
It is noted that he order of $||\bDs_i(\bbe_{g_i})-\overline\bDs_i(\bbe_{g_i})||_F$ is at most $O_p(T)$.
Then, form Lemma \ref{lem:gc}, $\sup_{\bbe\in\Bc_{nT}, \bga\in\Ga}|I_{222}|=o_p(T^{-\de})$, which implies that $I_{22}$ is dominated by $I_{21}$.
Thus, $(\bbe_g-\bbe_g^0)^\top \S_g(\bbe_g)$ on $\bbe\in\Bc_{nT}$ and $\bga\in\Ga$ is asymptotically dominated in probability by $I_{11}+I_{21}=C\tau-C^2\tau$, which is negative for $C$ large enough, which proves the first part of the Theorem.

Next, we show the second part of the theorem.
We have
\begin{align*}
P\Big(&\max_{1\leq i\leq n}|\gh_i(\bbeh)-g_i^0|>0\Big)\\
\leq&G\max_{1\leq g\leq G}P(\bbeh_g\notin\Bc_{nT})+n\max_{1\leq i\leq n}P\Big(\bbeh_g\in\Bc_{nT},\gh_i(\bbeh)\neq g_i^0\Big).
\end{align*}
The order of the first term is $o(1)$ from the first part of the Theorem.
We have $\sup_{\bbe\in\Bc_{nT}} \one\{\gh_i(\bbe)\neq g_i^0\}\leq\sum_{g=1}^G{\widetilde Z}_{ig}$.
Then,
\begin{align*}
\max_{1\leq i\leq n}P\Big(\bbeh_g\in\Bc_{nT},\gh_i(\bbeh)\neq g_i^0\Big)
=&
\max_{1\leq i\leq n}E[\one\{\bbeh_g\in\Bc_{nT}\}\one\{\gh_i\neq g_i^0\}]\\
\leq&
\max_{1\leq i\leq n}E\Big[\one\{\bbeh_g\in\Bc_{nT}\}\sum_{g=1}^G{\widetilde Z}_{ig}\Big]\\
\leq&
\max_{1\leq i\leq n}\sum_{g=1}^G P({\widetilde Z}_{ig}=1)=o(T^{-\de}),
\end{align*}
which proves the theorem.

\section{Proof of Theorem 2}\label{sec:pan}
To show Theorem 2, we need to show the next lemmas.

Let $\widetilde\bbe_g$ denote a root of $\overline\S_g^{*}(\bbe_g)=0$.
The next result shows that the grouped GEE estimator and the infeasible estimator with known population groups are asymptotically equivalent.

\begin{lemma}\label{lem:ae}
Suppose the Assumptions (A1)-(A9) hold.
As $n$ and $T$ tend to infinity such that $n/T^\nu\to0$ for some $\nu>0$, we have $\bbeh_g=\widetilde\bbe_g+o_p(1)$ for $g=1,\ldots,G$.
\end{lemma}
\begin{proof}
We have
\begin{align*}
\sup_{\bbe\in\Bc_{nT},\bga\in\Ga}&||\S_g(\bbe_g)-\overline\S_g^{*}(\bbe_g)||\\
\leq&
\sup_{\bbe\in\Bc_{nT}}||\S_g(\bbe_g)-\overline\S_g(\bbe_g)||+\sup_{\bbe\in\Bc_{nT},\bga\in\Ga}||\overline\S_g(\bbe_g)-\overline\S_g^{*}(\bbe_g)||.
\end{align*}
Then, we have $\sup_{\bbe\in\Bc_{nT},\bga\in\Ga}||\S_g(\bbe_g)-\overline\S_g^{*}(\bbe_g)||=O_p(T^2)$ from Lemmas \ref{lem:so} and \ref{lem:s}.
Since $\bbeh_g\in\Bc_{nT}$ for $\bga\in\Ga$ from Theorem 1 and $\widetilde\bbe_g\in\Bc_{nT}$ from Theorem 2 in \cite{XY2003}, this implies
\begin{align*}
\sup_{\bga\in\Ga}|(\bbeh_g-\widetilde\bbe_g)^\top\{\S_g(\bbeh_g)-\overline\S_g^{*}(\bbeh_g)\}|=|(\bbeh_g-\widetilde\bbe_g)^\top\overline\S_g^{*}(\bbeh_g)|=O_p(T^2).
\end{align*}
From Taylor expansion, for $\bbe_g^*$ between $\bbeh_g$ and $\widetilde\bbe_g$ we have
\begin{align*}
\overline\S_g^{*}(\bbeh_g)
=&
\overline\S_g^{*}(\widetilde\bbe_g)-\bDs_g^{*}(\bbe_g^*)(\bbeh_g-\widetilde\bbe_g)\\
=&
-\overline\H_g^{*}(\bbe_g^*)(\bbeh_g-\widetilde\bbe_g)-\{\overline\bDs_g^{*}(\bbe_g^*)-\overline\H_g^{*}(\bbe_g^*)\}(\bbeh_g-\widetilde\bbe_g).
\end{align*}
Then, we have, from Lemmas \ref{lem:hh} - \ref{lem:he},
\begin{align*}
|(\bbeh_g-\widetilde\bbe_g)^\top\{\S_g(\bbeh_g)-\overline\S_g^{*}(\bbeh_g)\}|
=&(\bbeh_g-\widetilde\bbe_g)^\top\overline\H_g^{*}(\bbe_g^*))(\bbeh_g-\widetilde\bbe_g)+o_p(1).
\end{align*}
Hence, we have
\begin{align*}
\sup_{\bga\in\Ga}\inf_{\bbe\in\Bc_{nT}}\la_{\min}(\overline\H_g^{*}(\bbe))||\bbeh_g-\widetilde\bbe_g||^2\leq O_p(T^2)+o_p(1),
\end{align*}
which implies $||\bbeh_g-\widetilde\bbe_g||=o_p(1)$, since the order of $\la_{\min}(\overline\H_g^{*}(\bbe))$ is at least $O_p(nT)$.
The Lemma follows from Lemma \ref{lem:gc}.
\end{proof}

Next lemma is almost the same with Lemma 2 in \cite{XY2003}.
\begin{lemma}\label{lem:ans}
Suppose the Assumptions (A1)-(A9) hold.
Moreover, suppose that, for all $g=1,\ldots,G$, there exists a constant $\zeta$ such that $(c^*T)^{1+\zeta}\ga^*\to0$ as $n\to\infty$.
Moreover, suppose the marginal distribution of each observation has a density of the form from (2.1) in the main text.
Then, when $n\to\infty$, we have
\begin{align*}
\{\overline\M_g^{*}(\bbe_g^0)\}^{-1/2}\overline\S_g^{*}(\bbe_g^0)\to N(0,\I_p) \quad  in \ distribution.
\end{align*}
\end{lemma}
\begin{proof}
For any $p\times1$ vector $\bla$ such that $||\bla||=1$, let $\bla^\top\{\overline\M_g^{*}(\bbe_g^0)\}^{-1/2}\overline\S_g^{*}(\bbe_g^0)=\sum_{i:g_i^0=g}Z_{nTi}$, where $Z_{nTi}=\bla^\top\{\overline\M_g^{*}(\bbe_g^0)\}^{-1/2}\X_i^\top\bDe_i(\bbe_g^0)\A_i^{1/2}(\bbe_g^0)\overline\R_i^{-1}(\bbe^0,\bga^0)\bep_i$.
To establish the asymptotic normality, it suffices to check the Lindeberg condition for $\bla^\top\{\overline\M_g^{*}(\bbe_g^0)\}^{-1/2}\overline\S_g^{*}(\bbe_g^0)$, that is, for any $\epsilon>0$, 
\begin{align*}
\sum_{i:g_i^0=g}E[Z_{nTi}^2\one\{|Z_{nTi}|>\epsilon\}]\to0,
\end{align*}
which is shown in the proof of Lemma 2 in \cite{XY2003}. 
\end{proof}

We will show
\begin{align*}
\{\overline\M_g^{*}(\bbe_g^0)\}^{-1/2}\overline\H_g^{*}(\bbe_g^0)(\widetilde\bbe_g-\bbe_g^0)\to N(0,\I_p) \quad  in \ distribution.
\end{align*}
The theorem follows from Lemma \ref{lem:ae}.

For $\bbe_g^*\in\Bc_{nT}$ between $\widetilde\bbe_g$ and $\bbe_g^0$, from Theorem 1, we have 
\begin{align*}
\{\overline\H_g^{*}&(\bbe_g^0)\}^{-1/2}\overline\S_g^{*}(\bbe_g^0)\\
=&
-\{\H_g(\bbeh_g^0)\}^{1/2}(\widetilde\bbe_g-\bbe_g^0)
+ \Big[\{\H_g(\bbeh_g)\}^{1/2}-\{\overline\H_g^{*}(\bbe_g^0)\}^{1/2}\Big](\widetilde\bbe_g-\bbe_g^0)\\
&-\Big[\{\overline\H_g^{*}(\bbe_g^0)\}^{-1/2}\overline\bDs_g^{*}(\bbe_g^*)\{\overline\H_g^{*}(\bbe_g^0)\}^{-1/2}-\I_p\Big]\{\overline\H_g^{*}(\bbe_g^0)\}^{1/2}(\widetilde\bbe_g-\bbe_g^0).
\end{align*}
From Lemmas \ref{lem:dhbe} and \ref{lem:hh} - \ref{lem:he}, the second term in the right hand side of the above equation is $o_p(1)$, which implies that $\{\overline\M_g^*(\bbe_g^0)\}^{-1/2}\overline\S_g^{*}(\bbe_g^0)$ and $\{\overline\M_g^*(\bbe_g^0)\}^{-1/2}\overline\H_g^*(\bbe_g^0)(\widetilde\bbe_g-\bbe_g^0)$ are asymptotically identically distributed.
Hence, the theorem follows from Lemma \ref{lem:ans}.

\section{Property of $\overline\R^{*}(\bbe,\bga)$}\label{sec:pr}

In this section, we denote the estimated unstructured working correlation matrix as $\widehat\R^*(\bbe,\bga)=\R^*(\balh(\bbe,\bga),\bbe,\bga)$ for $\balh(\bbe,\bga)$ given in (2.4) in the main text.
Then, it follows that 
\begin{align*}
&\overline\R^{*}(\bbe,\bga)=\frac1n\sum_{i=1}^n\A_i^{-1/2}(\bbe_{g_i})\A_i^{1/2}(\bbe_{g_i^0}^0)\R^0\A_i^{1/2}(\bbe_{g_i^0}^0)\A_i^{-1/2}(\bbe_{g_i})\\
& \ \ +\frac1n\sum_{i=1}^n\A_i^{-1/2}(\bbe_{g_i})\{m(\X_i\bbe_{g_i^0}^0)-m(\X_i\bbe_{g_i})\}\{m(\X_i\bbe_{g_i^0}^0)-m(\X_i\bbe_{g_i})\}^\top \A_i^{-1/2}(\bbe_{g_i}).
\end{align*}

The next lemma shows that $\overline\R^{*}(\bbe,\bga)$ satisfies Assumption (A5) (ii).
\begin{lemma}\label{lem:rs}
Suppose Assumptions (A1)-(A8) hold.
It holds that $\la_{\max}(\{\overline\R^{mo}(\bbe^0,\bga)\}^{-2}\R^0)=O_p(1)$ for any $\bga$.
\end{lemma}
\begin{proof}
Since the eigenvalues of $\overline\R^{*}(\bbe^0,\bga)(\R^0)^{-1/2}$ and $(\R^0)^{-1/4}\overline\R^{*}(\bbe^0,\bga)(\R^0)^{-1/4}$ are the same, we will show that $\la_{\min}((\R^0)^{-1/4}\overline\R^{*}(\bbe^0,\bga)(\R^0)^{-1/4})$ is bounded away from zero.
It can be written as
\begin{align*}
\la_{\min}(&(\R^0)^{-1/4}\overline\R^*(\bbe^0,\bga)(\R^0)^{-1/4})\\
\geq&
\la_{\min}\Big((\R^0)^{-1/4}\frac1n\sum_{i=1}^n\A_i^{-1/2}(\bbe_{g_i}^0)\A_i^{1/2}(\bbe_{g_i^0}^0)\R^0\A_i^{1/2}(\bbe_{g_i^0}^0)\A_i^{-1/2}(\bbe_{g_i}^0)(\R^0)^{-1/4}\Big)\\
&+\la_{\min}\Big((\R^0)^{-1/2}\frac1n\sum_{i=1}^n \A_i^{-1/2}(\bbe_{g_i}^0)\{m(\X_i\bbe_{g_i^0}^0)-m(\X_i\bbe_{g_i}^0)\}\\
&\ \ \ \ \ \ \ \ \ \ \ \ \ \ \ \ \ \ \ \ \ \ \ \ \ \ \ \ \ \ \ \ \ \ \ \times\{m(\X_i\bbe_{g_i^0}^0)-m(\X_i\bbe_{g_i}^0)\}^\top \A_i^{-1/2}(\bbe_{g_i}^0)(\R^0)^{-1/2}\Big).
\end{align*}
Since the smallest eigenvalue does not diverge to infinity, it is enough to show that the first term of the right-hand side of the above inequality is bounded away from zero.
Then, we have
\begin{align*}
\la_{\min}&\Big((\R^0)^{-1/4}\frac1n\sum_{i=1}^n\A_i^{-1/2}(\bbe_{g_i}^0)\A_i^{1/2}(\bbe_{g_i^0}^0)\R^0\A_i^{1/2}(\bbe_{g_i^0}^0)\A_i^{-1/2}(\bbe_{g_i}^0)(\R^0)^{-1/2}\Big)\\
\geq&
\frac1n\sum_{i=1}^n\la_{\min}\Big((\R^0)^{-1/4}\A_i^{-1/2}(\bbe_{g_i}^0)\A_i^{1/2}(\bbe_{g_i^0}^0)\R^0\A_i^{1/2}(\bbe_{g_i^0}^0)\A_i^{-1/2}(\bbe_{g_i}^0)(\R^0)^{-1/4}\Big)\\
\geq&
\frac1n\sum_{i=1}^n\la_{\min}^2\Big((\R^0)^{-1/4}\A_i^{-1/2}(\bbe_{g_i}^0)\A_i^{1/2}(\bbe_{g_i^0}^0)(\R^0)^{1/2}\Big)\\
\geq&
\frac1n\sum_{i=1}^n \min_{1\leq t \leq T} \{A_{it}^{-1}(\bbe_{g_i}^0)A_{it}(\bbe_{g_i^0}^0)\}\la_{\min}^{1/4}(\R^0)>0,
\end{align*}
where the last inequality follows from Assumption (A5) (i).
\end{proof}

The next lemma shows that $\widehat\R^*(\bbe,\bga)$ satisfies Assumption (A9) (i).
\begin{lemma}\label{lem:rr}
Under Assumptions (A1)-(A8), it holds that for any $\bga$,
\begin{align*}
\sup_{\bbe\in\Bc_{nT}} \max_{1\leq k,l\leq T}\{\widehat\R^*(\bbe,\bga)-\widehat\R^*(\bbe^0,\bga)\}_{kl}=O_p(\la_{\min}^{-1/2}(\overline\H^{*})\tau^{1/2}).
\end{align*}
\end{lemma}
\begin{proof}
For any $\bga$, we can write
\begin{align*}
&\widehat\R^{*}(\bbe,\bga)-\widehat\R^{*}(\bbe^0,\bga)\\
=&
\frac1n \sum_{i=1}^n\A_i^{-1/2}(\bbe_{g_i})\{\y_i-m(\X_i\bbe_{g_i})\}\{\y_i-m(\X_i\bbe_{g_i})\}^\top \A_i^{-1/2}(\bbe_{g_i})\\
&-\sum_{i=1}^n\A_i^{-1/2}(\bbe_{g_i}^0)\{\y_i-m(\X_i\bbe_{g_i}^0)\}\{\y_i-m(\X_i\bbe_{g_i}^0)\}^\top \A_i^{-1/2}(\bbe_{g_i}^0)\\
=&
\frac1n \sum_{i=1}^n\{\A_i^{-1/2}(\bbe_{g_i})-\A_i^{-1/2}(\bbe_{g_i}^0)\}\{\y_i-m(\X_i\bbe_{g_i})\}\\
&\ \ \ \ \ \ \ \ \ \  \times\{\y_i-m(\X_i\bbe_{g_i})\}^\top\{\A_i^{-1/2}(\bbe_{g_i})-\A_i^{-1/2}(\bbe_{g_i}^0)\}\\
&+\frac1n \sum_{i=1}^n\{\A_i^{-1/2}(\bbe_{g_i})-\A_i^{-1/2}(\bbe_{g_i}^0)\}\{\y_i-m(\X_i\bbe_{g_i})\}\{\y_i-m(\X_i\bbe_{g_i})\}^\top \A_i^{-1/2}(\bbe_{g_i}^0)\\
&+\frac1n \sum_{i=1}^n\A_i^{-1/2}(\bbe_{g_i}^0)\{\y_i-m(\X_i\bbe_{g_i})\}\{\y_i-m(\X_i\bbe_{g_i})\}^\top\{\A_i^{-1/2}(\bbe_{g_i})-\A_i^{-1/2}(\bbe_{g_i}^0)\}\\
&+\frac1n \sum_{i=1}^n\A_i^{-1/2}(\bbe_{g_i}^0)\Big[\{\y_i-m(\X_i\bbe_{g_i})\}\{\y_i-m(\X_i\bbe_{g_i})\}^\top\\
&\ \ \ \ \ \ \ \ \ \ \ \ \ \ \ \ \ \ \ \ \ \ \ \ \ \ \ -\{\y_i-m(\X_i\bbe_{g_i}^0)\}\{\y_i-m(\X_i\bbe_{g_i}^0)\}^\top\Big]\A_i^{-1/2}(\bbe_{g_i}^0)\\
\equiv&\sum_{j=1}^4I_j.
\end{align*}
From Taylor expansion, for $\bbe_{g_i}^*$ between $\bbe_{g_i}$ and $\bbe_{g_i}^0$, we have
\begin{align*}
1-A_{it}^{1/2}(\bbe_{g_i})A_{it}^{-1/2}(\bbe_{g_i}^0)
=&1-\sqrt{\frac {a''(\x_{it}^\top\bbe_{g_i})}{a''(\x_{it}^\top\bbe_{g_i}^0)}}\\
=& -\frac12 \{a''(\x_{it}^\top\bbe_{g_i}^*) a''(\x_{it}^\top\bbe_{g_i}^0)\}^{-1/2}u'(\x_{it}^\top\bbe_{g_i}^*)\x_{it}^\top(\bbe_{g_i}-\bbe_{g_i}^0).
\end{align*}
Then, the $(k,l)$-element of $I_1$ can be written as
\begin{align*}
\frac1n& \sum_{i=1}^n\{A_{ik}^{-1/2}(\bbe_{g_i})-A_{ik}^{-1/2}(\bbe_{g_i}^0)\}\{A_{il}^{-1/2}(\bbe_{g_i})-A_{il}^{-1/2}(\bbe_{g_i}^0)\}\\
&\ \ \ \ \ \times\{y_{ik}-m(\x_{ik}^\top\bbe_{g_i})\}\{y_{il}-m(\x_{il}^\top\bbe_{g_i})\}\\
=&
\frac1{4n} \sum_{i=1}^n
\{a''(\x_{ik}^\top\bbe_{g_i}^*) a''(\x_{ik}^\top\bbe_{g_i}^0)\}^{-1/2}u'(\x_{ik}^\top\bbe_{g_i}^*)\x_{ik}^\top(\bbe_{g_i}-\bbe_{g_i}^0)\\
&\ \ \ \ \ \ \ \ \ \ \times\{a''(\x_{il}^\top\bbe_{g_i}^*) a''(\x_{il}^\top\bbe_{g_i}^0)\}^{-1/2}u'(\x_{il}^\top\bbe_{g_i}^*)\x_{il}^\top(\bbe_{g_i}-\bbe_{g_i}^0)\\
&\ \ \ \ \ \ \ \ \ \ \times A_{ik}^{-1/2}(\bbe_{g_i})\{y_{ik}-m(\x_{ik}^\top\bbe_{g_i})\}\{y_{il}-m(\x_{il}\bbe_{g_i})\}A_{il}^{-1/2}(\bbe_{g_i})\\
\lesssim&
\Big(\frac1n \sum_{i=1}^n (\bbe_{g_i}-\bbe_{g_i}^0)^\top \x_{ik}\x_{ik}^\top(\bbe_{g_i}-\bbe_{g_i}^0)\{y_{ik}-m(\x_{ik}^\top\bbe_{g_i})\}^2\Big)^{1/2}\\
&\ \ \ \ \times \Big(\frac1n \sum_{i=1}^n (\bbe_{g_i}-\bbe_{g_i}^0)^\top \x_{il}\x_{il}^\top(\bbe_{g_i}-\bbe_{g_i}^0)\{y_{il}-m(\x_{il}^\top\bbe_{g_i})\}^2\Big)^{1/2}\\
\leq&\{\max_{1\leq t\leq T}\la_{\max}(\x_{it}\x_{it}^\top)\} \frac1n \sum_{i=1}^n ||\bbe_{g_i}-\bbe_{g_i}^0||^2\{y_{ik}-m(\x_{ik}^\top\bbe_{g_i})\}^2,
\end{align*}
where the second last inequality follows from Cauchy-Schwarz inequality.
Since we have for all $t=1,\ldots,T$, $\la_{\max}(\x_{it}\x_{it}^\top)=O_p(1)$ and $\frac1n \sum_{i=1}^n\{y_{it}-m(\x_{it}^\top\bbe_{g_i})\}^2=O_p(1)$, this implies that the order of $\{I_1\}_{k.l}$ is $O_p(\la_{\min}^{-1}(\overline\H^{*})\tau)$ for $\bbe\in\Bc_{nT}$.
Similarly, the order of $\{I_2\}_{kl}$ and $\{I_3\}_{kl}$ are $O_p(\la_{\min}^{-1/2}(\overline\H^{*})\tau^{1/2})$ for $\bbe\in\Bc_{nT}$.
For $I_4$, we can write
\begin{align*}
I_4=&
\frac1n \sum_{i=1}^nA_i^{-1/2}(\bbe_{g_i}^0)\{m(\X_i\bbe_{g_i}^0)-m(\X_i\bbe_{g_i})\}\{\y_i-m(\X_i\bbe_{g_i}^0)\}^\top A_i^{-1/2}(\bbe_{g_i}^0)\\
&+\frac1n \sum_{i=1}^nA_i^{-1/2}(\bbe_{g_i}^0)\{\y_i-m(\X_i\bbe_{g_i}^0)\}\{m(\X_i\bbe_{g_i}^0)-m(\X_i\bbe_{g_i})\}^\top A_i^{-1/2}(\bbe_{g_i}^0)\\
&+\frac1n \sum_{i=1}^nA_i^{-1/2}(\bbe_{g_i}^0)\{m(\X_i\bbe_{g_i}^0)-m(\X_i\bbe_{g_i})\}\{m(\X_i\bbe_{g_i}^0)-m(\X_i\bbe_{g_i})\}^\top A_i^{-1/2}(\bbe_{g_i}^0)\\
\equiv&\sum_{j=1}^3I_{4j}.
\end{align*}
By using (\ref{eqn:tem}) for $\bbe_{g_i}^*$ between $\bbe_{g_i}$ and $\bbe_{g_i}^0$, the $(k,l)$-element of $I_{41}$ can be written as from Cauchy-Schwarz inequality,
\begin{align*}
\frac1n& \sum_{i=1}^nA_{ik}^{-1/2}(\bbe_{g_i}^0)A_{il}^{-1/2}(\bbe_{g_i}^0)\{m(\x_{ik}^\top\bbe_{g_i}^0)-m(\x_{ik}^\top\bbe_{g_i})\}\{y_{il}-m(\x_{il}^\top\bbe_{g_i}^0)\}\\
=&
\frac1n\sum_{i=1}^nA_{ik}^{-1/2}(\bbe_{g_i}^0)A_{il}^{-1/2}(\bbe_{g_i}^0)\phi A_{ik}(\bbe_{g_i}^*)u'(\x_{ik}^\top\bbe_{g_i}^*)\x_{ik}^\top(\bbe_{g_i}^0-\bbe_{g_i})\{y_{il}-m(\x_{il}^\top\bbe_{g_i}^0)\}\\
\lesssim&
\Big(\frac1n\sum_{i=1}^n(\bbe_{g_i}^0-\bbe_{g_i})^\top \x_{ik}\x_{ik}^\top(\bbe_{g_i}^0-\bbe_{g_i})\Big)^{1/2}\\
&\times \Big(\frac1n\sum_{i=1}^nA_{ik}^{-1}(\bbe_{g_i}^0)A_{il}^{-1}(\bbe_{g_i}^0)A_{ik}^2(\bbe_{g_i}^*)\{u'(\x_{ik}^\top\bbe_{g_i}^*)\}^2\{y_{il}-m(\x_{il}^\top\bbe_{g_i}^0)\}^\top\{y_{il}-m(\x_{il}^\top\bbe_{g_i}^0)\}\Big)^{1/2},
\end{align*}
which implies that the order of $\{I_{41}\}_{kl}$ is $O_p(\la_{\min}^{-1/2}(\overline\H^{*})\tau^{1/2})$ for $\bbe\in\Bc_{nT}$.
Similarly, the order of $\{I_{42}\}_{kl}$ and $\{I_{43}\}_{kl}$ are $O_p(\la_{\min}^{-1/2}(\overline\H^{*})\tau^{1/2})$ and $O_p(\la_{\min}^{-1}(\overline\H^{*})\tau)$, respectively for $\bbe\in\Bc_{nT}$, which proves the lemma.
\end{proof}

The next lemma shows that $\widehat\R^*(\bbe,\bga)$ satisfies Assumption (A9) (ii).
\begin{lemma}\label{lem:rr0}
Under  Assumptions (A1)-(A8), it holds that for any $\bga$,
\begin{align*}
\sup_{\bbe\in\Bc_{nT}} \max_{1\leq k,l\leq T}|\{\widehat\R^*(\bbe,\bga)-\overline\R^*(\bbe,\bga)\}_{k.l}|=O_p(n^{-1/2}\vee\la_{\min}^{-1/2}(\overline\H^{*}(\bbe^0))\tau^{1/2}),
\end{align*}
\end{lemma}
\begin{proof}
From Lemma \ref{lem:rr}, it is enough to show that
\begin{align*}
\max_{1\leq k,l\leq T}\{\widehat\R^{*}(\bbe^0,\bga)-\overline\R^{*}(\bbe^0,\bga)\}_{kl}=O_p(n^{-1/2}).
\end{align*}
We can write
\begin{align*}
\widehat\R^{*}&(\bbe^0,\bga)-\overline\R^{*}(\bbe^0,\bga)\\
=&\frac1n\sum_{i=1}^n\A_i^{-1/2}(\bbe_{g_i}^0)\Big\{\{\y_i-m(\X_i\bbe_{g_i^0}^0)\}\{\y_i-m(\X_i\bbe_{g_i^0}^0)\}^\top-\bSi_i\Big\}\A_i^{-1/2}(\bbe_{g_i}^0)\\
&+\frac1n\sum_{i=1}^n\A_i^{-1/2}(\bbe_{g_i}^0)\{m(\X_i\bbe_{g_i^0}^0)-m(\X_i\bbe_{g_i}^0)\}\{\y_i-m(\X_i\bbe_{g_i}^0)\}^\top \A_i^{-1/2}(\bbe_{g_i}^0)\\
&+\frac1n\sum_{i=1}^n\A_i^{-1/2}(\bbe_{g_i}^0)\{\y_i-m(\X_i\bbe_{g_i}^0)\}\{m(\X_i\bbe_{g_i^0}^0)-m(\X_i\bbe_{g_i}^0)\}^\top \A_i^{-1/2}(\bbe_{g_i}^0)\\
=&
I_1+I_2+I_3.
\end{align*}
For $\si_{ikl}=\{\bSi_i\}_{kl}$, the $(k,l)$-element of $I_1$ can be written as
\begin{align*}
\{I_1\}_{kl}=\frac1n\sum_{i=1}^nA_{ik}^{-1/2}(\bbe_{g_i}^0)A_{il}^{-1/2}(\bbe_{g_i}^0)[\{y_{ik}-m(\x_{ik}^\top\bbe_{g_i^0}^0)\}\{y_{il}-m(\x_{il}^\top\bbe_{g_i^0}^0)\}-\si_{ikl}].
\end{align*}
Then, it is obvious $E[\{I_1\}_{kl}]=0$ and 
\begin{align*}
\Var(\{I_1\}_{kl})=&\frac1{n^2}\sum_{i=1}^nA_{ik}^{-1}(\bbe_{g_i}^0)A_{il}^{-1}(\bbe_{g_i}^0)A_{ik}(\bbe_{g_i^0}^0)A_{il}(\bbe_{g_i^0}^0)\Var(\ep_{ik}\ep_{il})\\
\leq&
\frac1{n^2}\sum_{i=1}^nA_{ik}^{-1}(\bbe_{g_i}^0)A_{il}^{-1}(\bbe_{g_i}^0)A_{ik}(\bbe_{g_i^0}^0)A_{il}(\bbe_{g_i^0}^0)(E[\ep_{ik}^4]E[\ep_{il}^4])^{1/2}
=O_p(1/n),
\end{align*}
where the last equality follows from Assumptions (A1) and (A4).
Then, this implies that the order of the $(k,l)$-element of $I_1$ is $O_p(n^{-1/2})$.
Similarly, both of the $(k,l)$-elements of $I_2$ and $I_3$ are $O_p(n^{-1/2})$, which implies the lemma.
\end{proof}

The next lemma shows that $\widehat\R^{*}(\bbe,\bga)$ satisfies Assumption (A9) (iii).
\begin{lemma}\label{lem:rri0}
Under Assumptions (A1)-(A8), it holds that for any $\bbe\in\Bc$, $\bga$ and $\bga_{i*}$ whose only $i$th component differs from that of $\bga$,
\begin{align*}
\max_{1\leq k,l\leq T}|\{\widehat\R^{*}(\bbe,\bga_{i*})-\widehat\R^{*}(\bbe,\bga)\}_{kl}|=O_p(1/n).
\end{align*}
\end{lemma}
\begin{proof}
The lemma immediately holds since we can write
\begin{align*}
\{\widehat\R^{*}(&\bbe,\bga_{i*})-\widehat\R^{*}(\bbe,\bga)\}_{kl}\\
=&\frac1n\Big\{A_{ik}^{-1/2}(\bbe_{g_i^*})A_{il}^{-1/2}(\bbe_{g_i^*})\{y_{ik}-m(\x_{ik}^\top\bbe_{g_i^*})\}\{y_{il}-m(\x_{il}^\top\bbe_{g_i^*})\}\\
&\ \ \ \ -A_{ik}^{-1/2}(\bbe_{g_i^*})A_{il}^{-1/2}(\bbe_{g_i^*})\{y_{ik}-m(\x_{ik}^\top\bbe_{g_i^*})\}\{y_{il}-m(\x_{il}^\top\bbe_{g_i^*})\},
\end{align*}
which is of order $O_p(1/n)$.
\end{proof}

The next lemma shows that $\widehat\R^{*}(\bbe,\bga)$ satisfies Assumption (A9) (iv).
\begin{lemma}\label{lem:rri0}
Under Assumptions (A1)-(A8), it holds that for any $\bbe\in\Bc$, any $\bga$ satisfying $\sup_{\bbe\in\Bc_{nT}} n^{-1}\sum_{i=1}^n \one\{g_i\neq g_i^0\}=o_p(T^{-\de})$ and all $\de>0$,
\begin{align*}
\max_{1\leq k,l\leq T}|\{\widehat\R^{*}(\bbe,\bga)-\widehat\R^{*}(\bbe,\bga^0)\}_{kl}|=o_p(T^{-\de}).
\end{align*}
\end{lemma}
\begin{proof}
From Cauchy-Schwarz inequality, we can write 
\begin{align*}
\{&\widehat\R^{*}(\bbe,\bga)-\widehat\R^{*}(\bbe,\bga^0)\}_{kl}\\
=&\frac1n\sum_{i=1}^n\one\{g_i\neq g_i^0\}\Big\{A_{ik}^{-1/2}(\bbe_{g_i})A_{il}^{-1/2}(\bbe_{g_i})\{y_{ik}-m(\x_{ik}^\top\bbe_{g_i})\}\{y_{il}-m(\x_{il}^\top\bbe_{g_i})\}\\
&\ \ \ \ \ \ \ \ \ \ \ \ \ \ \ \ \ \ \ \ \ \ \ \ -A_{ik}^{-1/2}(\bbe_{g_i^0})A_{il}^{-1/2}(\bbe_{g_i^0})\{y_{ik}-m(\x_{ik}^\top\bbe_{g_i^0})\}\{y_{il}-m(\x_{il}^\top\bbe_{g_i^0})\}\Big\}\\
\leq&
\Big(\frac1n\sum_{i=1}^n\one\{g_i\neq g_i^0\}\Big)^{1/2}\\
&\times\Big(\frac1n\sum_{i=1}^n\Big\{A_{ik}^{-1/2}(\bbe_{g_i})A_{il}^{-1/2}(\bbe_{g_i})\{y_{ik}-m(\x_{ik}^\top\bbe_{g_i})\}\{y_{il}-m(\x_{il}^\top\bbe_{g_i})\}\\
&\ \ \ \ \ \ \ \ \ \ \ \ \ \ \ \ \ -A_{ik}^{-1/2}(\bbe_{g_i^0})A_{il}^{-1/2}(\bbe_{g_i^0})\{y_{ik}-m(\x_{ik}^\top\bbe_{g_i^0})\}\{y_{il}-m(\x_{il}^\top\bbe_{g_i^0})\}\Big\}^2\Big)^{1/2}.
\end{align*}
Since we have
\begin{align*}
\frac1n&\sum_{i=1}^n\Big\{A_{ik}^{-1/2}(\bbe_{g_i})A_{il}^{-1/2}(\bbe_{g_i})\{y_{ik}-m(\x_{ik}^\top\bbe_{g_i})\}\{y_{il}-m(\x_{il}^\top\bbe_{g_i})\}\\
&\ \ \ \ \ \ \ -A_{ik}^{-1/2}(\bbe_{g_i^0})A_{il}^{-1/2}(\bbe_{g_i^0})\{y_{ik}-m(\x_{ik}^\top\bbe_{g_i^0})\}\{y_{il}-m(\x_{il}^\top\bbe_{g_i^0})\}\Big\}^2=O_p(1),
\end{align*}
the lemma follows from Lemma \ref{lem:gc}.
\end{proof}

\section{Additional numerical results}\label{sec:num}

\subsection{Details of competing methods in simulation studies}
We here provide details of competing methods used in the simulation study in Section~4. 
\begin{itemize}
\item[-]
(RC; random coefficient model) \ Fit the following logistic random coefficient model: 
$$
y_{it}\sim {\rm Ber}(p_{it}), \ \ \ \ {\rm logit}(p_{it})=\x_{it}^\top \bbe_i, \ \ \ \bbe_i\sim N(\bbe_0, \V).
$$
The model is fitted by using the R package ``lme4" \citep{Bates2016}.

\item[-]
(GMM; growth mixture model) \ Fit the following growth mixture model:
$$
f(y_{it}|\x_{it})=\sum_{\ell=1}^L \pi_\ell {\rm Be}(y_{it}; \x_{it}^\top \bbe_\ell), \ \ \ \ \sum_{\ell=1}^L \pi_\ell=1,
$$
where ${\rm Be}(y_{it}; x_{it}^\top \bbe_\ell)$ denotes the Bernoulli distribution with success probability being $1/\{1+\exp(-\x_{it}^\top \bbe_\ell)\}$, and $L$ is set to the same number of groups used in the GGEE method.  
The model parameters are estimated via an EM algorithm.
The subject-specific estimates of coefficients are given by $\bbeh_i=\sum_{\ell=1}^L \ph_{i\ell}\bbeh_\ell$, where $\ph_{i\ell}$ is the posterior probability that the $i$th subject is classified to the $\ell$th group.

\item[-]
(PWP; pair-wise penalization method) \  Consider the subject-wise logistic regression, $y_{it}\sim {\rm Ber}(p_{it})$ with ${\rm logit}(p_{it})=\x_{it}^\top \bbe_i$, and estimate $\bbe_i$ by maximizing the following objective function: 
$$
\sum_{i=1}^n\sum_{t=1}^T\{y_{it}\log p_{it}+(1-y_{it})\log(1-p_{it})\} - \lambda \sum_{i\sim j}\sum_{k=1}^p |\beta_{ik}-\beta_{jk}|,
$$ 
where $i\sim j$ denotes contingency between $i$th and $j$th subjects and $\lambda$ is a tuning parameter. 
Based on the output of RC, we first computed the pair-wise difference of estimated regression coefficients and obtained a minimum spanning tree over $n$ subjects.
Then, pairs of connected subjects in the minimum spanning tree are regarded as ``adjacent" in the above penalty term.
The above objective function is easily optimized, and $\lambda$ can be selected via cross-validation by using the R package ``glmnet" (Friedman et al., 2010). 
This method can be regarded as an alternative and scalable version of the pair-wise penalization method by \cite{Zhu2018}.     
\end{itemize}

\subsection{Performance of confidence intervals}

We carry out simulation studies to investigate the performance of the Wald-type confidence intervals based on the estimated variance-covariance matrices using the form given in Theorem~2 (plug-in method) and the clustered bootstrap. 
We adopted the same data generating process used in the first simulation study in Section~4. 
We estimate variance-covariance matrices of $\bbe_g$ for $g=1,2,3$, based on the plug-in and clustered bootstrap (with 100 bootstrap samples) methods, and then obtain Wald-type $95\%$ confidence intervals, denoted by ${\rm CI}_{gk}$ for $k=1,\ldots,p$.
The performance of the intervals are evaluated by coverage probability (CP), $(pG)^{-1}\sum_{g=1}^G\sum_{k=1}^p{\rm I}(\beta_{gk}\in {\rm CI}_{gk})$, and average length (AL), $(pG)^{-1}\sum_{g=1}^G\sum_{k=1}^p|{\rm CI}_{gk}|$, which are averaged over 500 Monte Carlo replications. 
The results are shown in Table~\ref{tab:CI}.
It shows that the plug-in method tends to exhibits under-coverage probability when $T$ is small. 
On the other hand, the bootstrap approach produces desirable confidence intervals with coverage probability close to the nominal level and longer interval lengths than those of the plug-in method.

\begin{table}[htbp]
\caption{Coverage probability (CP) and average length (AL) of $95\%$ confidence intervals of group-specific parameters based on the plug-in and clustered bootstrap methods under exchangeable correlation (EX), first-order autoregressive (AR) and unstructured (US) working correlation matrices, averaged over 500 Monte Carlo replications. }
\label{tab:CI}
\begin{center}
\begin{tabular}{ccccccccccccc}
\hline
&&& \multicolumn{3}{c}{Plug-in} && \multicolumn{3}{c}{Bootstrap}\\
$(n, T)$ &  &  & EX & AR & US &  & EX & AR & US \\
\hline
(180, 10) & CP &  & 90.7 & 87.3 & 88.4 &  & 95.3 & 93.8 & 95.3 \\
 & AL &  & 0.67 & 0.66 & 0.65 &  & 0.95 & 1.04 & 0.95 \\
 \hline
(180, 20) & CP &  & 92.9 & 90.4 & 88.0 &  & 95.2 & 94.6 & 96.5 \\
 & AL &  & 0.56 & 0.56 & 0.55 &  & 0.68 & 0.74 & 1.08 \\
\hline
(270, 10) & CP &  & 90.5 & 86.0 & 88.5 &  & 94.7 & 92.4 & 94.5 \\
 & AL &  & 0.55 & 0.54 & 0.54 &  & 0.71 & 0.78 & 0.73 \\
\hline
(270, 20) & CP &  & 93.1 & 91.1 & 89.7 &  & 95.4 & 95.1 & 95.7 \\
 & AL &  & 0.46 & 0.46 & 0.46 &  & 0.54 & 0.60 & 0.66 \\

\hline
\end{tabular}
\end{center}
\end{table}

\subsection{Additional results in Section 5}

\begin{figure}
\centering
\includegraphics[width=10cm,clip]{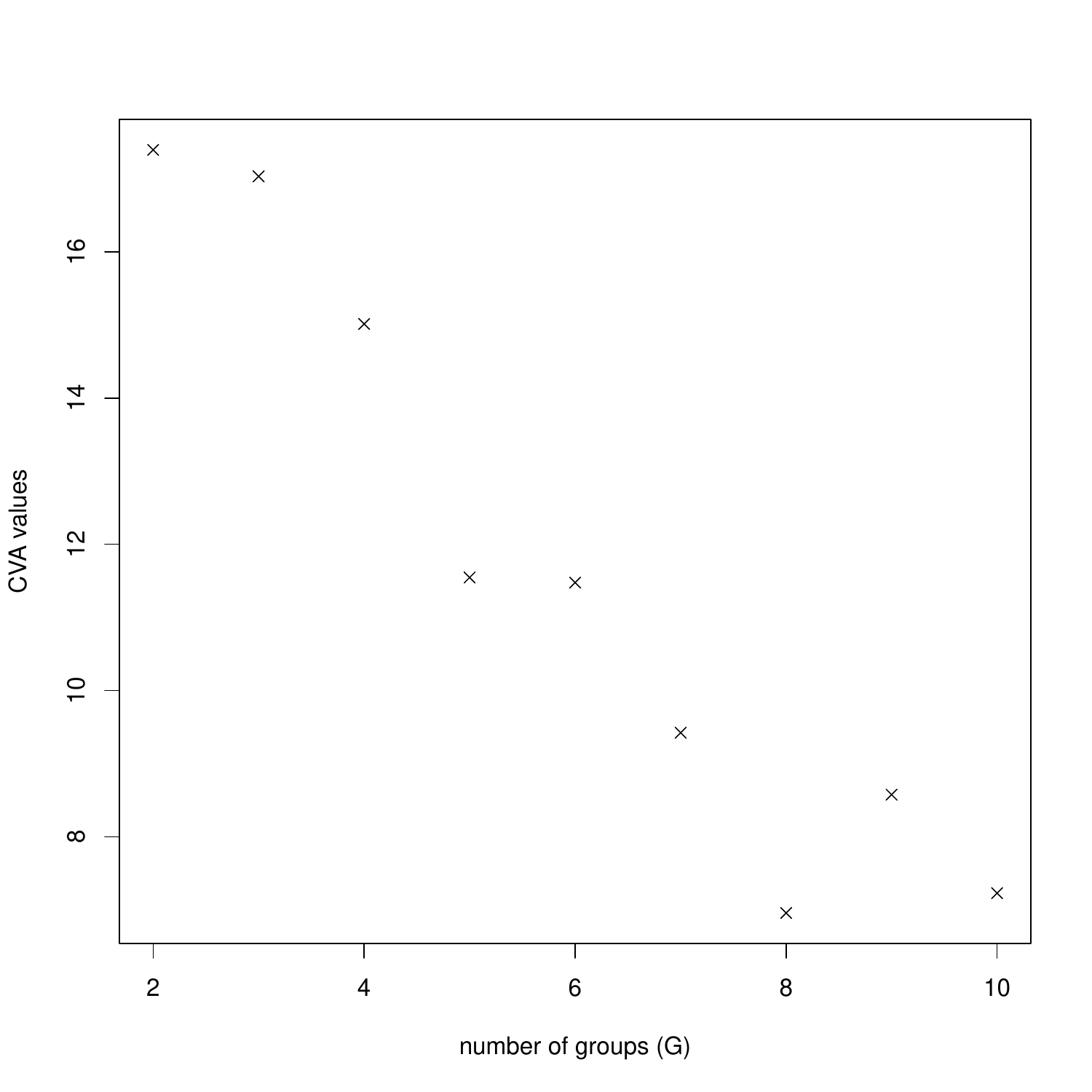}
\caption{The CVA value for each $G$ (the number of groups). }
\label{fig:app-CVA}
\end{figure}

In Figure~\ref{fig:app-CVA}, we provided the CVA values for candidate values of $G$.
It shows that the CVA value basically decreases from $G=2$ and attains the minimum value at $G=8$.

\end{document}